\documentclass[twoside,11pt]{article}

\usepackage{jmlr2e}
\usepackage{amssymb}
\usepackage{amsmath} 
\usepackage{bm}
\usepackage{xcolor}
\usepackage{bbm}
\usepackage{bbold}
\usepackage[T1]{fontenc}
\usepackage{subfigure}
\usepackage[english]{babel}
\usepackage[ruled,vlined]{algorithm2e}
\usepackage{algorithmic}
\usepackage[titletoc]{appendix}
\jmlrheading{ }{ }{ }{ }{ }{Qinghua Liu, Valeria Vitelli, Carlo Mannino, Arnoldo Frigessi and Ida Scheel}

\ShortHeadings{Pseudo-Mallows for Efficient Preference Learning} 
{Liu, Vitelli, Mannino, Frigessi and Scheel}
\firstpageno{1}

\begin{document}

\title{Pseudo-Mallows for Efficient Probabilistic Preference Learning}

\author{\name Qinghua\ Liu \email qinghual@math.uio.no \\
       \addr Department of Mathematics\\
       University of Oslo\\
       Moltke Moes vei 35\\
       Niels Henrik Abels hus\\
       0851 Oslo, Norway
       \AND
       \name Valeria\ Vitelli \email valeria.vitelli@medisin.uio.no \\
       \addr Oslo Centre for Biostatistics and Epidemiology\\
       University of Oslo\\
       Sognsvannsveien 9\\
       0372 Oslo
       \AND
       \name Carlo\ Mannino \email carlo.mannino@sintef.no \\
       \addr Mathematics and Cybernetics\\
       Sintef Digital\\
       Forskningsveien 1\\
       0373 Oslo
       \AND
       \name Arnoldo\ Frigessi \email arnoldo.frigessi@medisin.uio.no \\
       \addr Oslo Centre for Biostatistics and Epidemiology\\
       University of Oslo\\
       Sognsvannsveien 9\\
       0372 Oslo
       \AND
       \name Ida\ Scheel \email idasch@math.uio.no \\
       \addr Department of Mathematics\\
       University of Oslo\\
       Moltke Moes vei 35\\
       Niels Henrik Abels hus\\
       0851 Oslo, Norway
       }

\editor{}

\maketitle

\begin{abstract}%   <- trailing '%' for backward compatibility of .sty file
We propose the Pseudo-Mallows distribution over the set of all permutations of $n$ items, to approximate the posterior distribution with a Mallows likelihood. The Mallows model has been proven to be useful for recommender systems where it can be used to learn personal preferences from highly incomplete data provided by the users. Inference based on MCMC is however slow, preventing its use in real time applications. The Pseudo-Mallows distribution is a product of univariate discrete Mallows-like distributions, constrained to remain in the space of permutations. The quality of the approximation depends on the order of the $n$ items used to determine the factorization sequence.
In a variational setting, we optimise the variational order parameter by minimising a marginalized KL-divergence. We propose an approximate algorithm for this discrete optimization, and conjecture a certain form of the optimal variational order that depends on the data. Empirical evidence and some theory support our conjecture. Sampling from the Pseudo-Mallows distribution allows fast preference learning, compared to alternative MCMC based options, when the data exists in form of partial rankings of the items or of clicking on some items.
Through simulations and a real life data case study, we demonstrate that the Pseudo-Mallows model learns personal preferences very well and makes recommendations much more efficiently, while maintaining similar accuracy compared to the exact Bayesian Mallows model.
\end{abstract}

\begin{keywords}
  {Scalable recommender systems, Variational Bayes for discrete distributions, Bayesian Mallows, Probability measures on permutations} 
\end{keywords}

\section{Introduction and motivation}
Preference learning consists of estimating the unknown permutation of $n$ items that represents the comparative opinion about these items by an individual or user. Usually, the users have expressed only incomplete information about their ranking. For instance, they might have compared some pairs of items to express which is the preferred one. In a digital setting, they might have clicked on some chosen items, showing positive interest for these. At a population level, partial data is available from $N$ users who on the one hand share some level of preference, and on the other hand show individual opinions. The task is to infer on the shared consensus ranking of the items and on each individual complete ranking. Probabilistic preference learning recognises and quantifies the uncertainty of inference, by means of (posterior) distributions over the set $\mathcal{P}_n$ of all permutations of $n$ items. 

Probabilistic preference learning is a key component of a recommender system, whose aim is to present to each user a selection of items based on his/her personal preference. Recommender systems are iterative algorithms, which combine steps of learning of the individual preferences obtained by making suggestions to the users, taking their reactions into account (explore), with steps of exploitation, where the aim is to please the users by suggesting the items they wish the most. Inferring on the unknown consensus and individual distributions over $\mathcal{P}_n$ as done in this work, is an essential part of a recommender system.   

The study of distributions over $\mathcal{P}_n$ started with the work by Diaconis \citep{diaconis1988group}. Recently, \cite{shah20218} collected several approaches to build probability distributions over permutations. \cite{liu2019diverse} present additional options, in particular the Mallows model \citep{mallows1957non}. The difficulty of course originates from the dimension of $\mathcal{P}_n$ for larger $n$. In inference, the incompleteness of the data means that appropriate conditional distributions over $\mathcal{P}_n$ need to be constructed. A typical case for a recommender system could have hundreds of items, ten-thousands or even more users, each expressing preferences by clicking/liking some few items. 

The use of the Mallows model in a Bayesian setting for probabilistic preference learning was recently proposed in \cite{vitelli2017probabilistic}. The Mallows is an exponential family distribution, which assumes that the latent individual permutations of the items, which we call rankings, are centered around a group consensus $\bm{\rho} \in \mathcal{P}_n$. A distance between permutations measures the closeness between each user's ranking and $\bm{\rho}$. The dispersion of the users' rankings from the consensus, their individual inclination, is controlled by a scale parameter $\alpha$. Several right invariant distances are considered in \cite{vitelli2017probabilistic}, among which the $l_1$-norm, called foot-rule, and the $l_2$-norm, called the Spearman distance. Given a ranking of some of the $n$ items (partial ranking), a set of pair comparisons or a collection of clicked items \citep{liu2019diverse}, Bayesian inference can be performed by augmentation and Markov Chain Monte Carlo (MCMC). These papers demonstrate that accurate and diverse personalized recommendations can be made, with interpretable and well calibrated uncertainty estimation. 
However, both \cite{vitelli2017probabilistic} and \cite{liu2019diverse} use MCMC for inference, which shows slow convergence for large $n$ and $N$. The discrete posterior distribution is also highly multimodal.
Hence the Bayesian Mallows approach requires a fast alternative to MCMC to scale to industrial cases. In this paper we use variational inference to approximate the posterior probability densities \citep{blei2017variational, zhang2018advances}.  

We denote the vector of data as $\mathbf{R}$, the vector of model parameters as $\bm{\rho}$ and the target posterior distribution as $p(\bm{\rho}|\mathbf{R})$. The idea behind variational inference is to approximate this posterior with a simpler distribution  $q(\bm{\rho})$ belonging to a family of distributions $\mathcal{Q}$. The task of determining the $q(\bm{\rho})$ that approximates $p(\bm{\rho}|\mathbf{R})$ best is based on a choice of distance measure between the two distributions. The  Kullback-Leibler (KL) divergence \citep{kullback1951information} between $p(\bm{\rho}|\mathbf{R})$ and $q(\bm{\rho})$ is often used for this purpose. The optimal variational approximation $q^*(\bm{\rho})$  minimises the  KL divergence

\begin{equation}
	q^*(\bm{\rho}) =  \operatorname*{argmin}\limits_{q(\bm{\rho}) \in \mathcal{Q}} KL(q(\bm{\rho})||p(\bm{\rho}|R)).
\end{equation}
 When the best distribution $q^*(\bm{\rho})$ is found, it is used to replace the original posterior distribution $p(\bm{\rho}|\mathbf{R})$ for inference. Usually $\mathcal{Q}$ is a family of parametrized distributions, and its parameters, which are then optimized upon, are called variational parameters. 
Typically, instead of minimizing the KL divergence directly, it is more convenient to maximize the evidence lower bound (ELBO) \citep{hoffman2016elbo}. Algorithms that are commonly used to optimize the ELBO are the Coordinate Ascent Variational Inference (CAVI) algorithm \citep{bishop2006pattern} and the Stochastic Variational Inference (SVI) algorithm \citep{hoffman2013stochastic}.
These algorithms typically converge fast, and variational inference has therefore been a popular alternative to MCMC for Bayesian inference \citep{salimans2015markov}. However, variational inference is difficult for discrete distributions, like the Mallows over $\mathcal{P}_n$, and the most recent developments of variational inference have focused on continuous, exponential family distributions \citep{hensman2012fast, wainwright2008graphical}. Furthermore, variational inference is often built upon the so-called mean-field assumption, i.e., the assumption that $q(\bm{\rho})$ factors into univariate distributions in the components of the vector $\bm{\rho}$ \citep{xing2012generalized}. The mean-field assumption makes it difficult to capture the dependencies between the various parameters \citep{wang2005inadequacy}, an aspect that is essential when $\bm{\rho} \in \mathcal{P}_n$. \cite{bouchard2010variational} proposed a framework that extends variational inference to some classes of discrete problems such as matchings in the combinatorial space, yet, the application of variational inference to approximate other discrete distributions remains a challenge. 

This paper contributes as follows to the current literature. We propose the Pseudo-Mallows distribution as an efficient variational approximation to the Bayesian Mallows model's posterior distribution, without following the mean-field construction. Our proposed distribution allows sampling in the space of permutations, conditional on incomplete data. The Pseudo-Mallows distribution inherits most favorable traits of the Bayesian Mallows model, but at the same time, drawing independent samples from it is straight-forward and efficient, without a need for MCMC sampling. We propose a marginalised version of the KL-divergence and derive its equivalent ELBO. The variational parameter of the Pseudo-Mallows distribution is a permutation itself. We conjecture the optimal form of this variational permutation, which depends on the data. We support our conjecture i) by empirical evidence obtained in small dimensions, ii) through a new optimization algorithm that minimises the marginalised KL-divergence by transforming the problem into a bipartite matching problem, and iii) by theory valid for some special cases.
The form of the optimal variational permutation is interesting and not immediately intuitive. With the optimal variational permutation, we develop the Pseudo-Mallows distribution to enable learning from incomplete individual rankings as well as from clicking data. Through extensive simulations and a case study based on real-life preference data, we show that compared to MCMC, our proposed Pseudo-Mallows method is highly efficient so that accurate inference and personalized recommendations can be made in a timely manner.  To our knowledge, our proposed method is the first to develop variational inference for a discrete complex distribution over the set of permutation.

The rest of this paper is organized as follows: in Section \ref{sec:pseudo_full}, we give a brief introduction of the Mallows ranking model for full rankings, and of the Bayesian Mallows model introduced by \cite{vitelli2017probabilistic}. It is followed by a detailed explanation of the Pseudo-Mallows method for full rankings, wherein we introduce the Pseudo-Mallows distribution, our conjecture of the optimal variational parameter that achieves the best approximation to the Bayesian Mallows model, and a sampling scheme for inference. We also include the intuition of the optimal variational parameter as well as an empirical study to support it. In Section \ref{sec:clicking_data}, we introduce the extension of the Pseudo-Mallows method to learn individual rankings from partial rankings and make personalized recommendations from clicking data. The methodology is then supported by detailed simulations and case studies in Section \ref{sec:simulation_case_study}. In particular, in Section \ref{sec: full_sim}, we demonstrate the Pseudo-Mallows' ability to make fast inference from full ranking data, and Section \ref{sec:sim_clicking} contains a simulation study where we compare the Pseudo-Mallows with the Bayesian Mallows in terms of their abilities to learn individual rankings and make personalized recommendations from clicking data. In Section \ref{sec:NRK}, we test the Pseudo-Mallows on a real life clicking dataset to showcase its ability to make accurate and fast personalized recommendations given a time limitation. Finally, we collect some concluding comments in Section \ref{sec:summary}.

\section{Pseudo-Mallows: a fast approximation for the Bayesian Mallows posterior distribution}\label{sec:pseudo_full}
\subsection{The Mallows ranking model} \label{sec:Mallows_full}
Suppose we have a group of $N$ users, and each user has given a full ranking of $n$ items. We denote the ranking of user $j$ as $\mathbf{R}^j$ = \{$R^j_1, ..., R^j_n$\}, where $R^j_i = k$ indicates that user $j$ gives to item $i$ rank $k$, and hence $R^j_i \in \{1, ..., n\}$, $\forall i = 1, ..., n$. For a group of homogeneous users that share their preferences to some degree, the Mallows model assumes that their rankings are distributed around a consensus parameter $\bm{\rho}$, with a scale parameter $\alpha$. Assuming that all users' rankings are mutually independent given $\bm{\rho}$ and $\alpha$, the Mallows likelihood is given by
\begin{equation}\label{eq:Mallows_likelihood}
	P(\mathbf{R}^1, ..., \mathbf{R}^N|\alpha, \bm{\rho}) = \prod\limits_{j=1}^{N} \frac{\text{exp}\{- 
		\frac{\alpha}{n}\bm{d}(\mathbf{R}^j, \bm{\rho})\}}{Z_n(\alpha)} = \frac{\text{exp}\{- \frac{\alpha}{n}\sum\limits_{j=1}^{N}\bm{d}(\mathbf{R}^j, \bm{\rho})\}}{Z_n(\alpha)^N},
\end{equation}
where $\bm{d}( ,)$ is a right-invariant distance \cite{diaconis1988group} between two rankings, and $Z_n(\alpha)$ is the normalizing constant, which does not depend on $\bm{\rho}$ when the distance is right-invariant. We use the right-invariant foot-rule distance, defined as $\bm{d}(\mathbf{R}^j, \bm{\rho}) = \sum\limits_{i=1}^{n} d(\mathbf{R}^j_i, \rho_i) = \sum\limits_{i=1}^{n} |\mathbf{R}^j_i- \rho_i|$. The parameter of interest for inference is the consensus $\bm{\rho}$. Given $\alpha$, the Mallows posterior distribution of the consensus parameter is
\begin{equation}\label{eq:MallowsPosterior}
	P(\bm{\rho}|\alpha, \mathbf{R}^1, ..., \mathbf{R}^N) = \frac{P(\mathbf{R}^1, ..., \mathbf{R}^N|\alpha, \bm{\rho})\pi (\bm{\rho})}{\sum\limits_{\bm{r}\in \mathcal{P}_n}P(\mathbf{R}^1, ..., \mathbf{R}^N|\alpha, \bm{r})\pi (\bm{r})},
\end{equation} where $\mathcal{P}_n$ is the space of permutation of $n$ items, and $\pi(\bm{\rho})$ is a prior distribution for $\bm{\rho}$. For simplicity, we assume a non-informative uniform prior for $\pi (\bm{\rho})$, i.e., $\pi (\bm{\rho}) = \frac{1}{n!}$. Then, 
\begin{equation}\label{eq:MallowsPost}
P(\bm{\rho}|\alpha, \mathbf{R}^1, \ldots, \mathbf{R}^N) = \frac{\exp\left\{-\frac{\alpha}{n} \sum_{j=1}^N d(\mathbf{R}^j,\bm{\rho})\right\}}{Z_n(\alpha,\mathbf{R}^1, \ldots, \mathbf{R}^N)},
\end{equation} where 
\begin{equation}\label{eq:MallowsZ}
Z_n(\alpha,\mathbf{R}^1, \ldots, \mathbf{R}^N) = \sum_{\mathbf{r}\in\mathcal{P}_n} \exp\left\{ -\frac{\alpha}{n} \sum_{j=1}^N d(\mathbf{R}^j,\mathbf{r})\right\}.
\end{equation}
For simplicity, we first assume that $\alpha$ fixed; we will later remove this assumption. 

Markov Chain Monte Carlo (MCMC) can be used to make inference on $\bm{\rho}$ as in \cite{vitelli2017probabilistic}. Summary statistics such as the Maximum A Posteriori (MAP) estimators can be computed based on the MCMC samples. The MCMC scheme proposed by \cite{vitelli2017probabilistic} can perform well when there are not many items and users. However, the MCMC suffers from slow convergence. In order to achieve good mixing and precise inferential results, the MCMC chain needs to be run for very long. Because the posterior distribution has multiple peaks, the MCMC is ``sticky''  \citep{gilks1996strategies}, i.e., the chain gets easily stuck at a local optima. The ``stickiness'' problem is pronounced because the $\bm{\rho}$ parameter is discrete.

\subsection{The Pseudo-Mallows Distribution}

With the aim to resolve the above limitations, we introduce a new probability distribution on the space of permutations to approximate the Mallows posterior distribution (\ref{eq:MallowsPosterior}). In order to obtain independent samples of $\bm{\rho}$ conveniently, our new distribution is a product of $n$ conditional distributions, which can be easily sampled from. We name this distribution the Pseudo-Mallows distribution.

We introduce the notation $\{i_1, ..., i_n\} \in \mathcal{P}_n$ to indicate a ranking of the $n$ integers, so that $i_k$ is the rank of the integer item $k$. Vice versa, we use the notation $\bm{o} = \{o_1, ..., o_n\} \in \mathcal{P}_n$ to indicate an ordering, so that $o_m$ is the integer item that has rank $m$. It holds that $i_{o_k} = k$. 

The Pseudo-Mallows distribution is defined as follows:

\begin{equation}\label{eq:pseudoMallows}
	Q(\bm{\rho}|\alpha, \mathbf{R}^1, ..., \mathbf{R}^N) = \sum\limits_{\{i_1, ..., i_n\}\in \mathcal{P}_n}q(\bm{\rho}|i_1, ..., i_n,\alpha, \mathbf{R}^1, ..., \mathbf{R}^N)g(i_1, ..., i_n),
\end{equation} where

\begin{equation}\label{eq:factorization}
\begin{aligned}
    q({\bm{\rho}}|i_1, ..., i_n, \alpha, \mathbf{R}^1, ..., \mathbf{R}^N)= & q({\bm{\rho}}|o_1, ..., o_n, \alpha, \mathbf{R}^1, ..., \mathbf{R}^N)  \\
	=& q({\rho}_{o_1}|\alpha,o_1, R^1_{o_1},...,R^N_{o_1}) \cdot
	q({\rho}_{o_2}|\alpha,o_2,{\rho}_{o_1} R^1_{o_2},...,R^N_{o_2}) \cdot
	...  \\
	&\cdot q({\rho}_{o_{n-1}}|\alpha,o_{n-1}, {\rho}_{o_1},...,{\rho}_{o_{n-2}}, R^1_{n-1},...,R^N_{n-1}) \\
	&\cdot q({\rho}_{o_{n}}|\alpha,o_{n}, {\rho}_{o_1},...,{\rho}_{o_{n-1}}, R^1_n, ..., R^N_n).
\end{aligned}
\end{equation}

The first and the $k$-th factors have the form
\begin{equation}\label{eq:factorization_with_denom}
	\begin{gathered}
	q({\rho}_{o_1}|\alpha, o_1,R^1_{o_1}, ...,R^N_{o_1}) 
	= \frac{\text{exp}\{- \frac{\alpha}{n}\sum\limits_{j=1}^{N}d(R^j_{o_1}, {\rho}_{o_1})\}}
	{\sum\limits_{{r}_{o_1}\in \{1, .., n\}}\text{exp}\{- \frac{\alpha}{n}\sum\limits_{j=1}^{N}d(R^j_{o_1}, {r}_{o_1})\}}\mathbb{1}_{ {\rho}_{o_1}\in \{1, ...,n\}} \\
	{q({\rho}_{o_k}|\alpha, o_k, {\rho}_{o_1}, ..., {\rho}_{o_{k-1}},R^1_{o_k}, ...,R^N_{o_k}) }
	= \frac{\text{exp}\{- \frac{\alpha}{n}\sum\limits_{j=1}^{N}d(R^j_{o_k}, {\rho}_{o_k})\}}
	{\sum\limits_{{r}_{o_k}\in \{1, .., n\}\setminus \{{\rho}_{o_1}, ..., {\rho}_{o_{k-1}}\}}\text{exp}\{- \frac{\alpha}{n}\sum\limits_{j=1}^{N}d(R^j_{o_k}, {r}_{o_k})\}}\\\cdot\mathbb{1}_{{\rho}_{o_k}\in\{1, .., n\}\setminus \{{\rho}_{o_1}, ..., {\rho}_{o_{k-1}}\}}, \text{ for }k = 2, ..., n
\end{gathered}
\end{equation}
Equation (\ref{eq:factorization}) is a factorization in $n$ terms, each being a univariate distribution of one consensus ranking item. Within each term, on the numerator, we measure the distance between $\rho_{o_k}$ and the rankings of item $o_k$ given by the users. In the denominator, we sum over all possible values that $\rho_{o_k}$ can still take. The scale parameter $\alpha$ plays a similar role as in the Mallows distribution controlling the ``peakness'' of the distribution. Take note that the product of all terms on the numerator is effectively identical to the numerator of the Mallows distribution, which makes it plausible that the Pseudo-Mallows distribution shares similar characteristics of the Mallows distribution. 

In order to ensure that this is a distribution over $\mathcal{P}_n$, for the $k$-th term in the factorization, the values taken by  all previous $k-1$ terms need to be excluded. This implies that the sequence of the terms matter.  We use the ranking $\{i_1, ..., i_n\}$, or its corresponded ordering $\{o_1, ..., o_n\}$ to indicate this sequence. We assume the ranking $\{i_1, ..., i_n\}$ used in the factorization follows an arbitrary distribution $g(i_1, ..., i_n)$, defined on $\mathcal{P}_n$. This can be a $\delta$-distribution with mass 1 on one single permutation, or the uniform distribution or any other distribution. By construction using the multiplication rule of distributions and appropriate conditional independence,  the Pseudo-Mallows is a proper distribution on $\mathcal{P}_n$. This completes the definition of the Pseudo-Mallows distribution. 

For convenience, we introduce the notation $Z^{PM}_n(\bm{\rho},\alpha,\mathbf{R}_1, \ldots, \mathbf{R}_N, \mathbf{o})$ to indicate the product of all denominators in (\ref{eq:factorization_with_denom}), i.e.,

\begin{equation}\label{eq:Z_PM}
    \begin{split}
    Z^{PM}_n(\bm{\rho},\alpha,\mathbf{R}_1, \ldots, \mathbf{R}_N, \mathbf{o})   = 
    & \Bigg({\sum\limits_{r=1}^n \text{exp}\{- \frac{\alpha}{n} \sum\limits_{j=1}^N |R^j_{o_1} - r|}\} \Bigg) 
	\cdot \Bigg({\sum\limits_{r\in \{1, ..., n\}\setminus \rho_{o_1}} \text{exp}\{- \frac{\alpha}{n} \sum\limits_{j=1}^N |R^j_{o_2} - r|}\} \Bigg) \\
	& \cdot  \Bigg( {\sum\limits_{{r}_{o_n}\in \{1, .., n\}\setminus \{{\rho}_{o_1}, ..., {\rho}_{o_{n-1}}\}}\text{exp}\{- \frac{\alpha}{n}\sum\limits_{j=1}^{N}|R^j_{o_n} - r|\}} \Bigg)
    \end{split}.
\end{equation}
We can then write (\ref{eq:factorization}) to be:
\begin{equation}\label{eq:PLpost}
q(\bm{\rho}|\alpha, \mathbf{R}_1, \ldots, \mathbf{R}_N, \mathbf{o}) = \frac{\exp\left\{-\frac{\alpha}{n} \sum\limits_{j=1}^N d(\mathbf{R}_j,\bm{\rho})\right\}}{Z^{PM}_n(\bm{\rho},\alpha,\mathbf{R}_1, \ldots, \mathbf{R}_N, \mathbf{o})}.
\end{equation}

Given a ranking $\{i_1, ..., i_n\}$, or equivalently an ordering $\{o_1, ..., o_n\}$, as well as $\alpha = \alpha^0$ and the full ranking data $\{\mathbf{R}^1, ..., \mathbf{R}^N\} $, we obtain samples from the Pseudo-Mallows distribution of $\bm{\rho}$ by using Algorithm \ref{algo:full_data_given_ordering}.

\subsection{Derivation of the Evidence Lower Bound (ELBO)}
To obtain the best approximation to the Mallows posterior, we sought for the distribution $g(i_1, ..., i_n)$ that minimizes the Kullback-Leibler (KL) divergence between the Mallows posterior (\ref{eq:MallowsPosterior}) and the Pseudo-Mallows distribution (\ref{eq:pseudoMallows}). For simplicity, we first consider the special case where $g(i_1, ...,i_n)$ is a $\delta$-distribution with mass 1 on one ranking $\{i_1, ..., i_n\}$, which translates this optimization problem to searching for a ranking $\{i_1, ...,i_n\} \in \mathcal{P}_n$ that minimizes the KL-divergence. We write the objective function as:
\begin{equation}\label{eq:KL_full_data}
	 \operatorname*{argmin}\limits_{\{i_i, ..., i_n\}\in \mathcal{P}_n} KL\Big( q ({\bm{\rho}}|\alpha, \mathbf{R}^1,..., \mathbf{R}^N, i_1,...,i_n)||P(\bm{\rho}|\alpha, \mathbf{R}^1, ...,\mathbf{R}^N) \Big),
\end{equation}

or equivalently,
\begin{equation}\label{eq:KL}
\operatorname*{argmin}\limits_{\{o_i, ..., o_n\}\in \mathcal{P}_n} KL \left(  q(\bm{\rho}|\alpha,\mathbf{R}^1, \ldots, \mathbf{R}^N, o_1, ..., o_n)||P(\bm{\rho}|\alpha, \mathbf{R}^1, \ldots, \mathbf{R}^N)\right).
\end{equation}

As usual, instead of minimizing the KL divergence, it is convenient to maximize the {Evidence Lower Bound (ELBO)} $\mathcal{L},$ defined as the lower bound on the log marginal probability of the data \citep{zhang2018advances}. Let us now derive the ELBO expression for the Pseudo-Mallows approximation of the Mallows posterior distribution.

We can write the KL divergence in equation (\ref{eq:KL}) as
$$
KL \left(q^\mathbf{o}(\bm{\rho}|\text{data}) || P(\bm{\rho}|\text{data}) \right) = \sum_{\bm{\rho}\in\mathcal{P}_n}\left[ \log \left( \frac{q^\mathbf{o}(\bm{\rho}|\text{data})}{P(\bm{\rho}|\text{data})}\right) \cdot q^\mathbf{o}(\bm{\rho}|\text{data})\right]
$$
where for the ease of notation we simply write $P(\bm{\rho}|\text{data})$ for $P(\bm{\rho}|\alpha, \mathbf{R}^1, \ldots, \mathbf{R}^N)$, and $q^\mathbf{o}(\bm{\rho}|\text{data})$ for $q(\bm{\rho}|\alpha,\mathbf{R}^1, \ldots, \mathbf{R}^N, \mathbf{o})$. Using (\ref{eq:MallowsPost}) and (\ref{eq:PLpost}), this translates to
\begin{eqnarray*}
KL \left(q^\mathbf{o}(\bm{\rho}|\text{data}) \right || P(\bm{\rho}|\text{data}))
& = & \sum\limits_{\bm{\rho}\in\mathcal{P}_n}\left[ \log \left( \frac{\exp\left\{-\frac{\alpha}{n} \sum\limits_{j=1}^N d(\mathbf{R}^j,\bm{\rho})\right\}}{Z^{PL}_n(\bm{\rho},\alpha,\mathbf{R}^1, \ldots, \mathbf{R}^N, \mathbf{o})}\cdot \frac{Z_n(\alpha,\mathbf{R}^1, \ldots, \mathbf{R}^N)}{\exp\left\{-\frac{\alpha}{n} \sum\limits_{j=1}^N d(\mathbf{R}^j,\bm{\rho})\right\}} \right) \right. \\
&  & \left.\cdot\frac{\exp\left\{-\frac{\alpha}{n} \sum\limits_{j=1}^N d(\mathbf{R}^j,\bm{\rho})\right\}}{Z^{PL}_n(\bm{\rho},\alpha,\mathbf{R}^1, \ldots, \mathbf{R}^N, \mathbf{o})} \right].
\end{eqnarray*}

This expression can be re-written as
\begin{eqnarray*}
KL \left( q^\mathbf{o}(\bm{\rho}|\text{data}) \right || P(\bm{\rho}|\text{data}))
& = &  \sum_{\bm{\rho}\in\mathcal{P}_n}\left[\log(Z_n(\alpha,\mathbf{R}^1, \ldots, \mathbf{R}^N))\cdot \frac{\exp\left\{-\frac{\alpha}{n} \sum\limits_{j=1}^N d(\mathbf{R}^j,\bm{\rho})\right\}}{Z^{PL}_n(\bm{\rho},\alpha,\mathbf{R}^1, \ldots, \mathbf{R}^N, \mathbf{o})} \right] +\\
& & - \sum_{\bm{\rho}\in\mathcal{P}_n}\left[ \exp\left\{-\frac{\alpha}{n} \sum\limits_{j=1}^N d(\mathbf{R}^j,\bm{\rho})\right\}\cdot\frac{\log(Z^{PL}_n(\bm{\rho},\alpha,\mathbf{R}^1, \ldots, \mathbf{R}^N, \mathbf{o}))}{Z^{PL}_n(\bm{\rho},\alpha,\mathbf{R}^1, \ldots, \mathbf{R}^N, \mathbf{o})} \right].
\end{eqnarray*}

We observe that the term $\log(Z_n(\alpha,\mathbf{R}_1, \ldots, \mathbf{R}_N))$ does not depend on $\bm{\rho}$ thanks to right-invariance of the distance. Therefore it can be taken out of the summation, leaving us with a summation of the Pseudo-Mallows posterior over all possible permutations in $\mathcal{P}_n,$ which equals 1.

We thus obtain the ELBO as

\begin{equation}\label{eq:ELBO}
\text{ELBO} = \sum_{\bm{\rho}\in\mathcal{P}_n}\left[ \exp\left\{-\frac{\alpha}{n} \sum_{j=1}^N d(\mathbf{R}^j,\bm{\rho})\right\}\cdot\frac{\log(Z^{PM}_n(\bm{\rho},\alpha,\mathbf{R}^1, \ldots, \mathbf{R}^N, \mathbf{o}))}{Z^{PM}_n(\bm{\rho},\alpha,\mathbf{R}^1, \ldots, \mathbf{R}^N, \mathbf{o})} \right],
\end{equation}
which we need to maximize in $\bm{o}\in \mathcal{P}_n$ to minimize (\ref{eq:KL}).

\subsection{Optimal construction for the Pseudo-Mallows distribution}
We define the ``V''-rankings and the $\mathcal{V}$-set of a permutation $\bm{r}$.

\begin{mydef}\label{def:V}{
		Fix a permutation $\bm{r} \in \mathcal{P}_n$. Define $o_i$ such that $r_{o_i} = i$, for $i = 1, ..., n$. Let $n = 2m - 1$ if $n$ is odd, and $n = 2m$ if $n$ is even. $\mathcal{V}_{\bm{r}}$ is the set of permutations determined by $\bm{r}$ in the following way:
		
		If $n$ is odd: $\mathcal{V}_{\bm{r}} = \{\bm{v}\in \mathcal{P}_n: v_{o_m} = 1; (v_{o_{m-k}},v_{o_{m+k}}) = (2k, 2k+1) $ or $ (2k+1, 2k)$, $k = 1, ..., m-1\}$. \\
		If $n$ is even,  $\mathcal{V}_{\bm{r}} = \{\bm{v}\in \mathcal{P}_n:(v_{o_{m-k}},v_{o_{m+k+1}}) = (2k+1, 2k+2) $ or $ (2k+2, 2k+1)$, $k = 0, ..., m-1\}$.
		
		We call $\mathcal{V}_{\bm{r}}$ the ``$\mathcal{V}$''-set of $\bm{r}$, and any permutation that belongs to the $\mathcal{V}$-set of $\bm{r}$ is called a ``V''-ranking of $\mathcal{V}_{\bm{r}}$.
	}
\end{mydef}

Given a permutation of $n$ items $\bm{r}$, the $\mathcal{V}$-set of $\bm{r}$ is the set of rankings such that the middle ranked item in $\bm{r}$ are top-ranked, and the top- and bottom-ranked items in $\bm{r}$ are bottom-ranked, for $n$ odd.
When $\bm{o}$ is the ordering corresponding to $\bm{r}$, we define $\mathcal{V}_{\bm{o}}$ as the set of orderings corresponding to the permutations in $\mathcal{V}_{\bm{r}}$.
With this definition we propose the optimal construction of the Pseudo-Mallows. 

\paragraph{Optimal Pseudo-Mallows Conjecture}\label{conjecture}
Assume $\mathbf{R}^j \sim$ Mallows ($\bm{\rho}^0, \alpha$), $j = 1, ...., N$. For any V-ranking $\{i_1, ..., i_n\}^* \in \mathcal{V}_{\bm{\rho}^0}$, or equivalently, any ordering $\bm{o}^* \in \mathcal{V}_{\bm{o}^0}$ where $\bm{o}^0$ is the ordering of $\bm{\rho}^0$, we conjecture that \\

\resizebox{\textwidth}{!}{
$ \bm{o}^* = \operatorname*{argmax}\limits_{\bm{o} \in \mathcal{P}_n} \lim\limits_{N \rightarrow \infty}\sum\limits_{\bm{\rho}\in\mathcal{P}_n}\left[ \exp\left\{-\frac{\alpha}{n} \sum_{j=1}^N d(\mathbf{R}^j,\bm{\rho})\right\}\cdot\frac{\log(Z^{PM}_n(\bm{\rho},\alpha,\mathbf{R}^1, \ldots, \mathbf{R}^N, \mathbf{o}))}{Z^{PM}_n(\bm{\rho},\alpha,\mathbf{R}^1, \ldots, \mathbf{R}^N, \mathbf{o})} \right].$ }

In other words, we first compute the $\mathcal{V}$-set based on $\bm{o}^0$. We conjecture that any ordering that belongs to $\mathcal{V}_{\bm{o}^0}$ is the optimal ordering that maximizes the ELBO, as described in Equation (\ref{eq:ELBO}), when $N$ diverges. This also means that under the simplification that $g(i_1, ..., i_n)$ in (\ref{eq:pseudoMallows}) is a $\delta$ distribution, the optimal choice for $g(i_1, ..., i_n)$ is a $\delta$ distribution with its density concentrated uniformly on all V-ranking that belong to $\mathcal{V}_{\bm{\rho}^0}$.

\subsubsection{Intuition of the Optimal Pseudo-Mallows Conjecture}
The reason why the ``V''- rankings are the best choice for the way to factorize the Pseudo-Mallows distribution is because this makes the Pseudo-Mallows distribution the most similar to the Mallows posterior distribution, hence maximize the ELBO, when the data comes from a  Mallows distribution. For two distributions to be identical, it is necessary that their modes coincide. Assuming this data generating process, i.e., $\mathbf{R}^j \sim$ Mallows ($\bm{\rho}^0, \alpha$), $j = 1, ..., N$, the mode of the Mallows posterior is $\bm{\rho}^0$. In order to make the Pseudo-Mallows similar to the Mallows posterior, we want the mode of the Pseudo-Mallows distribution to be as close to $\bm{\rho}^0$ as possible.

Recall that the $k$-th term in the Pseudo-Mallows distribution, i.e.,
\begin{center}
    ${q({\rho}_{o_k}|\alpha, o_k, {\rho}_{o_1}, ..., {\rho}_{o_{k-1}},R^1_{o_k}, ...,R^N_{o_k}) }$
\end{center} has the form (\ref{eq:factorization_with_denom}). The mode of this is the value $l^*$ that minimizes the distances in the numerator, i.e.,
\begin{equation}\label{eq:pseudo_mode}
   l^* = \operatorname*{argmin}\limits_{l\in \{1, .., n\}\setminus \{{\rho}_{o_1}, ..., {\rho}_{o_{k-1}}\}}\sum\limits_{j=1}^N d(R^j_{o_k}, l)\}. 
\end{equation}

We simplify the above minimization by considering the expected distance with respect to the Mallows distribution, i.e., $l^* = \operatorname*{argmin}\mathbb{E}[\sum\limits_{j=1}^N d(R^j_{o_k}, l)\}]$. For the $k$-th term in~\eqref{eq:factorization_with_denom}, this expected distance is minimized by the median of $P_{Mallows}(R_{o_k}|\alpha, \bm{\rho}^0)$, i.e., the marginal Mallows distribution of $R_{o_k}$, or more formally, 
\begin{equation}\label{eq:pseudo_mode_median}
   l^* = \operatorname*{argmin}\limits_{l\in \{1, .., n\}\setminus \{{\rho}_{o_1}, ..., {\rho}_{o_{k-1}}\}}\mathbb{E}[\sum\limits_{j=1}^N d(R^j_{o_k}, l)\}] = \text{median}\Big(P_{Mallows}(R_{o_k}|\alpha, \bm{\rho}^0) \mathbb{1}_{R_{o_k}\neq \rho_{o_1}, ..., \rho_{o_{k-1}}} \Big),
\end{equation}
as proven in Appendix \ref{sec:proof_l_1}, and hence we want $l^*$ to be similar to $\rho^0_{o_k}$.

Let us first consider the first term $k=1$ in (\ref{eq:factorization_with_denom}). Recall that item $o^0_k$ is the item with rank $k$ in $\bm{\rho}^0$. For the item with middle rank in $\bm{\rho}^0$, i.e., item $o^0_m$, its ranking $R_{o^0_m}$ is distributed symmetrically about $m$, therefore, the median of $R_{o^0_m}$ is equal to its expected value, which is exactly $m$. Detailed proof of this symmetry and expected value is shown in Appendix \ref{sec:one}. Based on this property, by choosing item $o^0_m$ as the first factor in (\ref{eq:factorization_with_denom}), we ensure that the mode of $\rho_{o^0_m}$ coincides with the wished target $m$. 

\begin{figure}[hbt!]
		\begin{minipage}[t]{.3\textwidth}
			\centering
			\includegraphics[width=\textwidth]{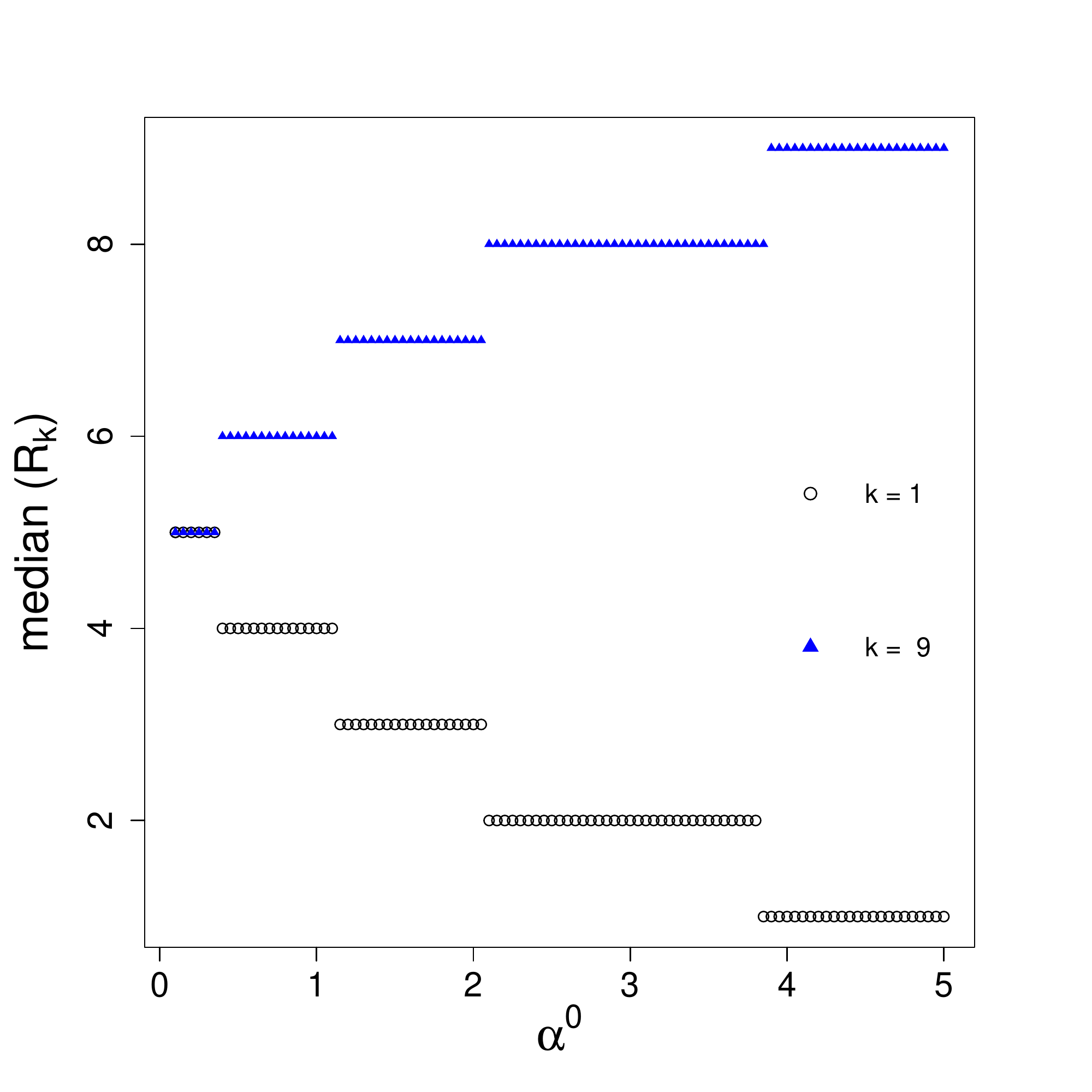}
		\end{minipage}
		\hfill
		\begin{minipage}[t]{.3\textwidth}
			\centering
			\includegraphics[width=\textwidth]{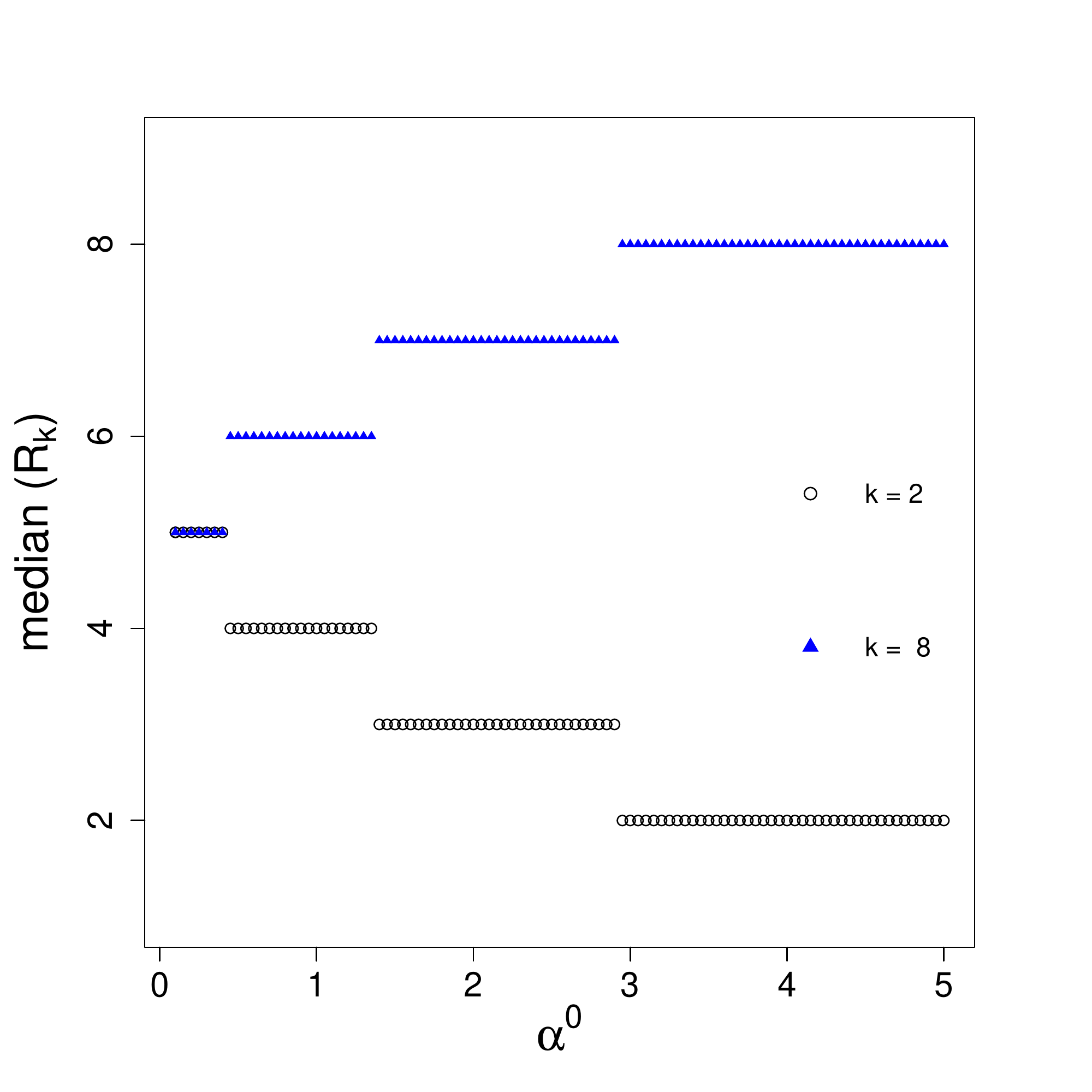}
			\end{minipage} 
		\hfill
		\begin{minipage}[t]{.3\textwidth}
			\centering
			\includegraphics[width=\textwidth]{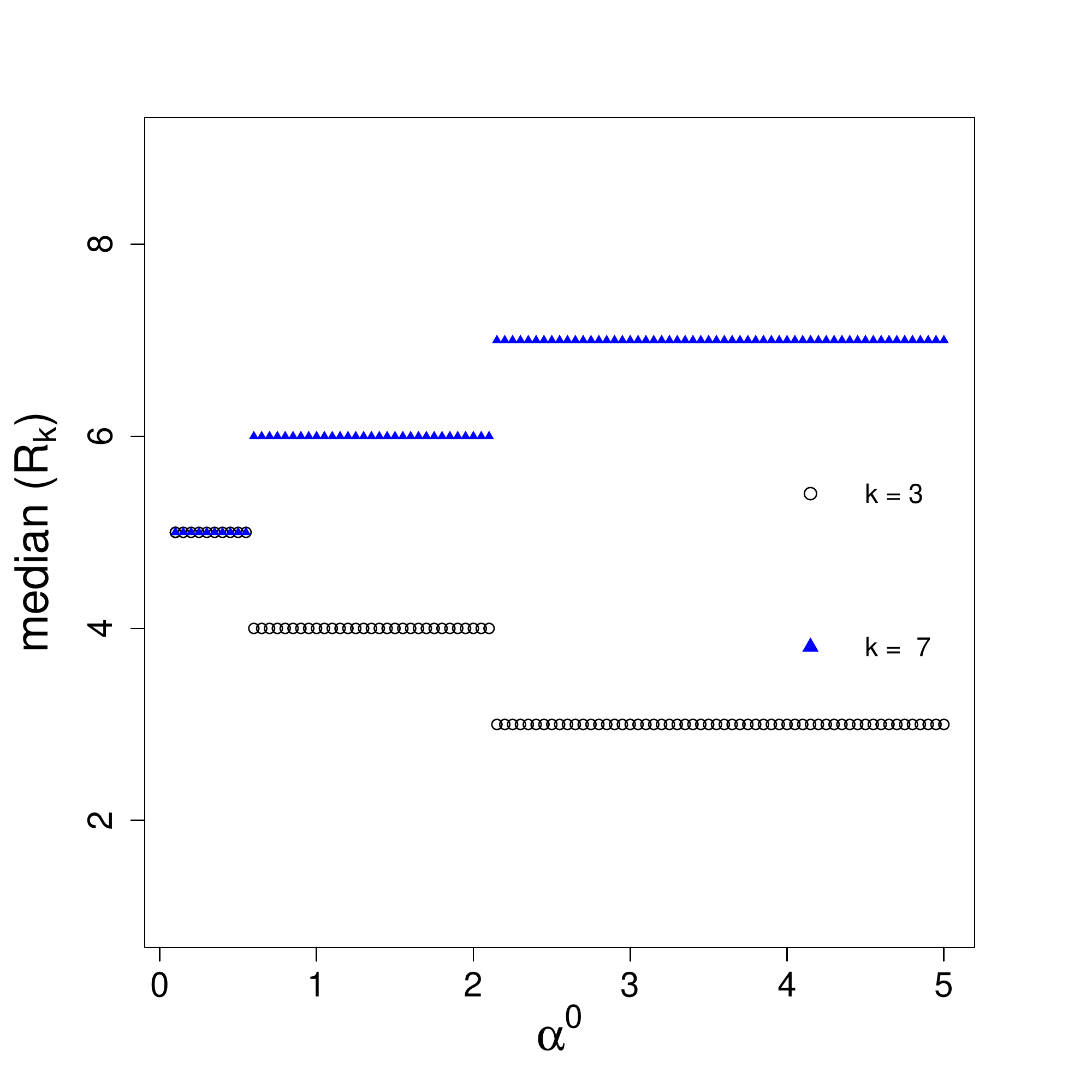}
		\end{minipage} 
		\hfill
		\begin{minipage}[t]{.3\textwidth}
			\centering
			\includegraphics[width=\textwidth]{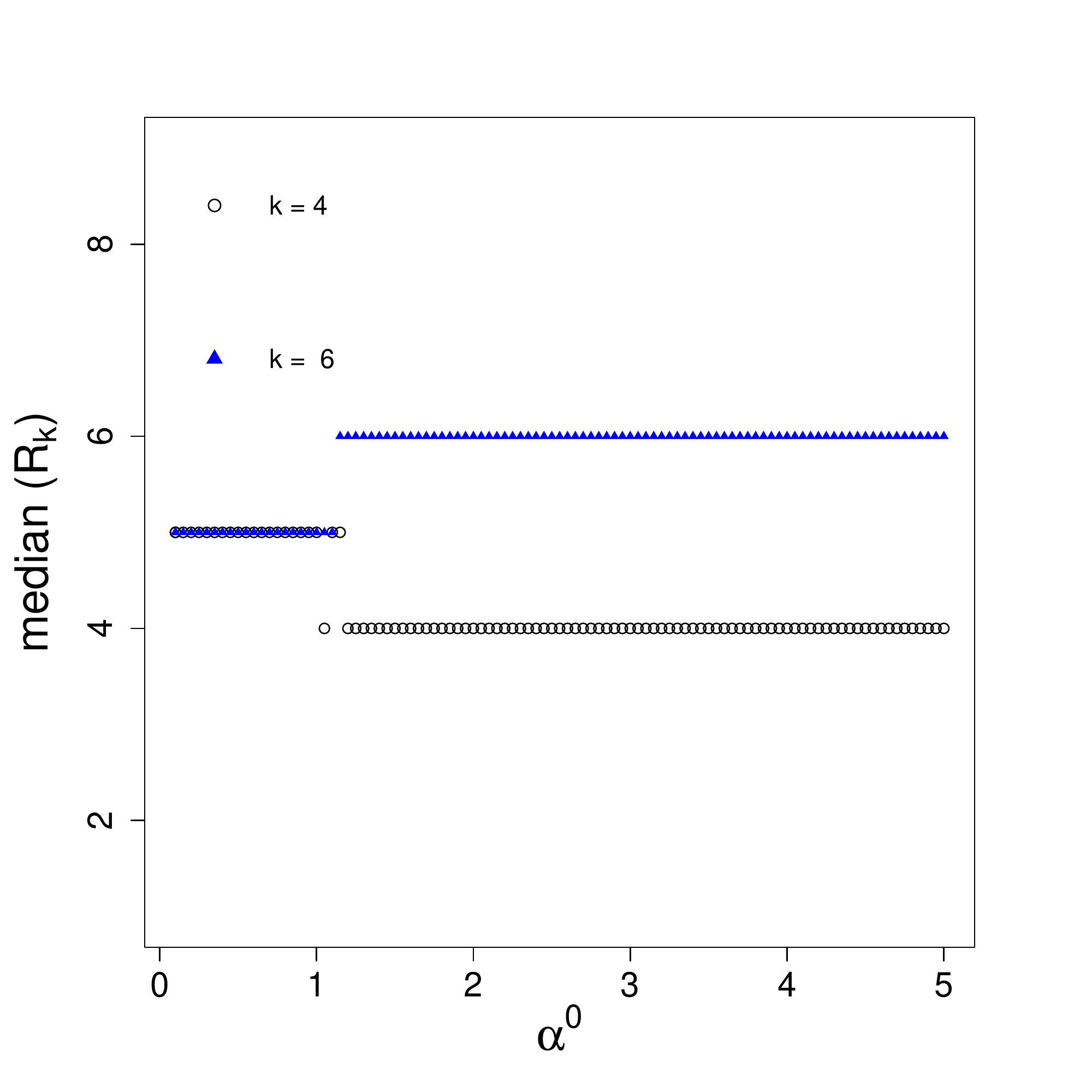}
		\end{minipage} 
		\hfill
	\begin{minipage}[t]{.3\textwidth}
	\centering
	\includegraphics[width=\textwidth]{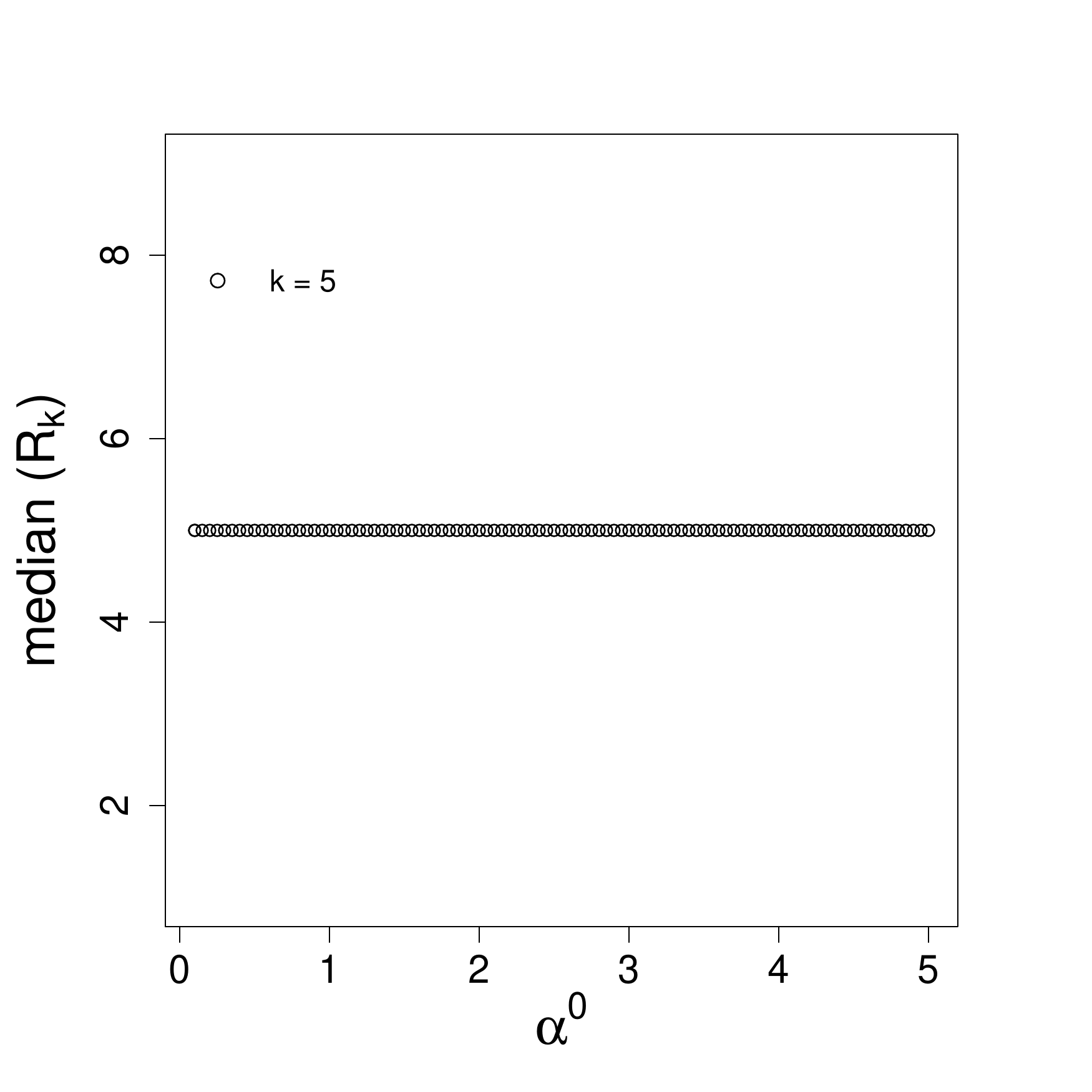}
	\end{minipage} 

	\caption{Marginal median of $R_{o^0_k}$ where $\mathbf{R}$ is drawn from a Mallows model centered at $\{1, 2, ..., n\}$. A similar figure can be obtained for any other values of $n$.}
	\label{fig:median}
\end{figure}

However, this special property of the median holds only for the middle item $o^0_m$. Let us take a closer look at the properties of the median of each $R_{o^0_k}$, $k = 1, ..., n$.  In {Figure} \ref{fig:median} we study the marginal median of $R_{o^0_k}$ as a function of $\alpha$, and compare them for all $k$. We approximate the medians by drawing 50000 independent samples from the Mallows distribution centered at $\bm{\rho}^0 = \{1,2, ..., 9\}$ (hence in this case, $o^0_k=k$ for $k=1, ..., n$ ), on a grid of $\alpha^0$ values. We plot the empirical marginal median of $R_{k}$ on the y-axis, against the corresponding $\alpha^0$ value. It can be observed the median for $R_{o_k}, k\neq 5$, is a function of $\alpha^0$, and converges to $k$ for $\alpha^0$ increasing, and towards the middle rank 5 as $\alpha^0$ decreases to 0. In other words, for all $k\neq 5$, for low enough values of $\alpha^0$, the median of $R_k$ is away from the corresponding rank in $\bm{\rho}^0$, i.e., $k$ in this case, and deviates further and further from $k$ (towards the middle value 5) as $\alpha^0$ decreases.

We also observe that the deviation of the median from its corresponding rank in $\bm{\rho}^0$ is the most pronounced for the items with top- and bottom-ranks in $\bm{\rho}^0$, i.e., item 1 and 9 in this example. In the case of very small $\alpha^0$, the medians of item 1 and 9 can deviate 4 positions from $\rho_1$ and $\rho_9$ respectively, whereas item 4 and 6's medians deviate only by 1 position. 

Based on these properties, we conjecture that the optimal sequence for the terms in (\ref{eq:factorization_with_denom}) is by starting with the middle item $o^0_m$, and then outwards i.e., first the middle item $o^0_m$, then the two second middle-ranked items (items $o^0_{m-1}$ and $o^0_{m+1}$) and so on until lastly the two edge items. By doing so, we first sample for the item whose marginal median (and hence mode) is the closest to the targeted value in $\bm{\rho}^0$. After a sample is drawn, the value that it takes on will be excluded for the subsequent items. As shown in (\ref{eq:pseudo_mode_median}), due to $l^*$ being the median of $R_{o_k} | R_{o_k}\neq \rho_{o_1},\ldots , \rho_{o_{k-1}})$, this forces the subsequent items to be centered closer to their corresponded rank in $\bm{\rho}^0$. This explains why a $\delta$ distribution with its mass concentrated on any ordering that belongs to the $\mathcal{V}_{\bm{\rho}^0}$-set helps the Pseudo-Mallows distribution to best approximate the Mallows posterior.

\subsubsection{Empirical study of the Optimal Pseudo-Mallows Conjecture}\label{sec:empirical_V_new}
To support the Optimal Pseudo-Mallows conjecture, we conduct an empirical study where the optimization problem (\ref{eq:KL_full_data}) is solved first by enumeration for a small number $n$ of items, and then by means of a local search algorithm for larger $n$. In these cases, we show that a ranking $\{i_1, ..., i_n\} \in \mathcal{V}_{\bm{\rho}^0}$ minimizes the KL-divergence between the Pseudo-Mallows distribution and the Mallows posterior. Computing the KL-divergence (\ref{eq:KL_full_data}) between two distributions over $\mathcal{P}_n$ is difficult for a large $n$ because of the exponential dimensions of the set of permutations. In this paper we use instead the sum of one-dimensional KL-divergences between the Mallows posterior's and the Pseudo-Mallows' marginal distributions of each item. More formally, we work with the sum of marginal KL-divergences defined as  
\begin{equation} \label{eq:KL_marginal}
	\sum\limits_{i=1}^{n}\text{KL}\left( q({{\rho}}_i|\alpha, \mathbf{R}^1, ..., \mathbf{R}^N, i_1, ..., i_n)|| P(\rho_i|\alpha, \mathbf{R}^1,..., \mathbf{R}^N)\right), 
\end{equation}
as measure of discrepancy. 

\begin{figure}[!ht]
	\begin{minipage}[t]{.31\textwidth}
		\centering
		\includegraphics[width=\textwidth]{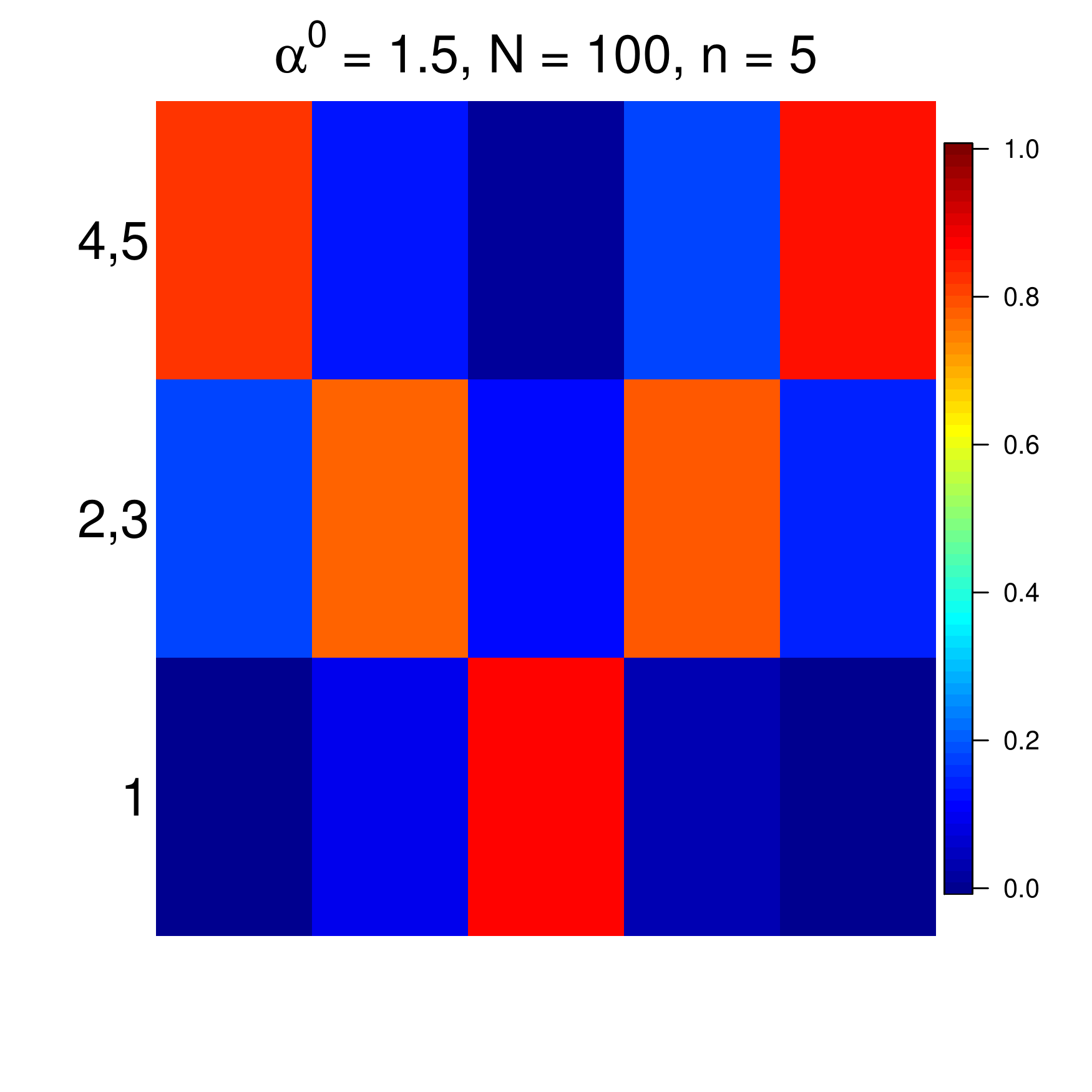}
	\end{minipage} 
	\hfill
	\begin{minipage}[t]{.31\textwidth}
		\centering
		\includegraphics[width=\textwidth]{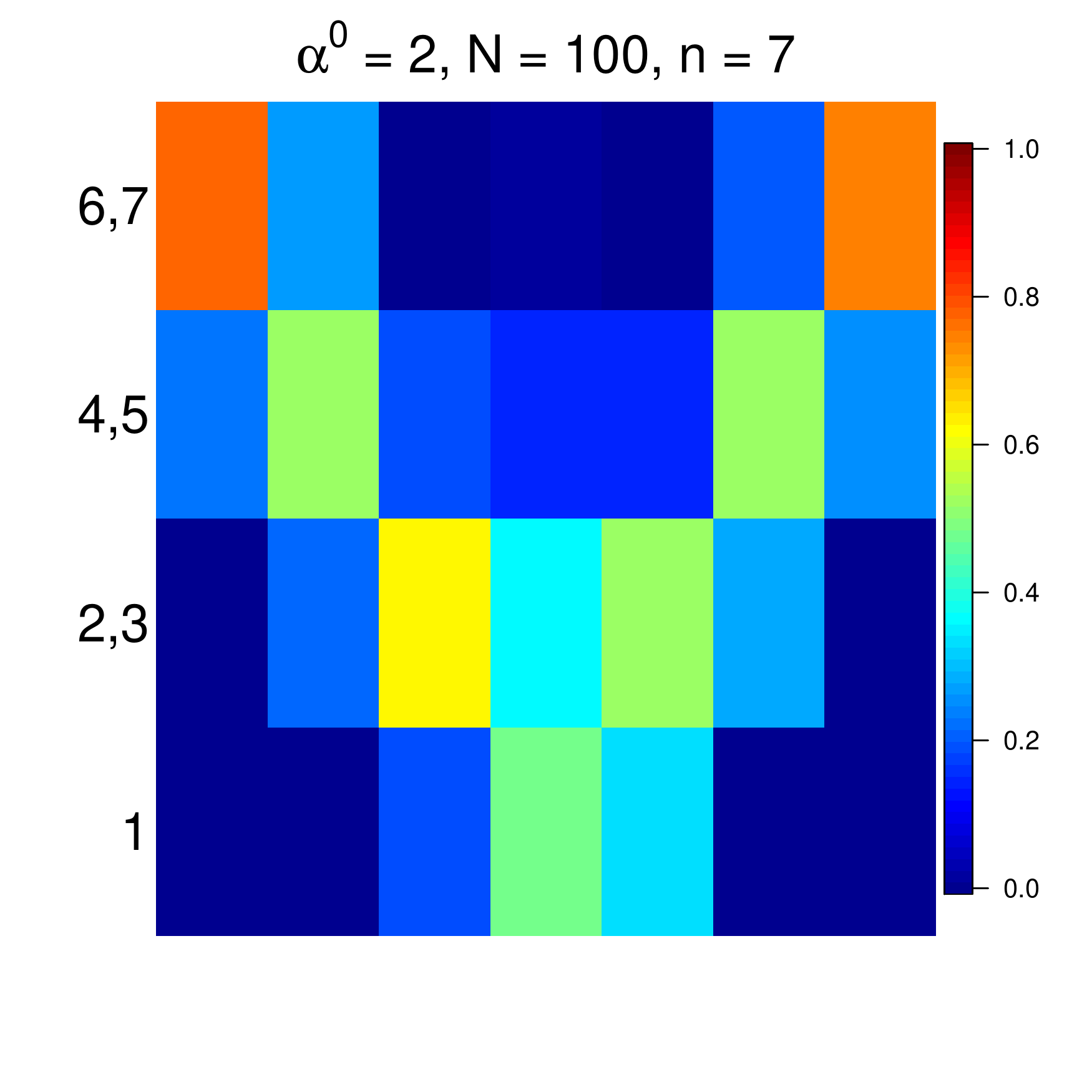}
	\end{minipage} 
	\hfill
	\begin{minipage}[t]{.31\textwidth}
		\centering
		\includegraphics[width=\textwidth]{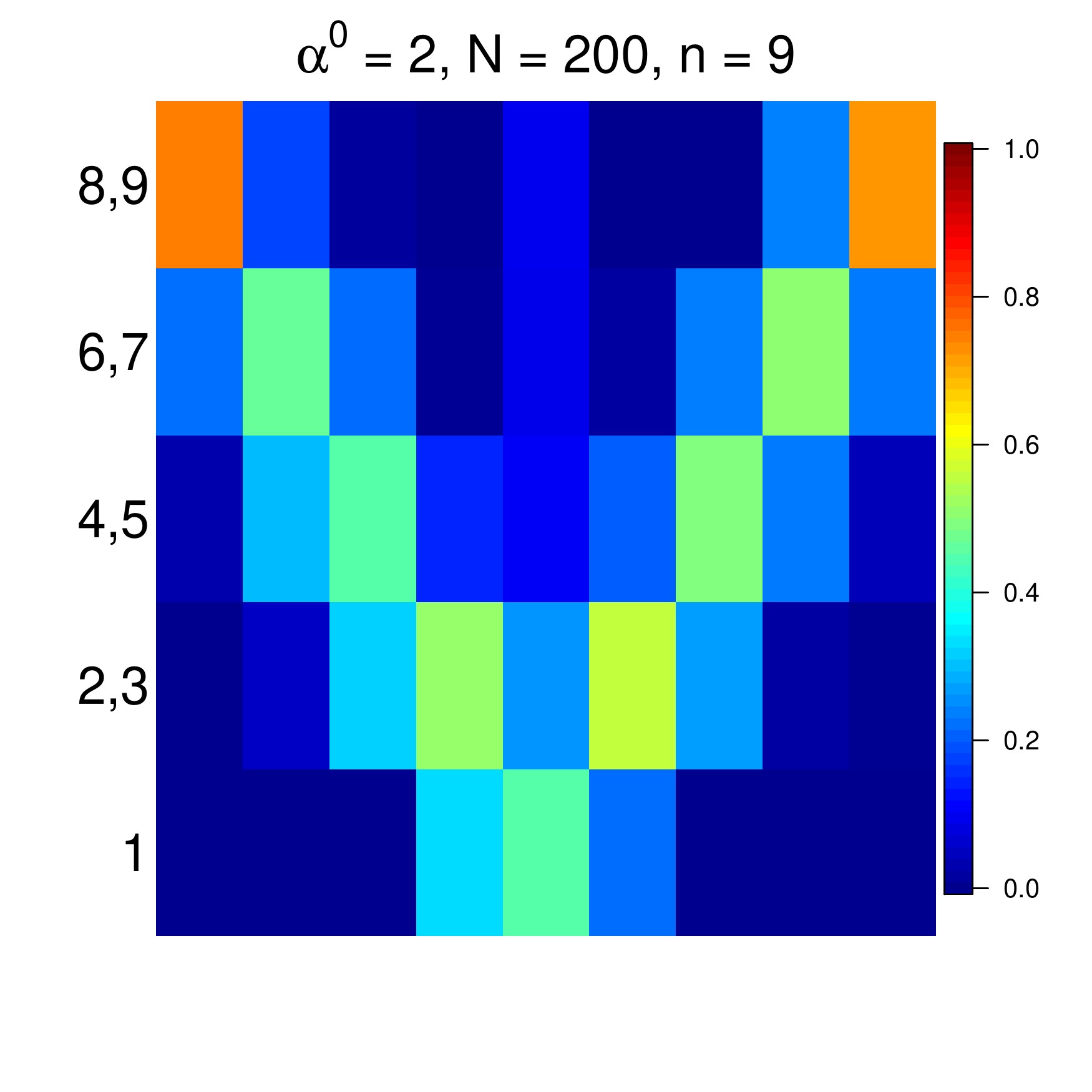}
	\end{minipage} 
	\caption{Heat plots of $\{i_1, ..., i_n\}$ rankings that result in the lowest KL-divergence. The bottom row represents all items' probability to be ranked as 1, the second row from the bottom represents all items' probability to be ranked as either 2 or 3, etc. }
	\label{fig:enumeration}
\end{figure}

\paragraph{Enumeration for $n\leq 9 $}
When the number of items $n$ is small, it is possible to enumerate all possible rankings $\{i_1, ..., i_n\} \in \mathcal{P}_n$, and compute the KL-divergence between the Pseudo-Mallows distribution and the Mallows posterior. We first generate full ranking datasets with $n$ items and $N$ users by drawing $N$ independent rankings from the Mallows distribution with $\alpha^0$ and $\bm{\rho}^0$ using the BayesMallows R package \citep{sorensen2019bayesmallows}. For convenience, we fix $\bm{\rho}^0 = \{1,2,...,n\}$. Next, we run the MCMC converging to  the Mallows posterior to obtain samples. Then for every ranking $\{i_1, ..., i_n\} \in \mathcal{P}_n$ we draw 200 independent samples from the Pseudo-Mallows distribution given $\{i_1, ..., i_n\}$ using Algorithm \ref{algo:full_data_given_ordering}. We then calculate the marginal KL-divergence (\ref{eq:KL_marginal}) between the Pseudo-Mallows distribution and the Mallows posterior. As the number of items $n$ is small, we carefully choose the $\{\alpha^0, N, n \}$ combination to avoid that the Mallows posterior distribution is too concentrated or too uniform. For each such combination of $\{\alpha^0, N, n \}$, we generate 20 datasets, and for each dataset, we record the ranking(s) $\{i_1, ..., i_n\}$ that lead to the lowest marginal KL-divergence during each run. The distribution of the optimal rankings from the 20 runs are shown in a heat plot in Figure \ref{fig:enumeration}. Item 1, ..., $n$ are shown on the x-axis, and their probabilities to be ranked as $\{1, (2i \text{ or } 2i+1), i = 1, ..., (n-1)/2 \}$ are shown on the y-axis. As demonstrated in Figure \ref{fig:enumeration},  rankings $\{i_1, ..., i_n\}$ such that the middle item is top-ranked, the two items next to the middle are ranked second and third, and so on, until lastly the items ``on the edge'' are bottom-ranked, appear to be the rankings that minimize the marginal KL-divergence between the Pseudo-Mallows and the Mallows posterior, as conjectured.  The heat plots show that these rankings present a ``V'' shape, which is the reason of the name ``V''-rankings.

\paragraph{Iterated search for moderate number of items.} For the case in which $n$ is such that enumeration of all possible $\{i_1, ..., i_n\}$ rankings is infeasible,  to minimize the marginal KL we propose a heuristic approach based on an iterative optimization algorithm. When we apply this algorithm to some examples we find that the ``V''-rankings consistently give rise to minimum KL-divergence.

The idea of our iterative algorithm is to compute a sequence of permutations $\rho^1, \rho^2, \dots$, trying to improve the marginal KL at each iteration. To this end, we define a sequence of objective functions $w^1, w^2, \dots: N\rightarrow \mathbb{R}$ such that, at iteration $l= 1,2,\dots$, ($i$) $\rho^l$ is the optimal permutation with respect to $w^l$ and ($ii$) $\rho^l$  can be computed efficiently because we are using the marginal KL. The function $w^l$ is defined by perturbing the original objective \eqref{eq:KL_marginal} according to the previous solution $\rho^{l-1}$ (we assume an initial solution $\rho^0$ is given). 

We first describe the form of the functions $w^1, w^2, \dots$ and how we can effectively find the optimal permutation $\rho^l = \{i_1^l, \dots, i_n^l\}$ at each iteration. We then specify how $w^l$ is built at iteration $l$ from $\rho^{l-1}$. 

For each item $j\in N$ and each iteration $l$, we let $w^l_{jr}$ represent some approximation of the marginal cost of assigning rank $r$ to $j$ (i.e. $i^l_r = j$). Then we define the cost of the permutation $\{i_1, \dots, i_n\}$ simply as

\begin{equation}\label{eq:matchingCost}
  w^l(\{i_1, \dots, i_n\}) = \sum_{r\in N} w^l_{i_r,r}  
\end{equation}
 
Computing the permutation which minimizes \eqref{eq:matchingCost} is equivalent to finding a minimum weighted perfect matching in a suitable (complete, balanced) bipartite graph \citep{cook1999computing}, and it can be performed in polynomial time (in $|N|$).    

We are now left with describing how $w^l_{j,r}$ is defined at each iteration $l= 1,2,\dots$, for any $j,r\in N$. As said, this quantity represents an approximation of the increase of cost (w.r.t. the previous permutation $\rho^{l-1}$) when item $j$ is assigned rank $r$ in $\rho^l$. Finding the exact value for $w^l_{j,r}$ is per se  as difficult as the original optimization problem, as it amounts to find the optimal permutation when $j$ has fixed rank $r$. So, we content ourselves with an approximate solution of this hard problem, following an idea developed in \citep{vitelli2017probabilistic}. Namely, we heuristically build a permutation $\mu^{l-1}$ from $\rho^{l-1}$ by assigning to $j$ rank $r$ and by "locally" re-adjusting the ranking of a subset of items. Then we let $w^l_{j,r}$ equal the difference in cost (computed as \eqref{eq:KL_marginal}) between $\mu^{l-1}$ from $\rho^{l-1}$. 

Namely, we let: 

$$
 w^l_{jr}= \text{KL}\Big(q\big(\cdot|L\&S(\{i^{l-1}_1, ..., i^{l-1}_n\}, j,r)\big)||p(\cdot)\Big) - \text{KL}\Big(q\big(\cdot|i^{l-1}_1, ..., i^{l-1}_n)\big)||p(\cdot)\Big), 
$$
where $L\&S (\{i^{l-1}_1, ..., i^{l-1}_n\}, j, r)$ is a variation of the ``Leap \& Shift'' proposal introduced in \cite{vitelli2017probabilistic}, described as follows: for a given ranking $\{i^{l-1}_1, ..., i^{l-1}_n\}$, the item $j$ to be perturbed with current ranking $q=i_j^{l-1}$ and destination ranking  $r = i^l_j$, we let $L\&S(\{i^l_1, ..., i^l_n\}, j, r) $ be the following ranking for $m = 1, ...,n$
 $$
 i^l_m=
\begin{cases}
j, \text{ if }  m =r\\
i^{l-1}_m - 1, \text{ if } q <i^{l-1}_m \leq r \text{ and } q < r \\
i^{l-1}_m + 1, \text{ if } r \leq i^{l-1}_m < q \text{ and } q > r\\
i^{l-1}_m, \text{ otherwise}\\
\end{cases} .
$$

At each iteration $l$, once all weights $w^l$ are computed as above, we use the Hungarian Method \citep{kuhn1955hungarian} to compute the minimum weight perfect matching, which provides us the new permutation $\rho^{l+1}$. The search algorithm will run iteratively until a stop criterion, such as a maximum number of iteration, is met. In practice, we recommend setting the maximum number of iterations of the algorithm  to a large enough value, depending on $n$. Clearly, the computational effort of the algorithm quickly increases with the number of items $n$. The iterated search algorithm is more formally described in Algorithm \ref{algo:localsearch}. 
%\begin{enumerate}
 %   \item {A maximum of 50 or 100 iterations depending on $n$;  or}
 %   \item {The algorithm remains at the same position for more than 3 iterations.}
%\end{enumerate}

In order to simplify the verification of the ``V''-ranking's optimality, we initialize the algorithm by generating a ``V''-ranking, and then conduct $n$ swaps steps with its neighbor so that the starting point is not a ``V''-ranking, but not too far away from the ``V''-ranking.

\begin{figure}[h!]
		\centering
		\includegraphics[width=\textwidth]{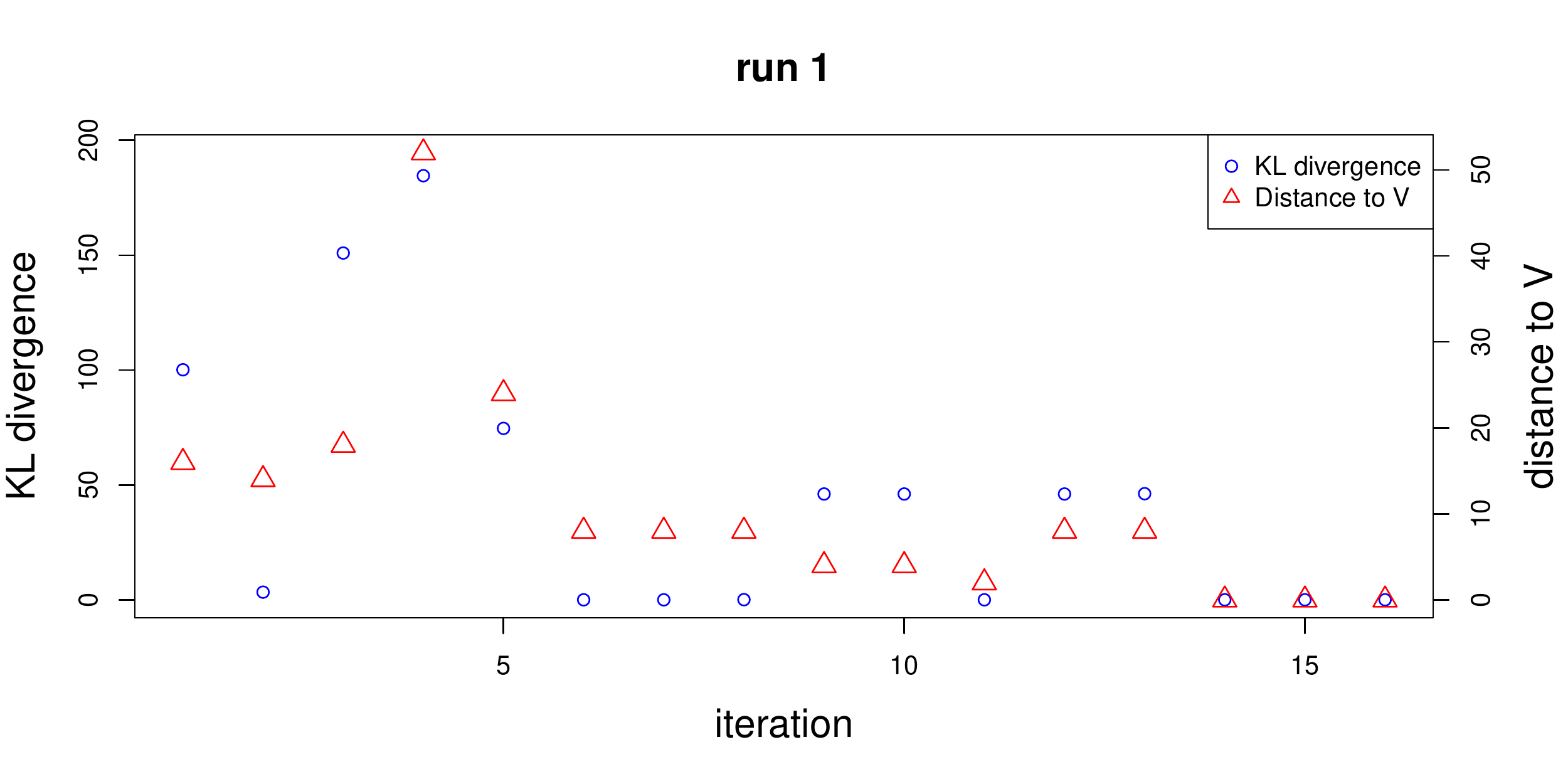}
		\includegraphics[width=\textwidth]{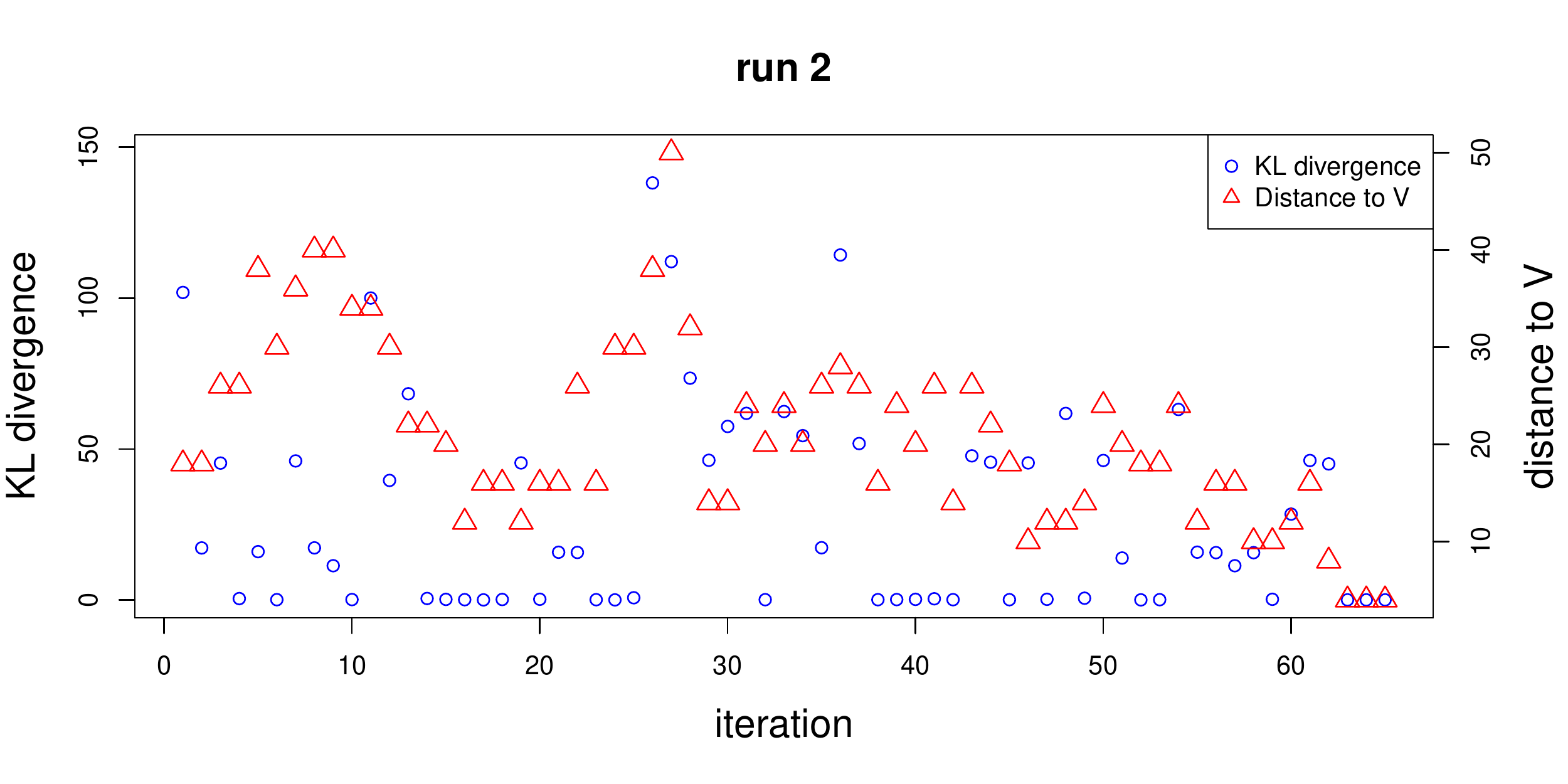}
	\caption{Selected runs of the iterative search algorithm, $n$= 15, $N = 500$, $\alpha^0 = 2.5$ }
	\label{fig:local_search_KL}
\end{figure}

Similarly to the full enumeration study, we generate 20 full ranking datasets for each combination of $\{ \alpha^0, N, n\}$, and run Algorithm \ref{algo:localsearch} on each dataset. The maximum number of iterations of the algorithm was set to 100 when $n$ was small, to 50 otherwise. 
In Figure \ref{fig:local_search_KL}, we have selected some runs to demonstrate the characteristics of the algorithm. The blue dots represent the marginal KL-divergence, while the red triangles represent the footrule distance between the current ranking and a ``V''-ranking. As ``V''-rankings are not unique, we use the distance between the current ranking and the ``V''-ranking closest to it. From Figure \ref{fig:local_search_KL}, it can be observed that the iterative search algorithm has a tendency to ``re-start'' occasionally: i.e., after finding a permutation that leads to a locally-minimum marginal KL, it can move away from it. However, consistently with the enumeration experiment, when the algorithm discovers a ``V''-ranking, the corresponding KL-divergence is typically very low.  

\begin{figure}[!ht]
	\begin{minipage}[t]{.3\textwidth}
		\centering
		\includegraphics[width=\textwidth]{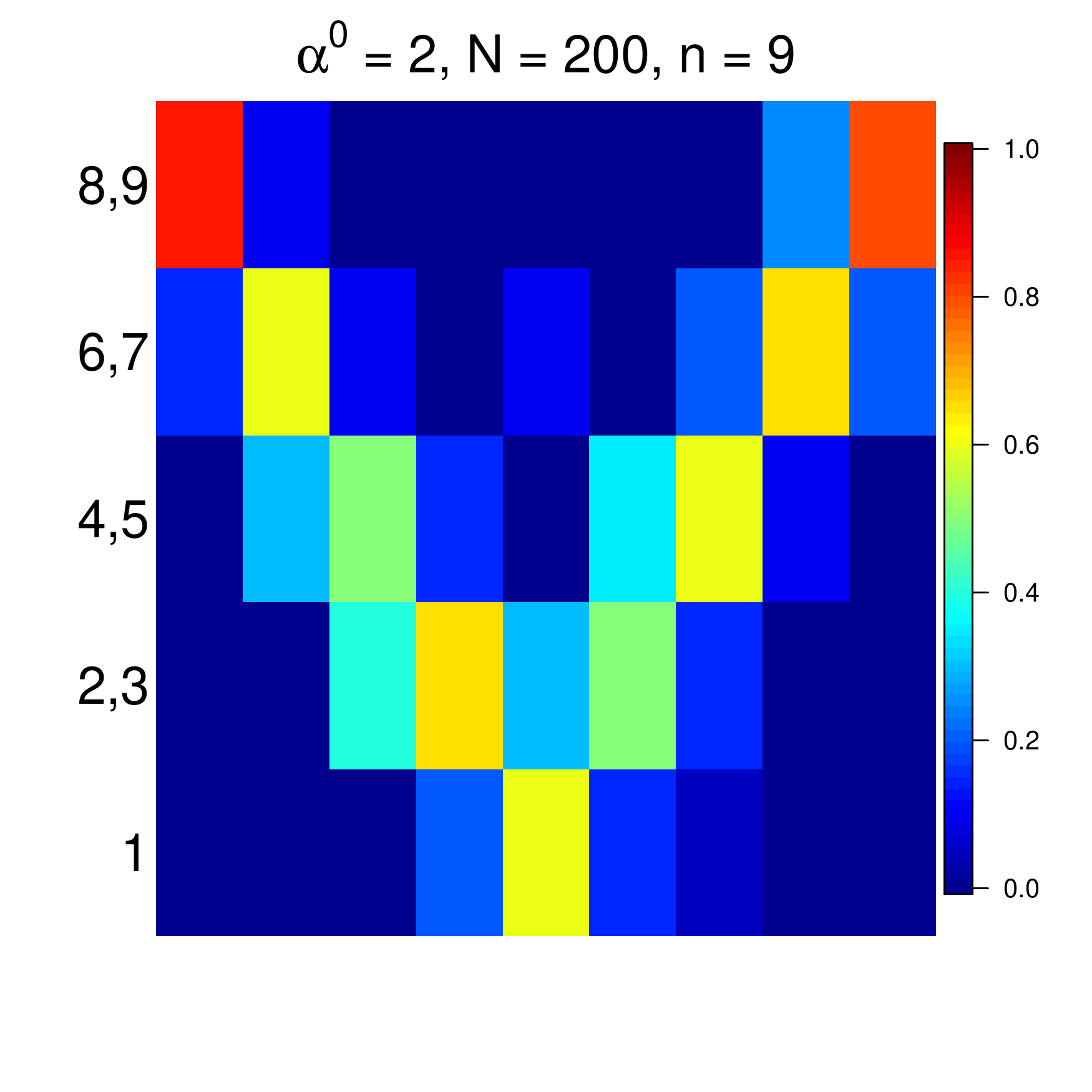}
	\end{minipage} 
	\hfill
	\begin{minipage}[t]{.3\textwidth}
		\centering
		\includegraphics[width=\textwidth]{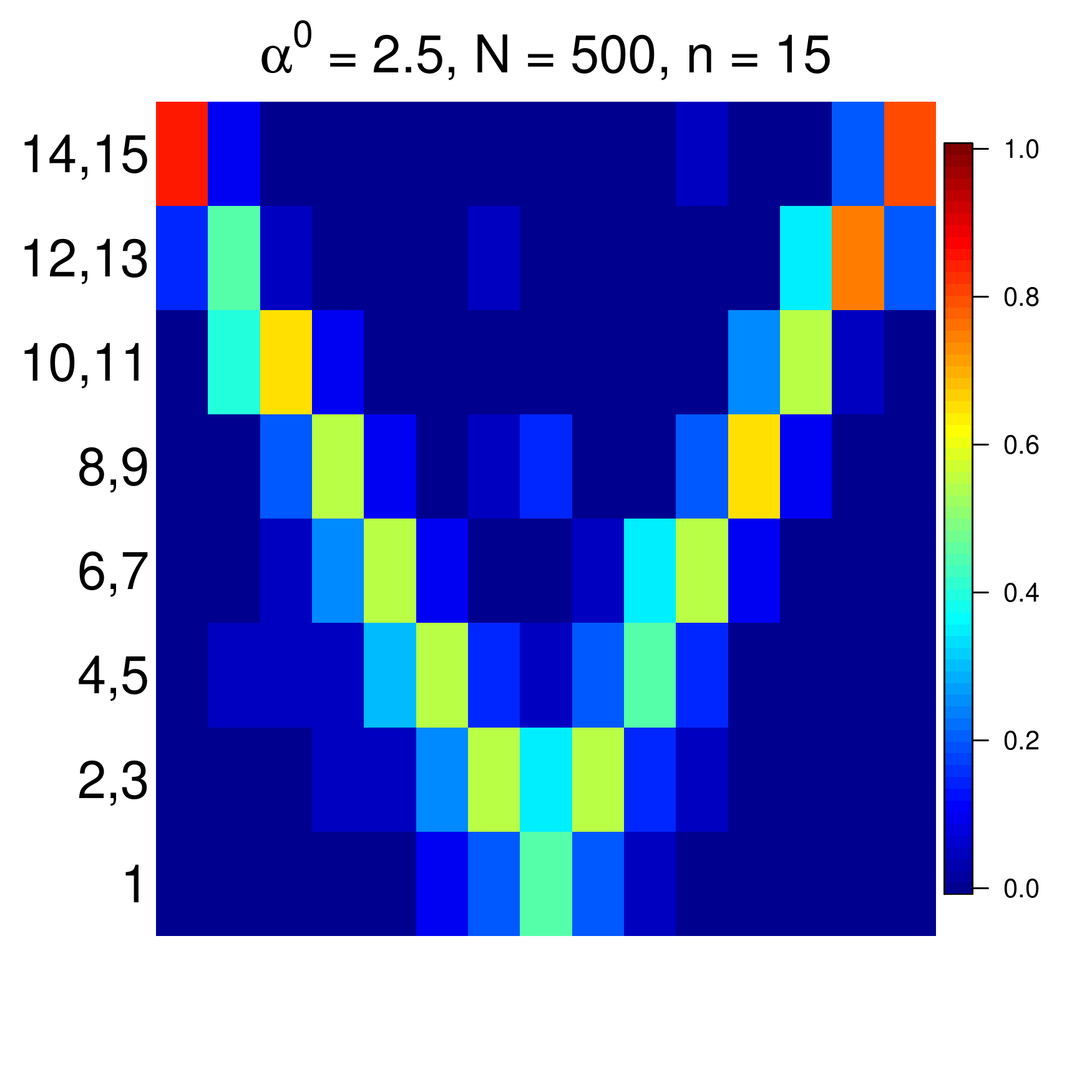}
	\end{minipage} 
	\hfill
	\begin{minipage}[t]{.3\textwidth}
		\centering
		\includegraphics[width=\textwidth]{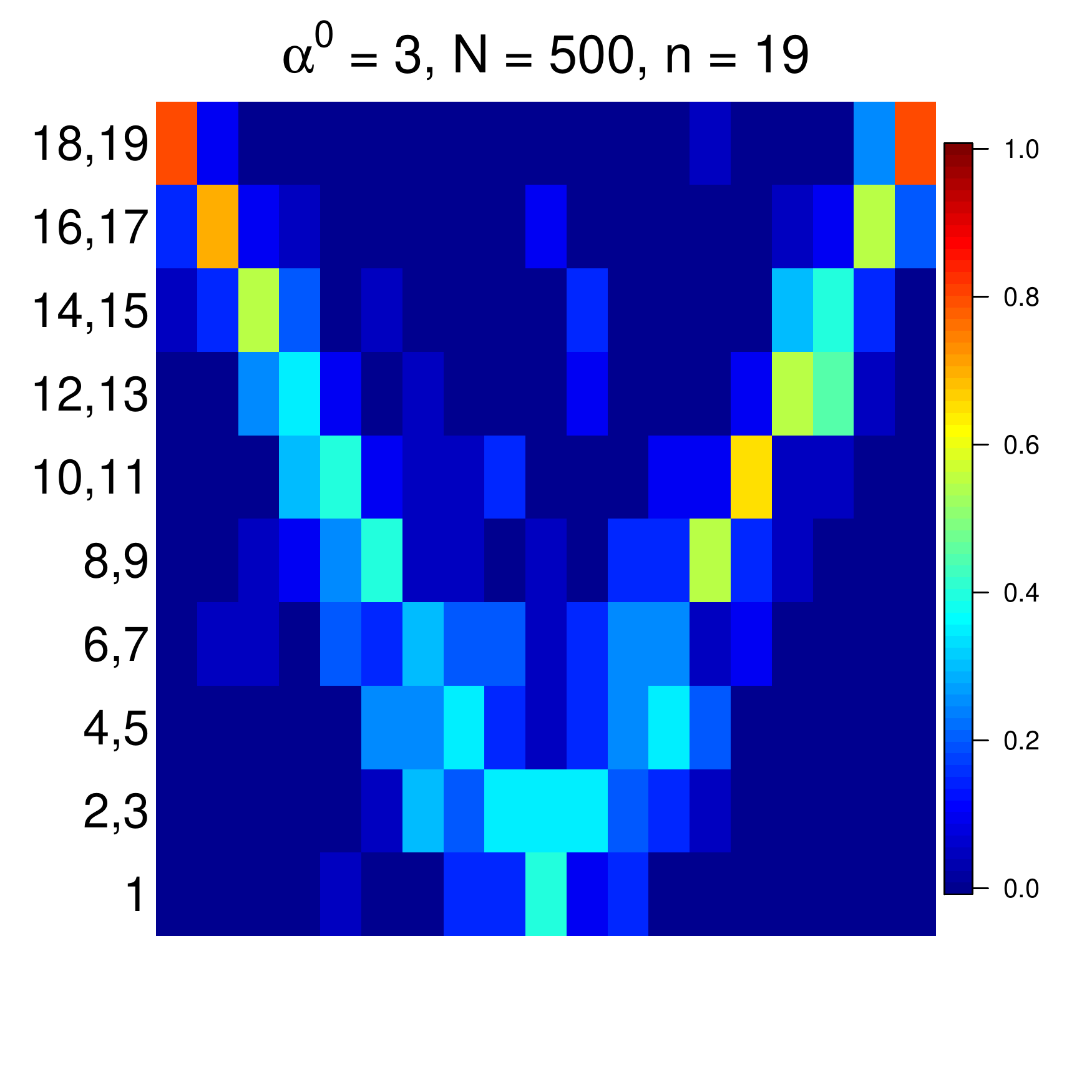}
	\end{minipage} 
	\caption{Iterative search algorithm's last iteration, aggregated over 20 runs }
	\label{fig:local_search_last_iter}
\end{figure}

We record the $\{i_1, ..., i_n\}$ ranking of the last iteration of the algorithm at each run, and aggregate these rankings in a heat plot, as shown in Figure \ref{fig:local_search_last_iter}. It can be clearly observed that the iterative search algorithm typically stops exactly at a ``V''-ranking, which is consistent with our conjecture. Sometimes however, the algorithm did not stop exactly at a ''V''-ranking, but with a short footrule distance therefrom, despite a very similar marginal KL-divergence. This is due to the approximation of the marginal KL-divergence and the finite sample situation.  

\subsubsection{Uniform distribution over the \texorpdfstring{$\mathcal{V}_{\bm{\rho}^0}$}{Lg}-set as optimal \texorpdfstring{$g(i_1, ...,i_n)$}{Lg}}\label{sec:improved_pseudo_mallows}

 So far, we have worked with a $\delta$ distribution for $g(i_1, ...i_n)$ for the Pseudo-Mallows ordering. By definition of the $\mathcal{V}_{\bm{\rho}^0}$-set,  ``V''-rankings are not unique. From the empirical study, we observed that all ``V''-rankings appear to be equivalently effective at achieving a good approximation of the Mallows posterior. Therefore, instead of using a $\delta$ distribution over one $\bm{v} \in \mathcal{V}_{\bm{\rho}^0}$ for the Pseudo-Mallows distribution, we use a uniform distribution $g^*({i_1, ..., i_n})$ over $\mathcal{V}_{\bm{\rho}^0}$. More formally, the Pseudo-Mallows distribution is then expressed as follows:
 
\begin{equation}\label{eq:pseudolikelihood}
Q(\bm{\rho}|\alpha, \mathbf{R}^1, ..., \mathbf{R}^N) = \sum\limits_{\{i_1, ..., i_n\}\in \mathcal{P}_n}q(\bm{\rho}|\alpha, \mathbf{R}^1, ..., \mathbf{R}^N, i_1, ..., i_n)g^*(i_1, ..., i_n),
\end{equation} where
$$g^*(i_1, ..., i_n|\mathcal{V}_{\bm{\rho}^0}) =
  \begin{cases}
    \frac{1}{|\mathcal{V}_{\bm{\rho}^0}|} & \text{if } \{i_1, ..., i_n\} \in \mathcal{V}_{\bm{\rho}^0}\\
    0 & \text{otherwise} .
  \end{cases} $$
  
\subsection{Inferring the \texorpdfstring{$\mathcal{V}$}{Lg}-set from data}
\subsubsection{The asymptotic  \texorpdfstring{$N \rightarrow \infty$}{Lg}  case}
Since $\bm{\rho}^0$ is unknown in practice, we need to infer the $\mathcal{V}_{\bm{\rho}^0}$ set from data. 

\begin{lemma} \label{lem:nocross}
	Assume $\mathbf{R} \in \mathcal{P}_n$ and $\mathbf{R} \sim \text{Mallows}(\bm{\rho}^0, \alpha)$. Let $o^0_i$ be such that $\rho^0_{o^0_i} = i$, for $i = 1, ..., n$. For a fixed $\bm{\rho}^0$ and $\alpha \in (0, \infty)$,  it holds that $\mathbb{E}[R_{o^0_j}|\bm{\rho}^0, \alpha] <\mathbb{E}[R_{o^0_{j+1}}|\bm{\rho}^0, \alpha]$.
\end{lemma}
See Section \ref{sec:appen_two} in the appendix for a proof. 
Of course, as $N \rightarrow \infty$, $\frac{1}{N}\sum\limits_{j=1}^{N}R^j_i \rightarrow \mathbb{E}[R_i|\bm{\rho}^0, \alpha ]$, $\forall i = 1, ..., n$. We now define the rank operator:
\begin{mydef}
	The rank vector of a real number vector $\{x_1, ..., x_n \}$, is defined by:\\
	
	$rank(x_1, ..., x_n)$ = $(r_1,..., r_n), \text{ s.t. } r_i = \sum\limits_{j=1}^{n}\delta (x_i - x_j)$ for $i = 1, ..., n$, where $\delta(x) = \left \{
	\begin{aligned}
	&1, && \text{if } x\geq 0 \\
	&0, && \text{if } x < 0
	\end{aligned} \right.
	$.
	
\end{mydef}  
When the ranking data comes from the Mallows distribution, the following theorem follows from \textbf{Lemma} \ref{lem:nocross}. 
\begin{theorem}\label{theorem:inferrho}
	As $N \rightarrow \infty$, and $\forall \alpha > 0$, \\ $rank(\frac{1}{N}\sum\limits_{j=0}^{N}R_1^j,...,\frac{1}{N}\sum\limits_{j=0}^{N}R_n^j) \rightarrow rank(\mathbb{E}[{R}_1|\bm{\rho}^0, \alpha],...,\mathbb{E}[{R}_n|\bm{\rho}^0, \alpha]) =\bm{\rho}^0 $
\end{theorem}
Summing up, $\mathcal{V}_{\bm{\rho}^0}$ can be inferred by the $\mathcal{V}$ set of $rank(\frac{1}{N}\sum\limits_{j=0}^{N}R_1^j,...,\frac{1}{N}\sum\limits_{j=0}^{N}R_n^j)$. 

\subsubsection{Finite number of users}
For finite $N$, $\bm{\rho}^0$ and therefore $\mathcal{V}_{\bm{\rho}^0}$, cannot be accurately inferred from the data. We define the estimate $\hat{\bm{\rho}}^0$ = $rank(\frac{1}{N}\sum\limits_{j=0}^{N}R_1^j, ..., \frac{1}{N}\sum\limits_{j=0}^{N}R_n^j)$ and its corresponding $\mathcal{V}$-set $\mathcal{V}_{\hat{\bm{\rho}}^0}$, which will differ from $\mathcal{V}_{{\bm{\rho}}^0}$. We introduce some variability around the set $\mathcal{V}_{\hat{\bm{\rho}}^0}$ to increase the precision of the Pseudo-Mallows approximation. Instead of simply using a uniform distribution over the $\mathcal{V}$-set $\mathcal{V}_{\bm{\hat{\rho}^0}}$, i.e., $g^*(i_1,\ldots , i_n|\mathcal{V}_{\bm{\hat{\rho}^0}})$, we sample from $g'(i_1, ..., i_n|{\hat{\bm{\rho}}^0}, \sigma) $ where
\begin{itemize}
		\item {$\hat{\bm{v}} \sim g^* (\hat{\bm{v}}|\mathcal{V}_{\bm{\hat{\rho}^0}})$}
		\item {$x_i\sim \mathcal{N}(x_i|\hat{v}_i, \sigma)$} for $i = 1,...,n $
		\item {$\{i_1, ..., i_n\} = rank(x_1, ..., x_n)$ }. 
	\end{itemize}

That is to say, we start from a ranking $\hat{\bm{v}}$ from the $\mathcal{V}$ set  $\mathcal{V}_{\bm{\hat{\rho}^0}}$, and for each item $i$, we introduce a perturbation by drawing from a Gaussian distribution centered on $\hat{\bm{v}_i}$ with some standard deviation $\sigma$. We then build a perturbed ``V''-ranking using the rank operator. In this way, a smaller KL-divergence of the Pseudo-Mallows from the Mallows posterior can usually be achieved in the case where the estimate of $\bm{\rho}^0$ is inaccurate, due to limited data points.

\begin{figure}[h!]
	\subfigure[]{
		\begin{minipage}[t]{.3\textwidth}
			\centering
			\includegraphics[width=\textwidth]{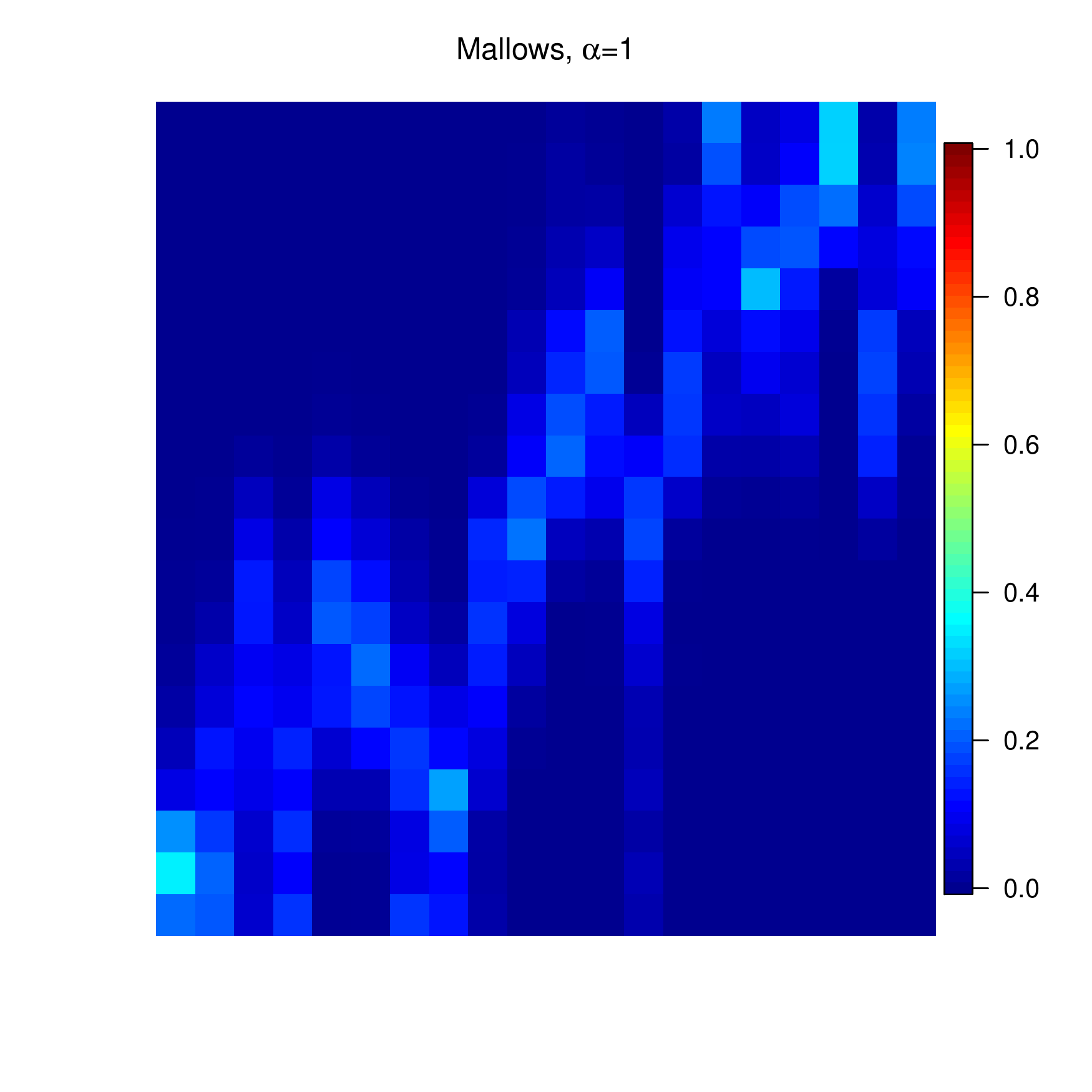}
			
		\end{minipage}
		\hfill
		\begin{minipage}[t]{.3\textwidth}
			\centering
			\includegraphics[width=\textwidth]{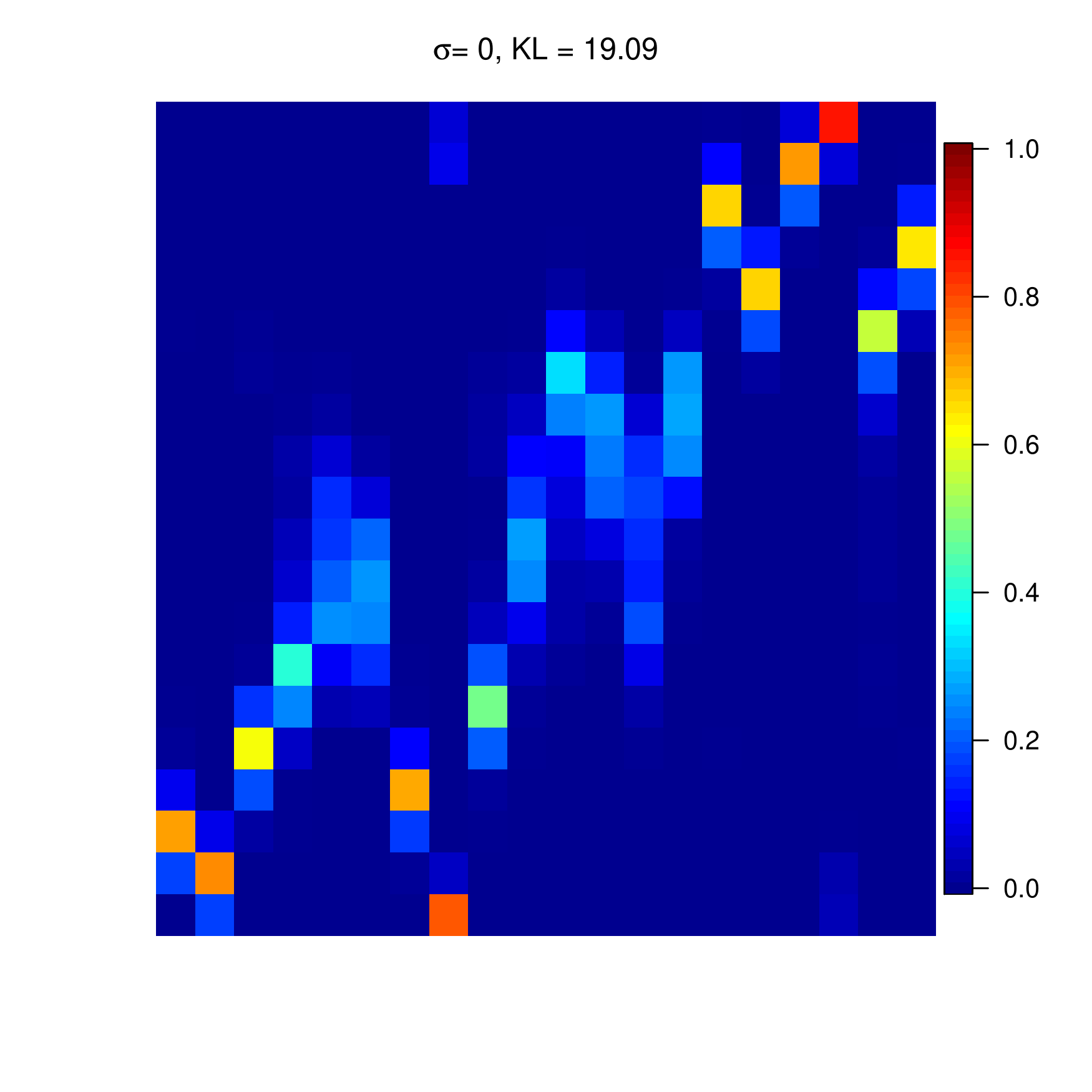}
			
		\end{minipage} 
		\hfill
		\begin{minipage}[t]{.3\textwidth}
			\centering
			\includegraphics[width=\textwidth]{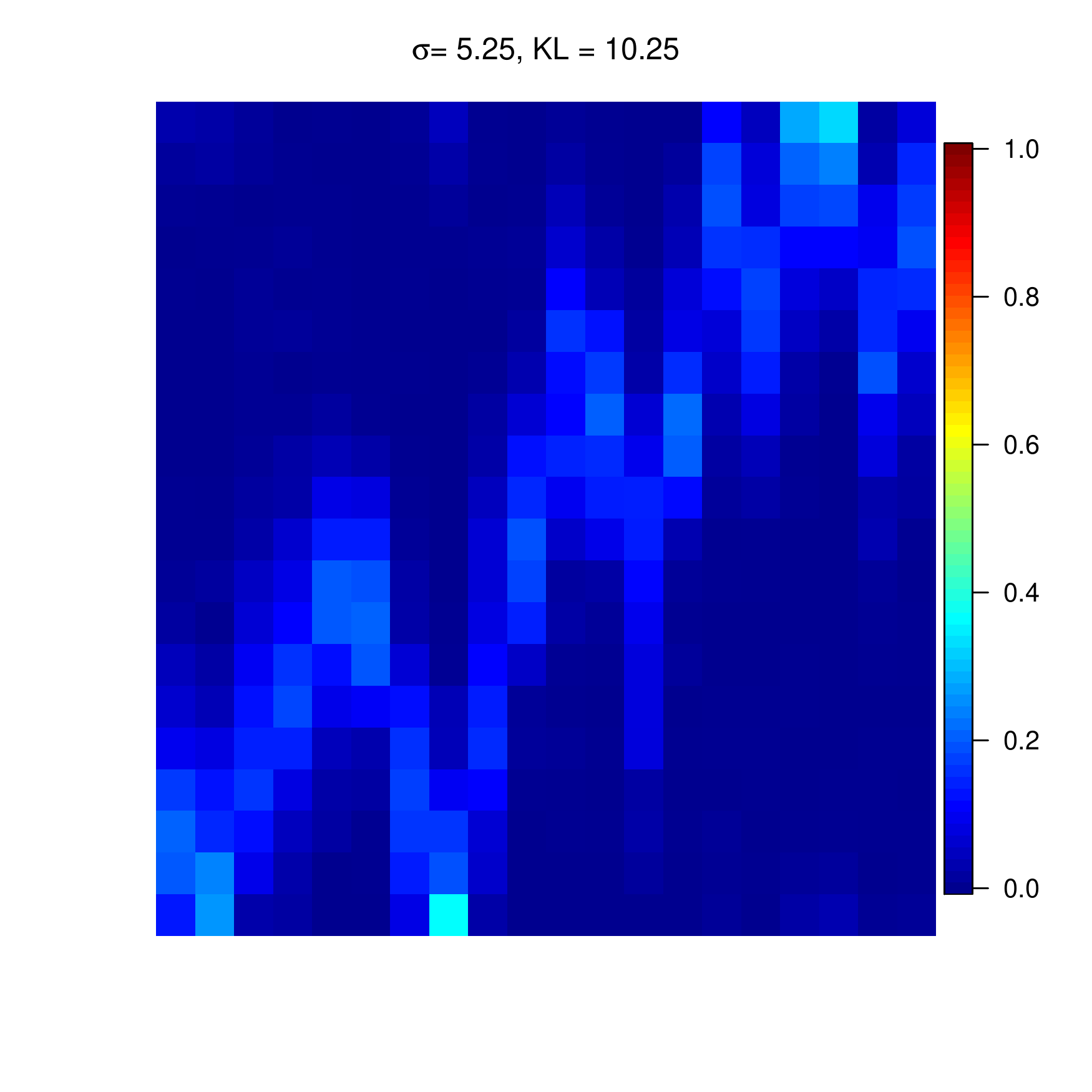}
			
		\end{minipage} 

}
	\subfigure[]{
		\begin{minipage}[t]{.3\textwidth}
			\centering
			\includegraphics[width=\textwidth]{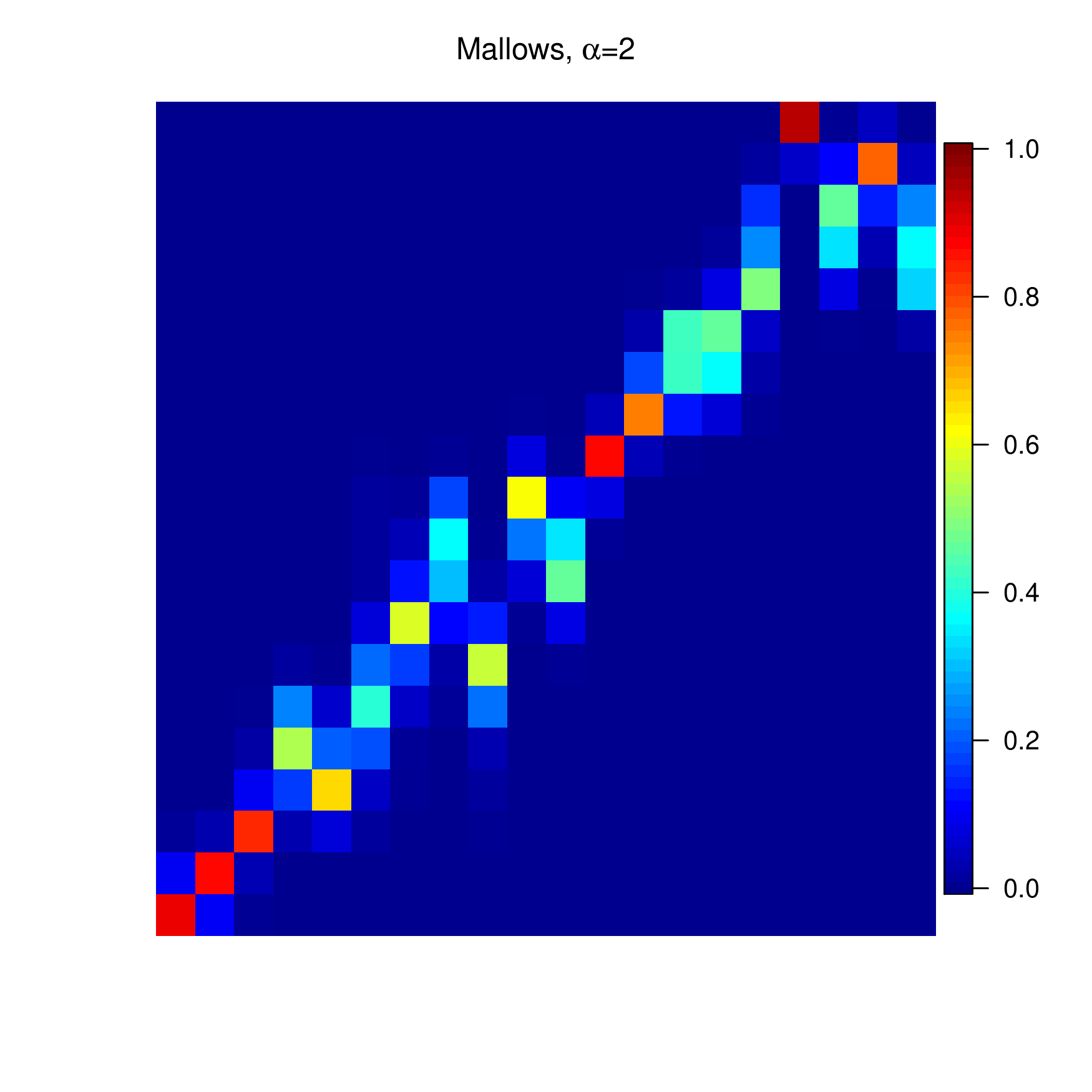}
			
		\end{minipage}
		\hfill
		\begin{minipage}[t]{.3\textwidth}
			\centering
			\includegraphics[width=\textwidth]{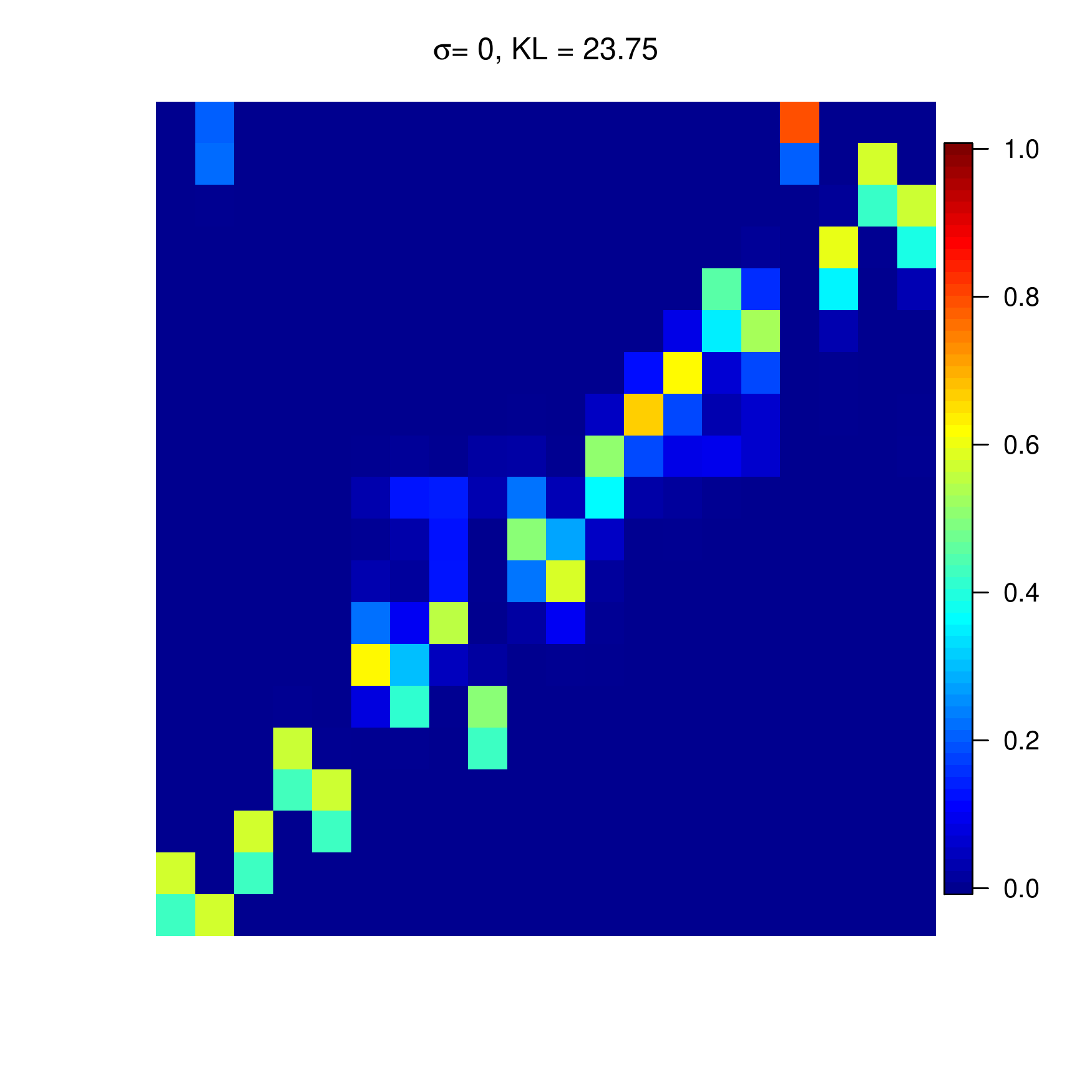}
			
		\end{minipage} 
		\hfill
		\begin{minipage}[t]{.3\textwidth}
			\centering
			\includegraphics[width=\textwidth]{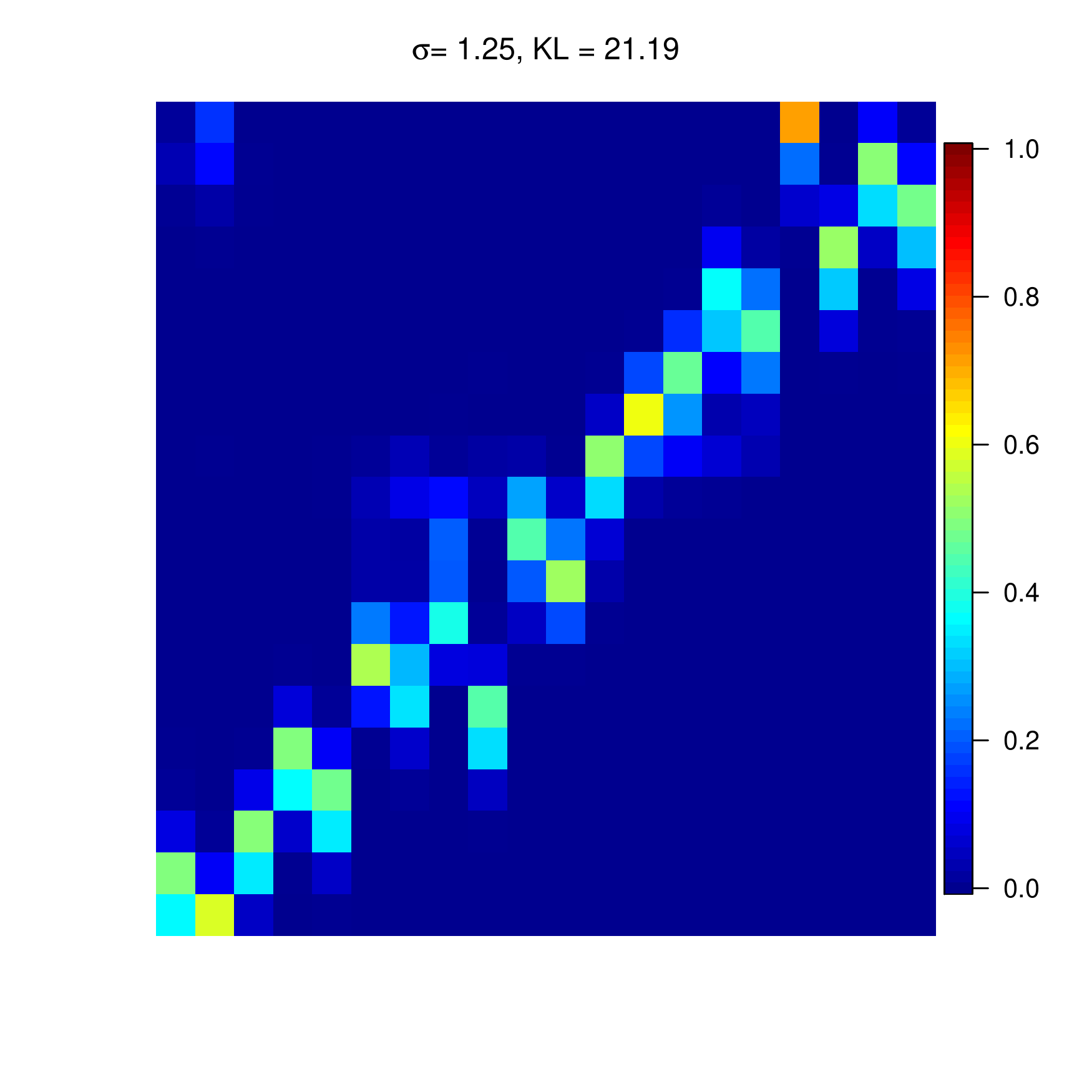}
			
		\end{minipage} 
		
	}	
	
	\caption{Left column: Mallows posterior; middle column: pseudo-Mallows with $\sigma = 0$; right column: pseudo-Mallows with $\sigma > 0$. $N = 100, n = 20$. (a) $\alpha^0 = 1$ and (b) $\alpha^0 = 2$.}
	\label{fig:heatPlot_comparison}
\end{figure}

The performance of this procedure is illustrated in Figure \ref{fig:heatPlot_comparison}. For both the cases of $\alpha^0 = 1$ and $\alpha^0 = 2$, introducing a perturbation with  $\sigma >0$, the marginal KL-divergence between the Pseudo-Mallows and the Mallows posterior is reduced. It can also be observed that as $\alpha^0$ increases, $N$ increases and $n$ decreases, the $\sigma$ value that resulted in a reduction of the marginal KL-divergence decreases.
In some situations, the optimal value for $\sigma$ can be 0. %It is typically the case when $N$ and $\alpha$ is large, and when $n$ is small. Moreover, as the number of users $N$ approaches infinity, $\sigma$ approaches 0. 

\begin{figure}[h!]
\subfigure[]{
		\begin{minipage}[t]{.3\textwidth}
			\centering
			\includegraphics[width=\textwidth]{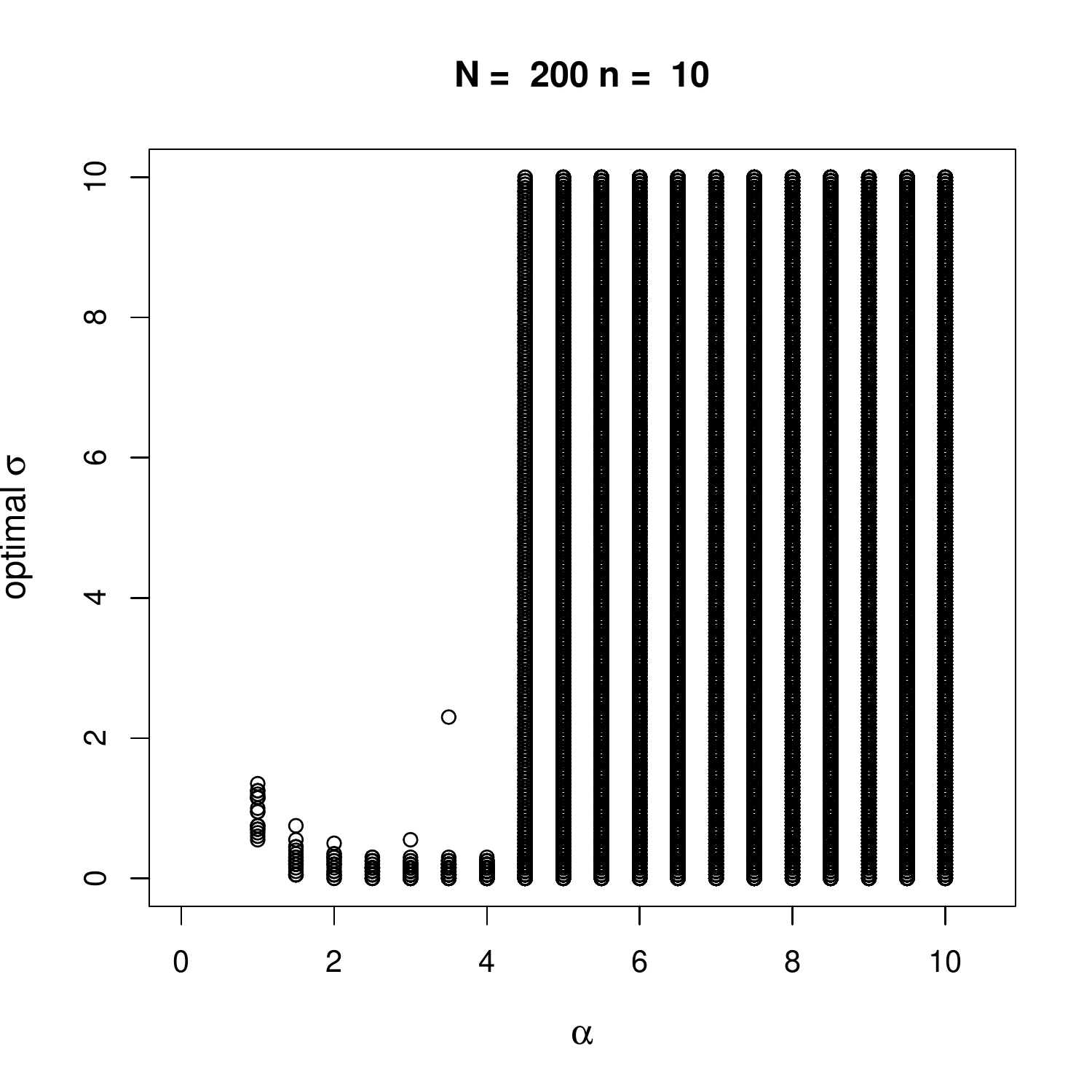}
			
		\end{minipage}
		\hfill
		\begin{minipage}[t]{.3\textwidth}
			\centering
			\includegraphics[width=\textwidth]{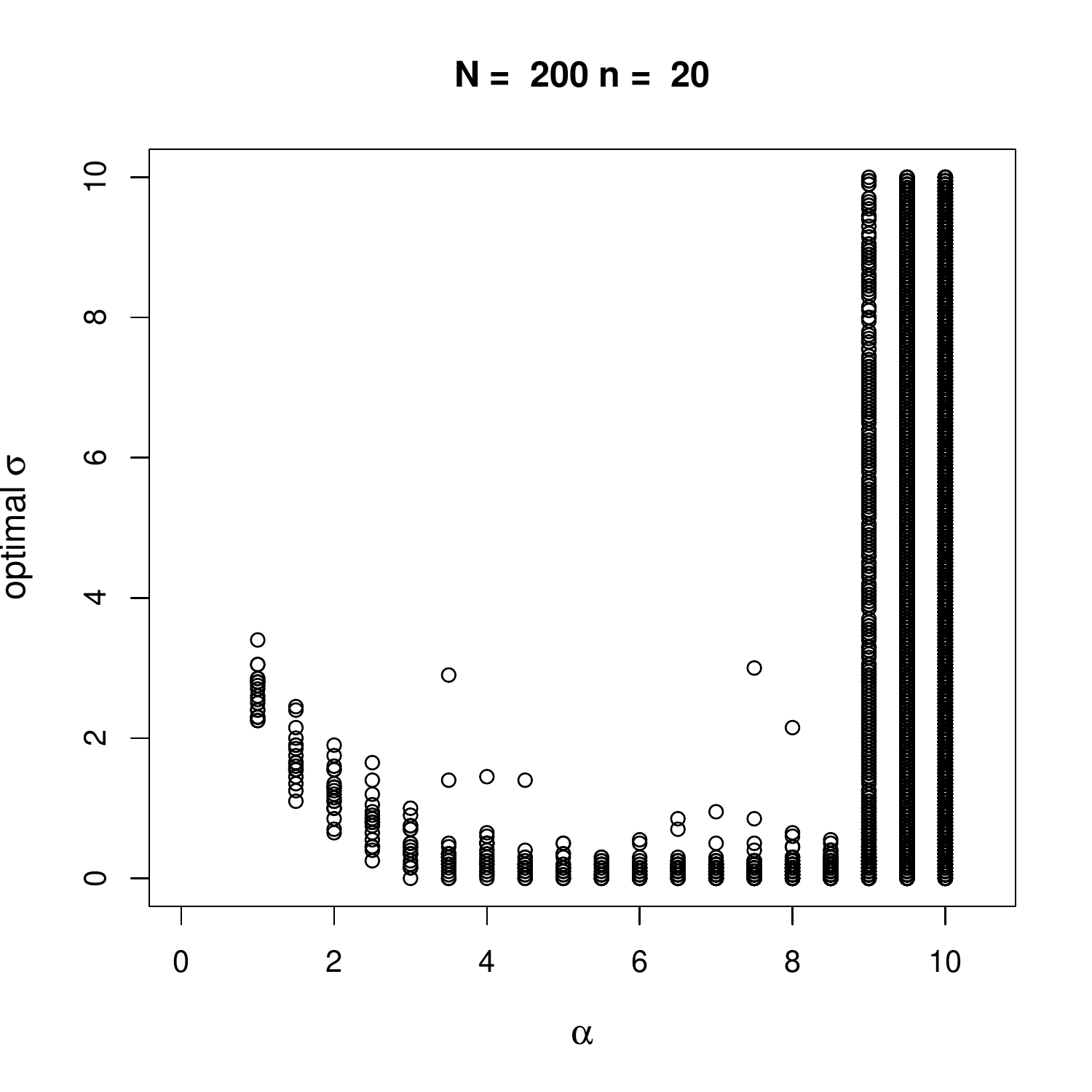}
			
		\end{minipage} 
		\hfill
		\begin{minipage}[t]{.3\textwidth}
			\centering
			\includegraphics[width=\textwidth]{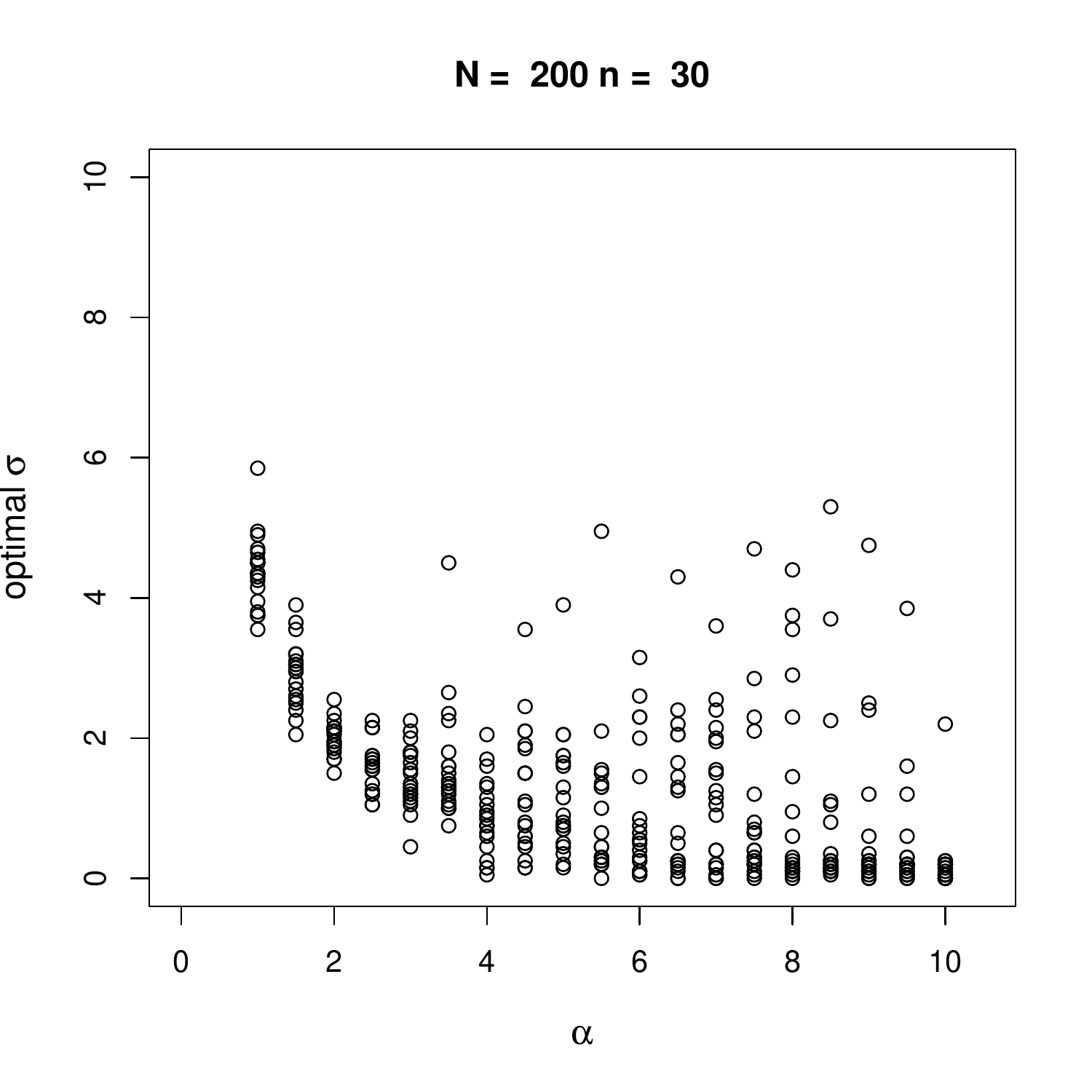}
			
		\end{minipage} 

}
	\subfigure[]{
		\begin{minipage}[t]{.3\textwidth}
			\centering
			\includegraphics[width=\textwidth]{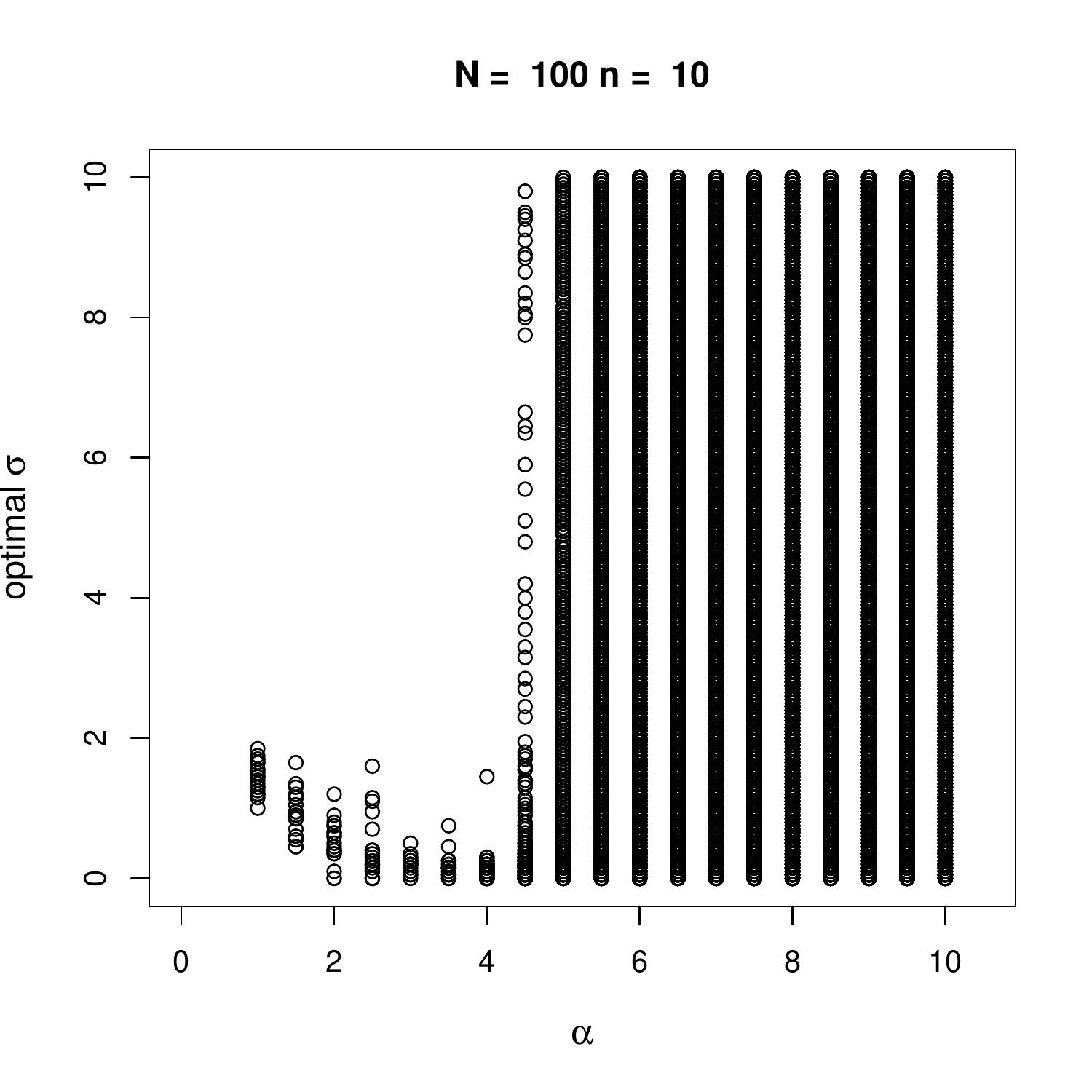}
			
		\end{minipage}
		\hfill
		\begin{minipage}[t]{.3\textwidth}
			\centering
			\includegraphics[width=\textwidth]{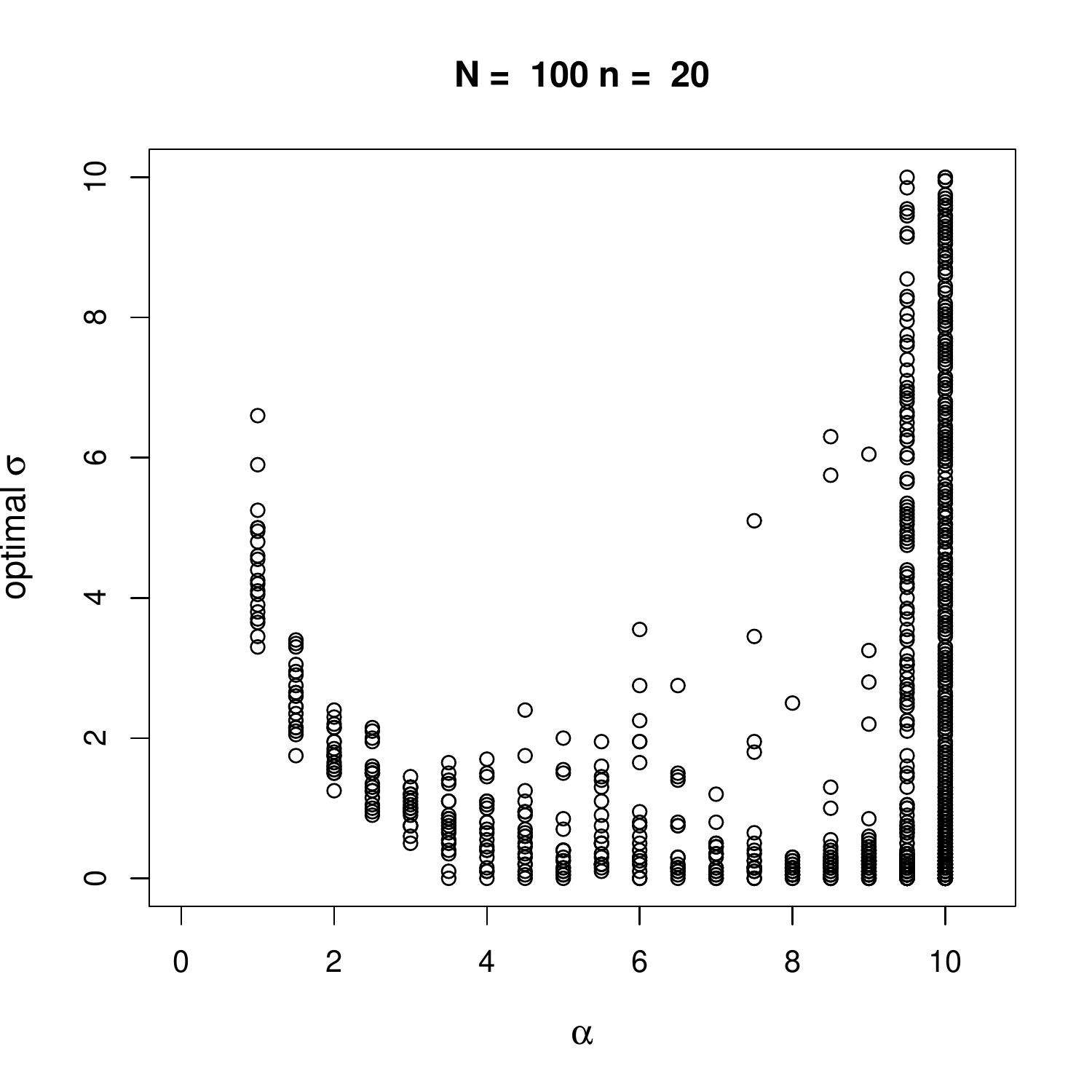}
			
		\end{minipage} 
		\hfill
		\begin{minipage}[t]{.3\textwidth}
			\centering
			\includegraphics[width=\textwidth]{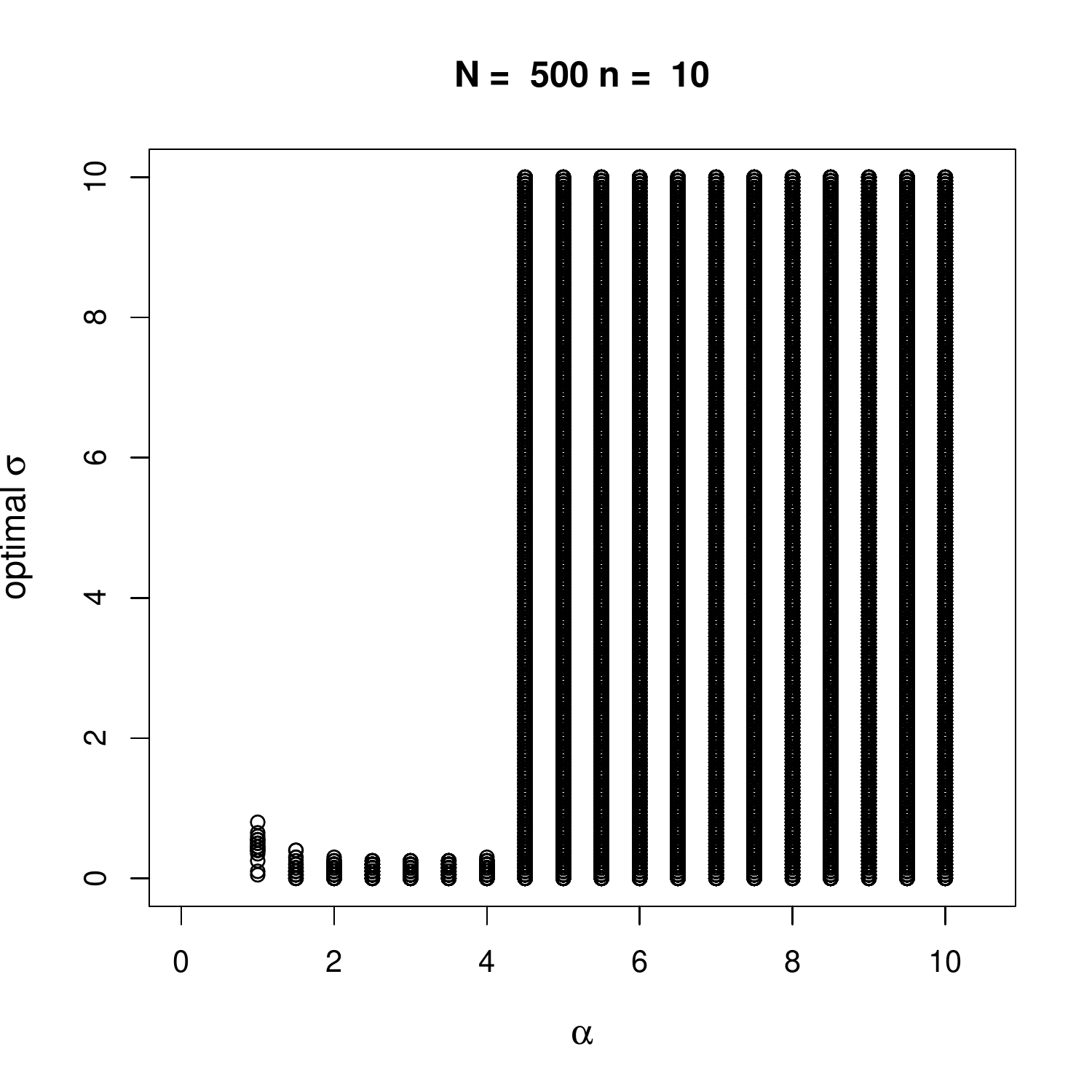}
			
		\end{minipage} 
		
	}	
	\caption{The optimal $\sigma$ value as a function of $\alpha$, for different combinations of values of $N$ and $n$.}
		\label{fig:optimSigma}
\end{figure}

Figure \ref{fig:optimSigma} illustrates how the optimal $\sigma$ changes with respect to $\alpha^0$, $N$ and $n$. In each panel, for the chosen $N$ and $n$, we generate datasets by sampling from the Mallows distribution, on a grid of different values of $\alpha^0$. Ten datasets are generated for each value of $\alpha^0$. For each of these datasets, a grid of $\sigma$ values is tested, and the $\sigma$ value that results in the smallest marginal KL-divergence is plotted on the y-axis- resulting in 10 values for each value of $\alpha$ on the x-axis. We see how the optimal $\sigma$ depends on $\alpha^0$, $N$ and $n$. In particular, $\sigma$ decreases towards 0 as $N$ increases and $\alpha^0$ increases, and as $n$ decreases. When $\alpha^0$ and $N$ are sufficiently large and/or $n$ is sufficiently small, the different choices of $\sigma$ do not affect the KL-divergence. We can in other words safely use $\sigma = 0$ in such situations.

A description of the complete Pseudo-Mallows algorithm for sampling $\bm{\rho}$ based on full ranking data is summarized in Algorithm \ref{algo:full_data}.

\subsection{Preliminary estimation of \texorpdfstring{$\alpha$}{Lg}} \label{sec:alpha_tuning_full}

When given a ranking dataset where $\alpha$ is unknown, an estimate of $\alpha$ is required before the Pseudo-Mallows method can be applied to make inference on the consensus parameter $\bm{\rho}$. In this section, we propose a method to  estimate $\alpha$ when given a ranking dataset.

The scale parameter $\alpha$ describes the ``peakness'' of the Mallows distribution: a larger value of $\alpha$ indicates that the individual rankings are more closely distributed around the group consensus, while a smaller value of $\alpha$ indicates a flatter distribution. In other words, when $\alpha$ is large, the individual rankings are more similar to each other compared to when $\alpha$ is smaller.  We can measure the pairwise user similarity between two vectors $\mathbf{R}^j$  and $\mathbf{R}^k$ using cosine similarity, defined as 

\begin{center} \label{eq:pairwise_sim_full}
	sim($\mathbf{R}^j$, $\mathbf{R}^k$) = $\frac{\mathbf{R}^j \cdot \mathbf{R}^k}{|\mathbf{R}^j|\cdot |\mathbf{R}^k|}$.
\end{center}

To obtain a quick estimation of $\alpha$, we assume that the mean pairwise user cosine similarity of a ranking dataset resembles that of a simulated ranking dataset generated by drawing independent samples from the Mallows distribution with the same number of items $n$, and the same value of $\alpha$, and any chosen $\rho^0$. The mean pairwise user cosine similarity for $N$ users is defined as:

\begin{center}\label{eq:mean_pairwise_sim_full}
	$Sim_{\alpha^0}$ = $\frac{1}{N(N-1)} \sum\limits_{j=1}^{N} \sum\limits_{k\neq j}$ 	sim($\mathbf{R}^j$, $\mathbf{R}^k$).
\end{center}

To find an approximation of $\alpha$, we first generate full ranking datasets of $N$ users and $n$ items by drawing from the Mallows distribution with a grid of $\alpha^0$ values, and a shared consensus parameter $\rho^0$, which is fixed as $\{1, ...,n\}$ for convenience. We then calculate the mean pairwise user cosine similarity of each simulated ranking dataset, and compare these with the mean pairwise user cosine similarity of the real data set to find the closest one, and use the corresponding $\alpha^0$ as estimate. Note that the number of users $N$ in each simulated dataset is unimportant, the only requirement for $N$ is that it should be large enough such that a good estimate of the mean pairwise user similarity can be obtained.

\section{Extension to clicking data} \label{sec:clicking_data}

We now move to incomplete data, and focus on clicking data, where a click on an item is interpreted as a preference for that item. Bayesian inference with the Mallows model has already been extended to clicking data \citep{liu2019model}. Here we extend the Pseudo-Mallows distribution to clicking data and study the variational approach to inference in this case. The goal here, in addition to estimating the group consensus $\bm{\rho}$, is to learn the full ranking $\mathbf{R}^j$ of each individual $j$, on the basis of clicking data.

\subsection{Mallows augmentation for clicking data}
Suppose user $j$ has clicked on a subset of the $n$ items $\mathcal{A} = \{A_1, ..., A_n\}$. For user $j$, let the binary vector $\bm{B}^j = \{b_1^j, ..., b_n^j\}$ be defined by $b_i^j = 1$ if item $i$ is clicked and $b_i^j = 0$ otherwise. We also denote the set of items clicked by user $j$ as $\mathcal{A}^j = \{A_k: b_k^j = 1 \}$, and the set of items not clicked by user $j$ as ${\mathcal{A}^j}^c$. Furthermore, we define $c_j = |\mathcal{A}^j|$ and ${c_j}^c := |{\mathcal{A}^j}^c| = n-c_j$.

Inference for the Mallows posterior in the clicking data case can be carried out through data augmentation. We first assume that each user $j$ has a latent ranking of all items $\{R^j_1, ..., R^j_n\}$ in mind, so that the user clicking behavior is a consistent manifestation of that latent ranking. This consistency is ensured by assuming in practice that, for each user $j$, all  clicked items  are ranked higher than those not clicked, i.e., $R^j_i < R^j_k$, $ \forall A_i \in \mathcal{A}^j$ and $ A_k \in {\mathcal{A}^j}^c$. Within the groups of clicked or non-clicked items, no ordering is assumed. In this way, for each user $j$, $R_i^j \leq c_j$, $\forall i: A_i \in \mathcal{A}^j$ and $R_k^j \geq c_j+1$, $\forall k: A_k \in {\mathcal{A}^j}^c$. We denote the set of rankings that are compatible with the binary clicks of user $j$ as $\mathcal{S}(\bm{B}^j) = \{\mathbf{R}^j \in \mathcal{P}_n: R^j_i \leq c_j$,  $\forall i: A_i \in \mathcal{A}^j; R^j_k \geq c_j + 1$, $ \forall k: A_k \in {\mathcal{A}^j}^c \}$. The posterior distribution given the binary clicking data can therefore be written as
\begin{equation} \label{eq: aug_Mallows}
P(\bm{\rho}|\alpha, \bm{B}^1, ...,\bm{B}^N ) =\sum\limits_{\mathbf{R}^1 \in \mathcal{S}(\bm{B}^1)}...\sum\limits_{\mathbf{R}^N \in \mathcal{S}(\bm{B}^N)} P(\bm{\rho}|\alpha, \mathbf{R}^1, ...,\mathbf{R}^N) P(\mathbf{R}^1, ...,\mathbf{R}^N|\bm{B}^1. ...,\bm{B}^N)
\end{equation}

To make inference on (\ref{eq: aug_Mallows}), the MCMC algorithm alternates between two steps: (i) estimate the full rankings for each user $\{\mathbf{R}^1, ..., \mathbf{R}^N\}$ based on the current estimation of $\bm{\rho}$ and on $\{\bm{B}^1, ..., \bm{B}^N\}$; (ii) update the estimate of $\bm{\rho}$ based on the current estimate of $\{\mathbf{R}^1, ..., \mathbf{R}^N\}$.
More precisely, in step (i), based on the current estimation of $\bm{\rho}$, the MCMC draws a sample for each user $j$ independently using the Metropolis-Hasting algorithm from the distribution
\begin{equation}\label{eq:pseudolikelihood_ind_Mallows}
P(\mathbf{R}^j|\alpha, \bm{\rho}, \bm{B}^j) = P(\mathbf{R}^j|\alpha, \bm{\rho})\mathbb{1}_{\mathbf{R}^j \in \mathcal{S}(\bm{B}^j)}.
\end{equation}
In step (ii), the sampling procedure is the same as when inferring for the group consensus $\bm{\rho}$ given full ranking data, as in Section \ref{sec:Mallows_full}.

\subsection{Pseudo-Mallows for clicking data}\label{sec:aug_ind_pseudo}
Inference for the Pseudo-Mallows for clicking data follows a similar augmentation scheme.
As mentioned, step (ii) in the augmentation scheme is practically the same as inferring the group consensus parameter $\bm{\rho}$ given the current estimation of $\{\mathbf{R}^1, ..., \mathbf{R}^N\}$. Therefore we follow Algorithm \ref{algo:full_data}, which samples for $\bm{\rho}$ based on full ranking data, to achieve the best approximation to step (ii). 

Regarding step (i), i.e., to approximate the Mallows posterior described in equation (\ref{eq:pseudolikelihood_ind_Mallows}), we draw a sample for each user $j$ independently from 
\begin{equation}\label{eq:pseudolikelihood_ind}
Q^R(\mathbf{R}^j|\alpha, \bm{\rho}, \bm{B}^j) = \sum\limits_{\{i_1, ..., i_n\}^j\in \mathcal{P}_n}f(\mathbf{R}^j|\alpha, \bm{\rho}, \bm{B}^j)%.g^j(\{i_1, ..., i_n\}).
\end{equation}
where $f(\cdot)$ is specified in the following. Given the assumption that all clicked items must be ranked higher than all unclicked items, $f(\cdot)$ factorizes into two independent, identical distributions, one for each group:
\begin{equation}\label{eq:aug_ind_factorization}
\begin{aligned} 
&f({\mathbf{R}^j}| \alpha, \bm{\rho},\bm{B}^j)  \\=
&{q^R({\mathbf{R}^j_{\mathcal{A}^j}}|\alpha,\bm{\rho}^{t,j},\bm{i}^{t,j}) }\cdot g(\bm{i}^{t,j}) \cdot q^R({\mathbf{R}^j_{{\mathcal{A}^j}^c}} - c_j|\alpha,\bm{\rho}^{b,j},\bm{i}^{b,j})\cdot g(\bm{i}^{b,j}), \end{aligned}
\end{equation}
where $\bm{\rho}^{t,j}$ := rank($\bm{\rho}_{\mathcal{A}^j})\in \mathcal{P}_{c_j}$,  $\bm{\rho}^{b,j}$ := rank($\bm{\rho}_{{\mathcal{A}^j}^c})\in \mathcal{P}_{c_j^c}$,  $\bm{i}^{t,j} := \{i_1^{t,j},..., i_{c_j}^{t,j}\} \in \mathcal{P}_{c_j}$, and $\bm{i}^{b,j} := \{i_1^{b,j},..., i_{c_j^c}^{b,j}\} \in \mathcal{P}_{{c_j}^c}$. Note that the input to $q^R(\cdot)$ for the unclicked items is the ranking vector $\mathbf{R}^j_{{\mathcal{A}^j}^c}$ shifted by $c_j$, which takes values in $\mathcal{P}_{c_j^c}$ after the shift.  
Here, $q^R(\cdot)$ follows the same type of item-by-item factorization form as $q(\cdot)$ in the Pseudo-Mallows for $\bm{\rho}$, and $g(\cdot)$ is some arbitrary distribution on the space of permutations defined on the respective number of items (see Section \ref{sec:g_distribution} for more details).
Following the strategy for defining $q(\cdot)$ in the Pseudo-Mallows for $\bm{\rho}$, each term in $q^R(\cdot)$ can be factorized as follows: 
\begin{equation}\label{eq:factorisation_ind_simplified}
\begin{aligned}
&q^R(\mathbf{R}^j_{{\mathcal{A}^j}}| \alpha, \bm{\rho}^{t,j}, \bm{i}^{t,j})  \\
= &q^R(\mathbf{R}^j_{{\mathcal{A}^j}}|\alpha, \bm{\rho}^{t,j}, \bm{o}^{t,j})  \\
=&q^R({R}^j_{o^{t,j}_1}|\alpha,{\rho^{t,j}_{o^{t,j}_1}},o^{t,j}_1) \cdot
q^R({R}^j_{o^{t,j}_2}|\alpha,{\rho^{t,j}_{o^{t,j}_2}},o^{t,j}_2,{R}^{j}_{o^{t,j}_1}) \cdot
... \cdot \\
&q^{R}({R}^j_{o^{t,j}_{c_j-1}}|\alpha,{\rho^{t,j}_{o^{t,j}_{c_j-1}}},o^{t,j}_{c_j-1}, {R}^j_{o^{t,j}_1},...,R^j_{o^{t,j}_{c_j-2}}) \cdot
q^{R}({R}^j_{o^{t,j}_{c_j}}|\alpha,\rho^{t,j}_{o^{t,j}_{c_j}},o^{t,j}_{c_j}, {R}^j_{o^{t,j}_1},...,R^j_{o^{t,j}_{c_j-1}});\\
&q^R(\mathbf{R}^j_{{\mathcal{A}^j}^c}-c_j| \alpha, \bm{\rho}^{b,j}, \bm{i}^{b,j})  \\
= &q^R(\mathbf{R}^j_{{\mathcal{A}^j}^c}-c_j|\alpha, \bm{\rho}^{b,j}, \bm{o}^{b,j})  \\
=&q^R({R}^j_{o^{b,j}_1}|\alpha,{\rho^{b,j}_{o^{b,j}_1}},o^{b,j}_1) \cdot
q^R({R}^j_{o^{b,j}_2}-c_j|\alpha,{\rho^{b,j}_{o^{b,j}_2}},o^{b,j}_2,{R}^{j}_{o^{b,j}_1}) \cdot
... \cdot \\
&q^{R}({R}^j_{o^{b,j}_{c_j^c-1}}-c_j|\alpha,{\rho^{b,j}_{o^{b,j}_{c_j^c-1}}},o^{b,j}_{c_j^c-1}, {R}^j_{o^{b,j}_1},...,R^j_{o^{b,j}_{c_j-2}}) \cdot
q^{R}({R}^j_{o^{b,j}_{c_j^c}}|\alpha,\rho^{b,j}_{o^{b,j}_{c_j^c}},o^{b,j}_{c_j^c}, {R}^j_{o^{b,j}_1},...,R^j_{o^{b,j}_{c_j^c-1}});\\
\end{aligned}
\end{equation}where 
\begin{equation}\label{eq:qtj}
\begin{aligned}
q^R({R}^j_{o^{t,j}_1}|\alpha,{\rho^{t,j}_{o^{t,j}_1}},o^{t,j}_1) 
= &\frac{\text{exp}\{- \frac{\alpha}{n}d(R^j_{o^{t,j}_1}, {\rho}^{t,j}_{o^{t,j}_1})\}}
{\sum\limits_{{r}\in \{1, .., c_j\}}\text{exp}\{- \frac{\alpha}{n}d( {r}, \rho^{t,j}_{o^{t,j}_1})\}}\mathbb{1}_{{R^j_{o^{t,j}_1}}\in \{1, ...,c_j\}}\\
q^R({R}^j_{o^{t,j}_k}|\alpha,{\rho}^{t,j}_{o^{t,j}_k}, o^{t,j}_k, {R}^{j}_{o^{t,j}_1}, ..., {R}^{j}_{o^{t,j}_{k-1}}) 
= &\frac{\text{exp}\{- \frac{\alpha}{n}d(R^j_{o^{t,j}_k}, {\rho}^{t,j}_{o^{t,j}_k})\}}
{\sum\limits_{{r}\in \{1, .., c_j\}\setminus \{{R}^j_{o^{t,j}_1}, ..., {R}^j_{o^{t,j}_{k-1}}\}}\text{exp}\{- \frac{\alpha}{n}d(r, {\rho}^{t,j}_{o^{t,j}_k})\}}\\\cdot &\mathbb{1}_{{R}^j_{o^{t,j}_k}\in\{1, .., c_j\}\setminus \{{R}^j_{o^{t,j}_1}, ..., {R}^j_{o^{t,j}_{k-1}}\}}, \text{ for }k = 2, ..., c_j; \\
q^R({R}^j_{o^{b,j}_1}-c_j|\alpha,{\rho^{b,j}_{o^{b,j}_1}},o^{b,j}_1) 
= &\frac{\text{exp}\{- \frac{\alpha}{n}d(R^j_{o^{b,j}_1}, {\rho}^{b,j}_{o^{b,j}_1})\}}
{\sum\limits_{{r}\in \{1, .., c_j^c\}}\text{exp}\{- \frac{\alpha}{n}d( {r}, \rho^{b,j}_{o^{b,j}_1})\}}\mathbb{1}_{{R^j_{o^{b,j}_1}}\in \{1, ...,c_j^c\}}\\
q^R({R}^j_{o^{b,j}_k}-c_j|\alpha,{\rho}^{b,j}_{o^{b,j}_k}, o^{b,j}_k, {R}^{j}_{o^{b,j}_1}, ..., {R}^{j}_{o^{b,j}_{k-1}}) 
= &\frac{\text{exp}\{- \frac{\alpha}{n}d(R^j_{o^{b,j}_k}, {\rho}^{b,j}_{o^{b,j}_k})\}}
{\sum\limits_{{r}\in \{1, .., c_j^c\}\setminus \{{R}^j_{o^{b,j}_1}, ..., {R}^j_{o^{b,j}_{k-1}}\}}\text{exp}\{- \frac{\alpha}{n}d(r, {\rho}^{b,j}_{o^{b,j}_k})\}}\\\cdot &\mathbb{1}_{{R}^j_{o^{b,j}_k}\in\{1, .., c_j^c\}\setminus \{{R}^j_{o^{b,j}_1}, ..., {R}^j_{o^{b,j}_{k-1}}\}}, \text{ for }k = 2, ..., c_j^c; \\
\end{aligned}
\end{equation}

\subsection{\texorpdfstring{$g(\cdot)$}{Lg} distributions for individual data augmentation}\label{sec:g_distribution}
In order to use (\ref{eq:aug_ind_factorization}) for performing inference on the users' individual rankings, we need to specify the user-specific distribution $g(\cdot)$ for the items ordering in the Pseudo-Mallows. We suggest to use the uniform distribution defined on $\mathcal{P}_{c_j}$ and $\mathcal{P}_{c_j^c},$ respectively, for the clicked and unclicked groups of items for each user $j$. Specifically:
\begin{center}
	$g(i_1, ..., i_n) = g(\bm{i}^{t,j})g(\bm{i}^{b,j})=\left \{
	\begin{aligned}
	&\frac{1}{c_j!}\frac{1}{c_j^c!}, &&\text{if } \bm{i}^{t,j} \in \mathcal{P}_{c_j} \text{ and } \bm{i}^{b,j} \in \mathcal{P}_{c_j^c}\\
	&0 , && \text{otherwise. }
	\end{aligned} \right.$
\end{center}

As clarified in the remainder of this section, this choice leads to an effective approximation of the Mallows posterior by means of the Pseudo- Mallows distribution in the augmentation step. 

Indeed, the uniform distribution for $g(\cdot)$ gives the best approximation to the Mallows posterior distribution for user $j$'s full ranking $\mathbf{R}^j$, in terms of maximizing the ELBO, given that the rankings underlying the clicks come from a Mallows distribution. The intuitive motivation is similar to the reasoning why the $\mathcal{V}$-set construction gives rise to the best approximation for the consensus parameter $\bm{\rho}$ when full individual rankings are available. However in this case, considering the clicked group, when inferring $\mathbf{R}^j$ using (\ref{eq:aug_ind_factorization}) the mode of the $k$-th term of \eqref{eq:qtj} is $\rho_{o_k}$, for all $k$. For the unclicked group, the argument follows the same logic. Hence, no systematic rules regarding the sequence following which items should be sampled for (such as the $\mathcal{V}$-set construction) are needed, and the uniform distribution is the most effective choice. In fact, systematic rules would actually introduce systematic bias, as will be shown in the following experiment.

To conduct this experiment, we draw independent full rankings $\mathbf{R}^j$ from (\ref{eq:pseudolikelihood_ind}), and to illustrate the impact of different $g(\cdot)$ distributions in the augmentation step,  we use different $g(\cdot)$ functions in the sampling, with known $\alpha^0$ and $\bm{\rho}^0$. We also draw independent full ranking samples using the Mallows likelihood with the same parameters $\alpha^0$ and $\bm{\rho}^0$, and monitor how the distribution of the full ranking samples resulting from different $g(\cdot)$ distributions differ from the distribution based on the Mallows samples.

For each $g(\cdot)$ distribution, we sample for $N=1000$ full rankings of $n = 20$ items based on $\alpha^0 = 15$ and $\bm{\rho}^0 = \{1, 2, ..., n\}$. The different $g(\cdot)$ distributions that we consider are:
\begin{itemize}
	\item{Uniform: for each $\mathbf{R}^j$ sample, we sample one random ranking uniformly from $\mathcal{P}_n$}
	\item {V: for each $\mathbf{R}^j$ sample, one V-ranking \{$i_1, ...,i_n$\} is sampled uniformly from $\mathcal{V}_{\bm{\rho}^0}$}
% 	\item {opp\_V: for each $\mathbf{R}^j$ sample, we first sample one V-ranking \{$i_1, ...,i_n$\} based on $\bm{\rho}^0$, then we generate the opposite V-ranking \{$n+1 - i_1, ..., n+1 - i_n$\}}
	\item{top\_bottom: \{$i_1, ..., i_n$\} = $\bm{\rho}^0$} for all $\mathbf{R}^j$ samples
\end{itemize}

\begin{figure}[hbt!]
		\begin{minipage}[t]{.45\textwidth}
			\centering
			\includegraphics[width=\textwidth]{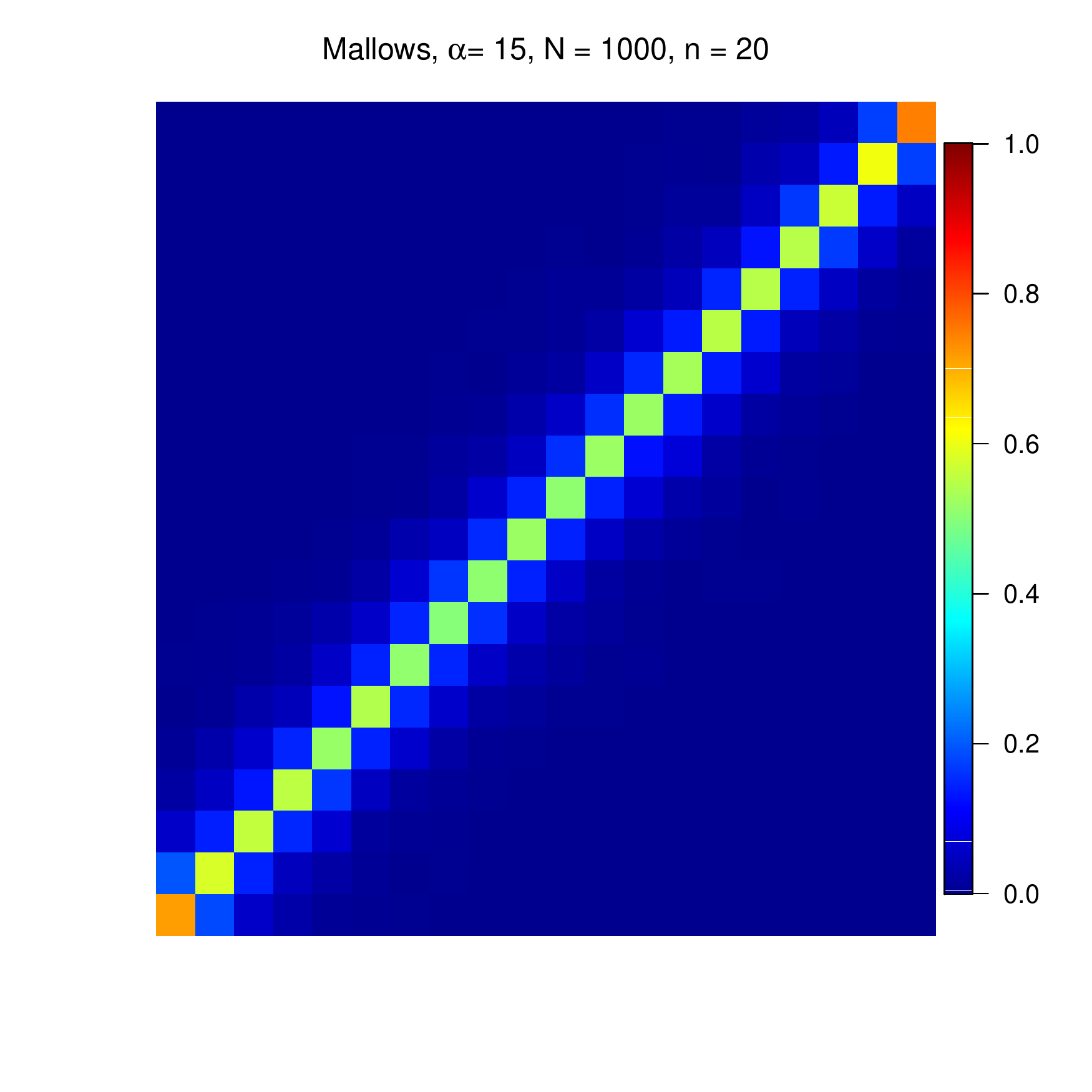}
			
		\end{minipage}
		\hfill
		\begin{minipage}[t]{.45\textwidth}
			\centering
			\includegraphics[width=\textwidth]{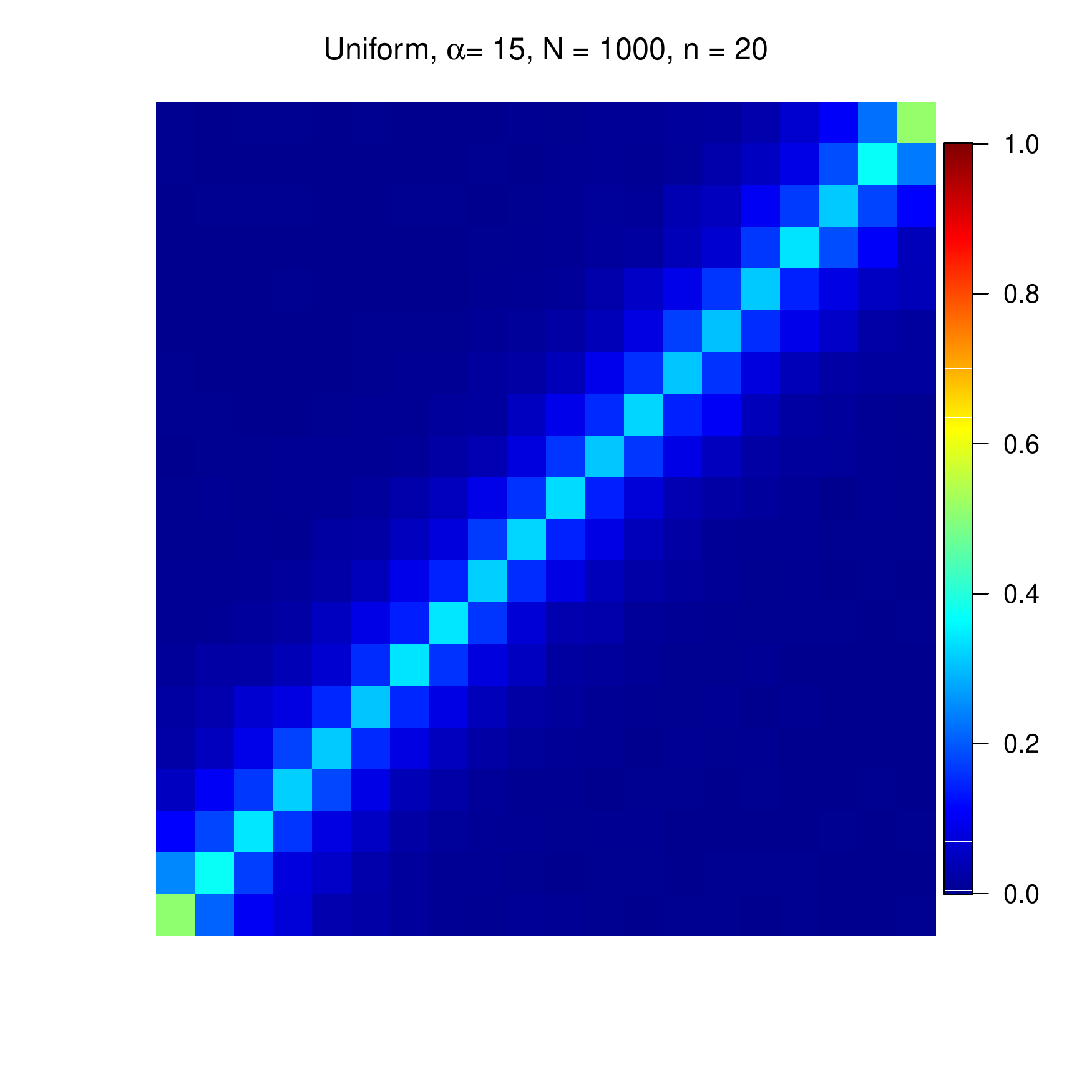}
			
		\end{minipage} 
		\hfill
		\begin{minipage}[t]{.45\textwidth}
			\centering
			\includegraphics[width=\textwidth]{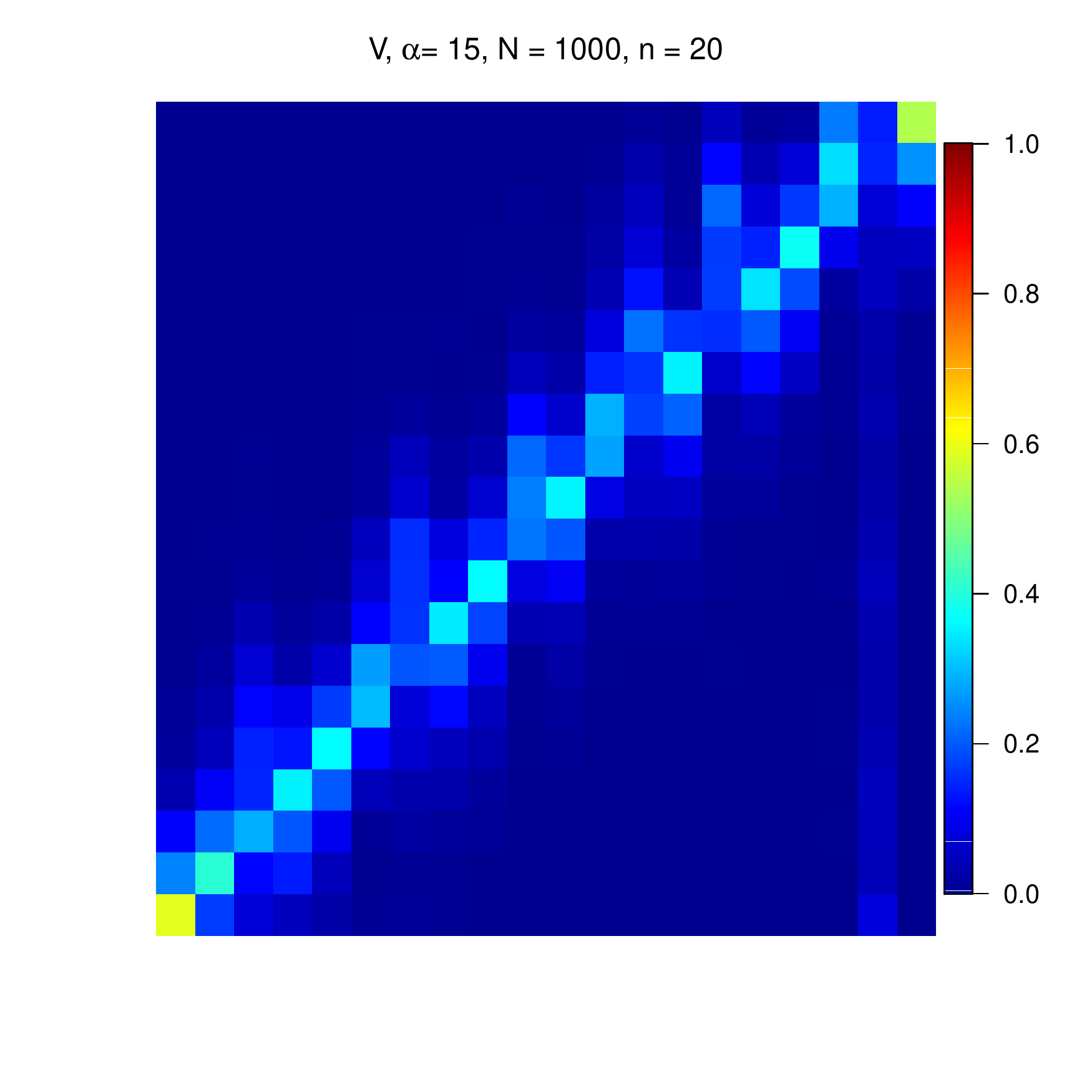}
		\end{minipage} 
% 		\hfill
% 		\begin{minipage}[t]{.3\textwidth}
% 			\centering
% 			\includegraphics[width=\textwidth]{figures/ind_ordering_full_sim/Rj_N1000n20alpha15methodOpp_V.pdf}
% 		\end{minipage} 
		\hfill
	\begin{minipage}[t]{.45\textwidth}
	\centering
	\includegraphics[width=\textwidth]{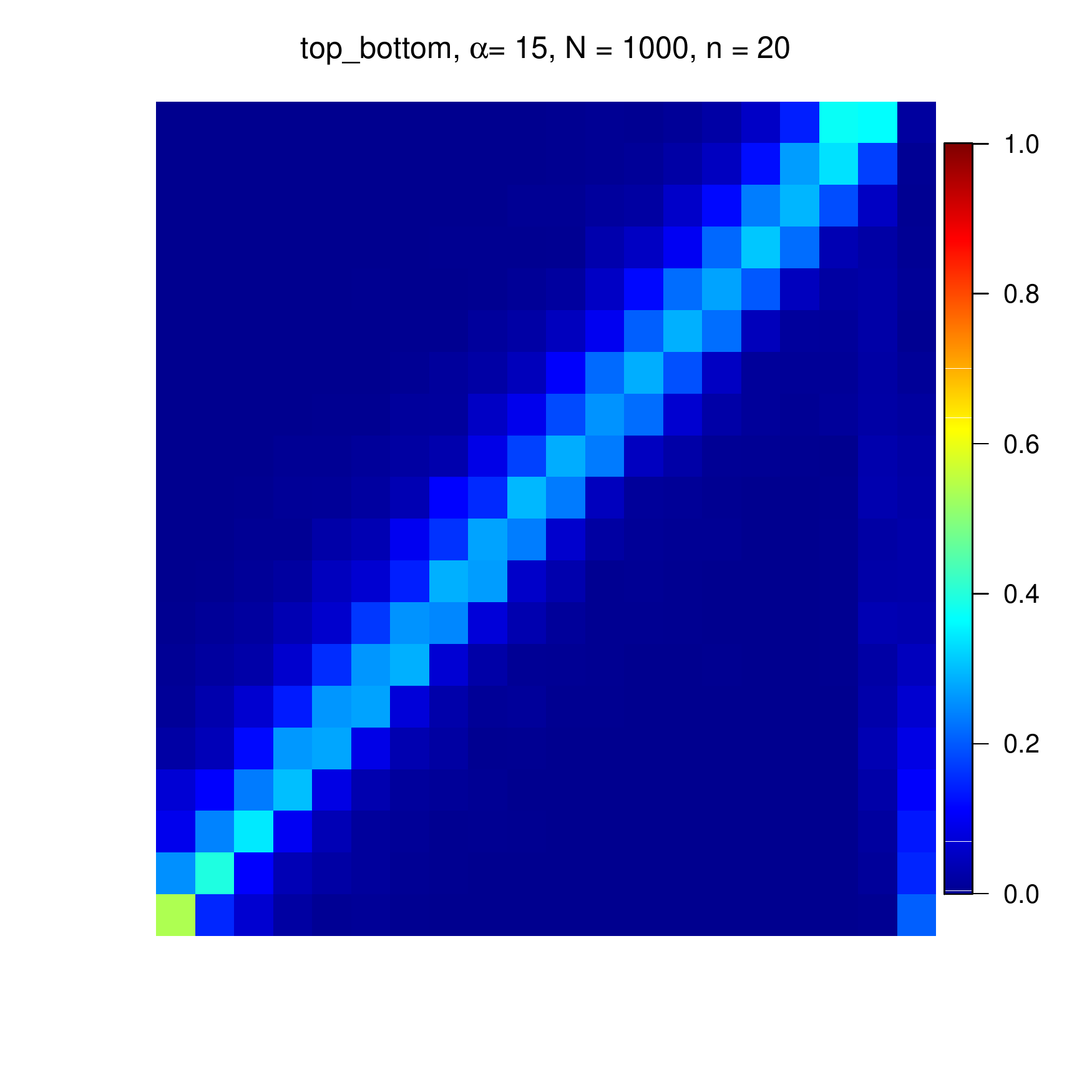}
	\end{minipage} 

	\caption{Heat plots of $P(R^j_i|\alpha^0, \bm{\rho}^0)$, when the $\mathbf{R}^j$'s are sampled from the Mallows (top-left), or from the Pseudo-Mallows with different $g(\cdot)$ (top-right: uniform; bottom-left: ``V''-rankings; bottom-right: top-bottom ranking). $N = 2000$ samples, $n = 20$ items, $\alpha^0 = 15$. }
	\label{fig:heatPlot_ind_full}
\end{figure}

Based on the $\mathbf{R}^j$ samples, we estimate the marginal distribution of each item $P(R^j_i|\alpha^0, \bm{\rho}^0)$, and we show the heat plots of these marginal distributions in Figure \ref{fig:heatPlot_ind_full}. The items are arranged according to their rankings in $\bm{\rho}^0$ on the x-axis.

It can be observed that when individual rankings are drawn from the Mallows distribution, the distribution of $\mathbf{R}^j$ is centred on $\bm{\rho}^0$, and the same is achieved when using the Pseudo-Mallows with uniform orderings for each sample. If we take a closer look, it can also be discovered that using any choice of the $g(\cdot)$ distribution other than the uniform distribution introduces some biases in the resulting Pseudo-Mallows distribution. As shown in Figure \ref{fig:heatPlot_ind_full}, the distribution of the items that tend to be the last sampled ones, is subject to the largest deviation from its position in $\bm{\rho}^0$. See for example the top- and bottom-ranked items using the ``V''-rankings, and the bottom-ranked items using the top-bottom ranking. 

A summary of the algorithm for estimating both individual rankings $\{R_1, ..., R_N\}$ and $\bm{\rho}$ from clicking data is described in Algorithm \ref{algo:sampleClicking}.

\subsection{Making personalized recommendations} \label{sec:personalRec}
Both the Bayesian Mallows and the Pseudo-Mallows can estimate, for each individual user $j$, the distribution of its full ranking $\mathbf{R}^j$ based on the clicking data. We can therefore make personalized recommendations, as well as estimating the recommendation uncertainty for each user. 

We continue with our assumption that for each user $j$, the $c_j$ clicked items are top ranked, and are not considered further for recommendation. In order to make $k$ recommendations for each user $j$, we need to infer the items that user $j$  would have ranked between $c_j+1$ and $c_j+k$. For each user $j$, we can calculate for each item $i$ its probability to be ranked between $c_j+1$ and $c_j+k$ given $\alpha$ and the clicking data $\bm{B}^1, ..., \bm{B}^N$, i.e., 
\begin{equation}\label{eq:tpp}
P(c_j \leq R^j_i \leq c_j+k|\alpha, \bm{B}^1, ..., \bm{B}^N).
\end{equation} 
The set of $k$ items with highest such probabilities can thus be recommended to the user.

\subsection{Preliminary estimation of \texorpdfstring{$\alpha$}{Lg} from clicking data} \label{sec:alpha_tuning}

Similar to the full data case, an estimate of $\alpha$ is needed before  the Pseudo-Mallows method can be applied to a clicking dataset to make inference on the group consensus parameter $\bm{\rho}$ and on the individual full ranking vector $\mathbf{R}^j$ for all users.

We continue with a similar assumption as in Section \ref{sec:alpha_tuning_full}, i.e., that the users' cosine similarity of the given clicking dataset resembles that of a clicking dataset with the same number of items $n$, generated by binarizing full Mallows rankings. As the individual data exists in the form of binary clickings, we define the pairwise user similarity between two binary vectors $\bm{B}^j$  and $\bm{B}^k$ as 

\begin{center} \label{eq:pairwise_sim}
	sim($\bm{B}^j$, $\bm{B}^k$) = $\frac{\bm{B}^j \cdot \bm{B}^k}{|\bm{B}^j|\cdot |\bm{B}^k|}$.
\end{center}

More specifically, we first generate full ranking datasets of $N$ users and $n$ items by drawing from the Mallows distribution with a grid of $\alpha^0$ values, and a shared consensus parameter $\bm{\rho}^0$, as in Section \ref{sec:alpha_tuning_full}. These datasets are then binarized by setting, for each user $j$, the top $c_j$ ranked items to 1 and the other items to 0. The number of clicked items $c_j$ for each user $j$ should come from a distribution that best reflects the number of clicks of the given dataset. Possible distributions to consider include the Poisson, the Gamma, and the exponential distributions. Lastly, we calculate the mean pairwise user cosine similarity of each simulated binary dataset, defined as

\begin{equation}\label{eq:mean_pairwise_sim}
	\text{Sim}_{\alpha^0} = \frac{1}{N(N-1)} \sum\limits_{j=1}^{N} \sum\limits_{k\neq j} 	\text{sim}(\bm{B}^j, \bm{B}^k).
\end{equation}

We can then calculate the mean pairwise user cosine similarity of the given dataset, and choose the $\alpha^0$ value of the simulated binary dataset that has the closest mean user cosine similarity. 
An example of $\alpha$ tuning using this method will be demonstrated in Section \ref{sec:NRK_result}. 

\section{Simulation and case studies} \label{sec:simulation_case_study}

\subsection{Infer \texorpdfstring{$\bm{\rho}$}{Lg} from full ranking data}\label{sec: full_sim}
In this section, we conduct experiments to compare the computation speed between the Mallows MCMC and the Pseudo-Mallows when full ranking data is given, and the task is to sample for the consensus parameter $\bm{\rho}$. We aim at assessing the Pseudo-Mallows' computation speed as well as its estimation accuracy compared to that of the Mallows MCMC.

To conduct this experiment, we first generate full ranking datasets by sampling from the Mallows distribution centered on a known consensus $\bm{\rho}^0 = \{1, ..., n\}$, with various \{$\alpha^0$, $N$, $n$\} settings. For each dataset, we then run the Mallows MCMC and the Pseudo-Mallows, as described in Algorithm \ref{algo:full_data}, for a grid of different numbers of iterations. After running the algorithms, we obtain a point estimate for the consensus parameter $\bm{\rho}$  by calculating the CP consensus \citep{sorensen2019bayesmallows} based on the samples obtained by the algorithm. Then we record the footrule distance between the CP consensus and the truth $\bm{\rho}^0$ in order to assess the estimation accuracy of both algorithms for a given computing time. 20 datasets are generated for each  setting of \{$\alpha^0$, $N$, $n$\}.

\begin{figure}[h!]
	\begin{minipage}[t]{.45\textwidth}
		\centering
		\includegraphics[width=\textwidth]{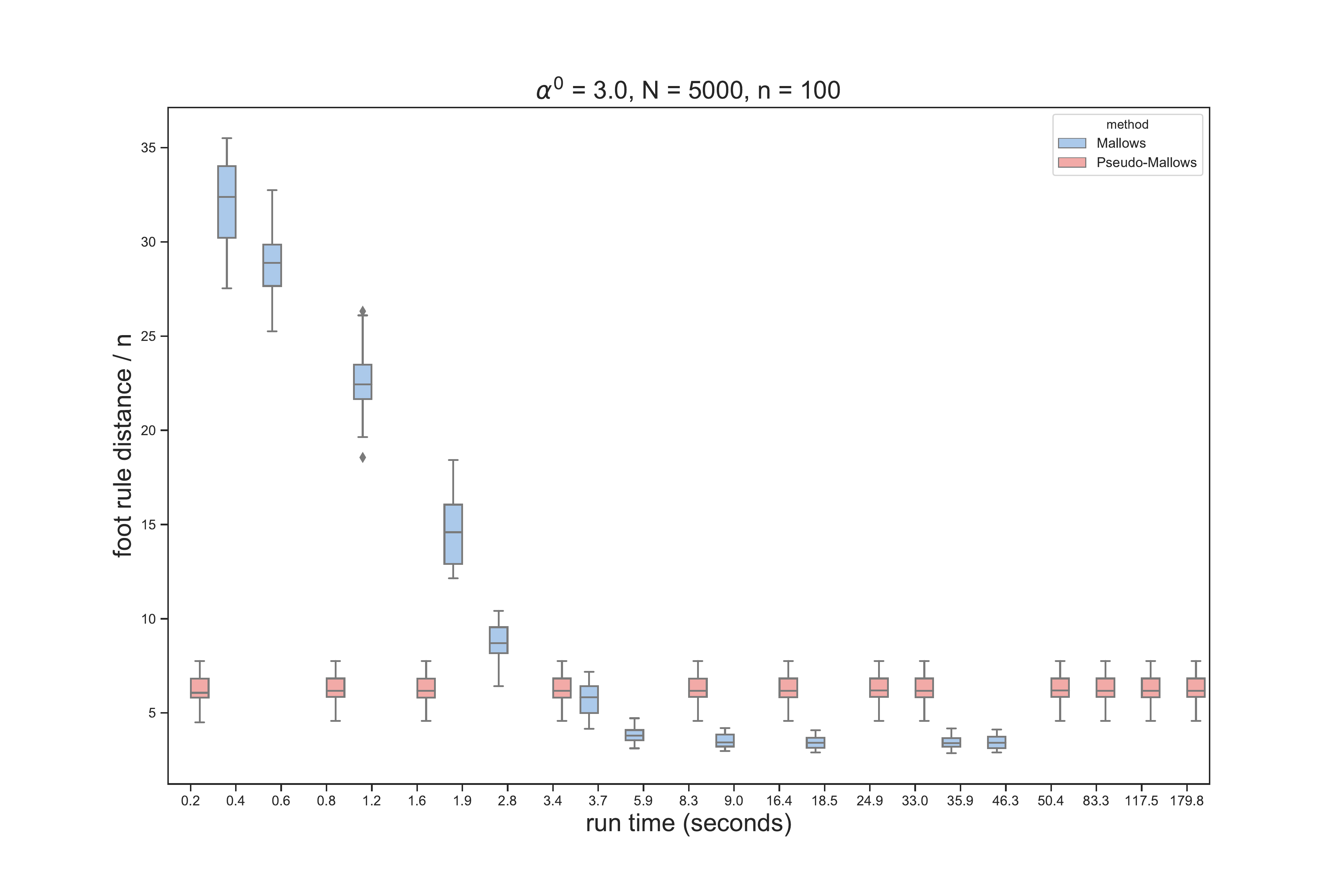}
	\end{minipage}
	\hfill
	\begin{minipage}[t]{.45\textwidth}
		\centering
		\includegraphics[width=\textwidth]{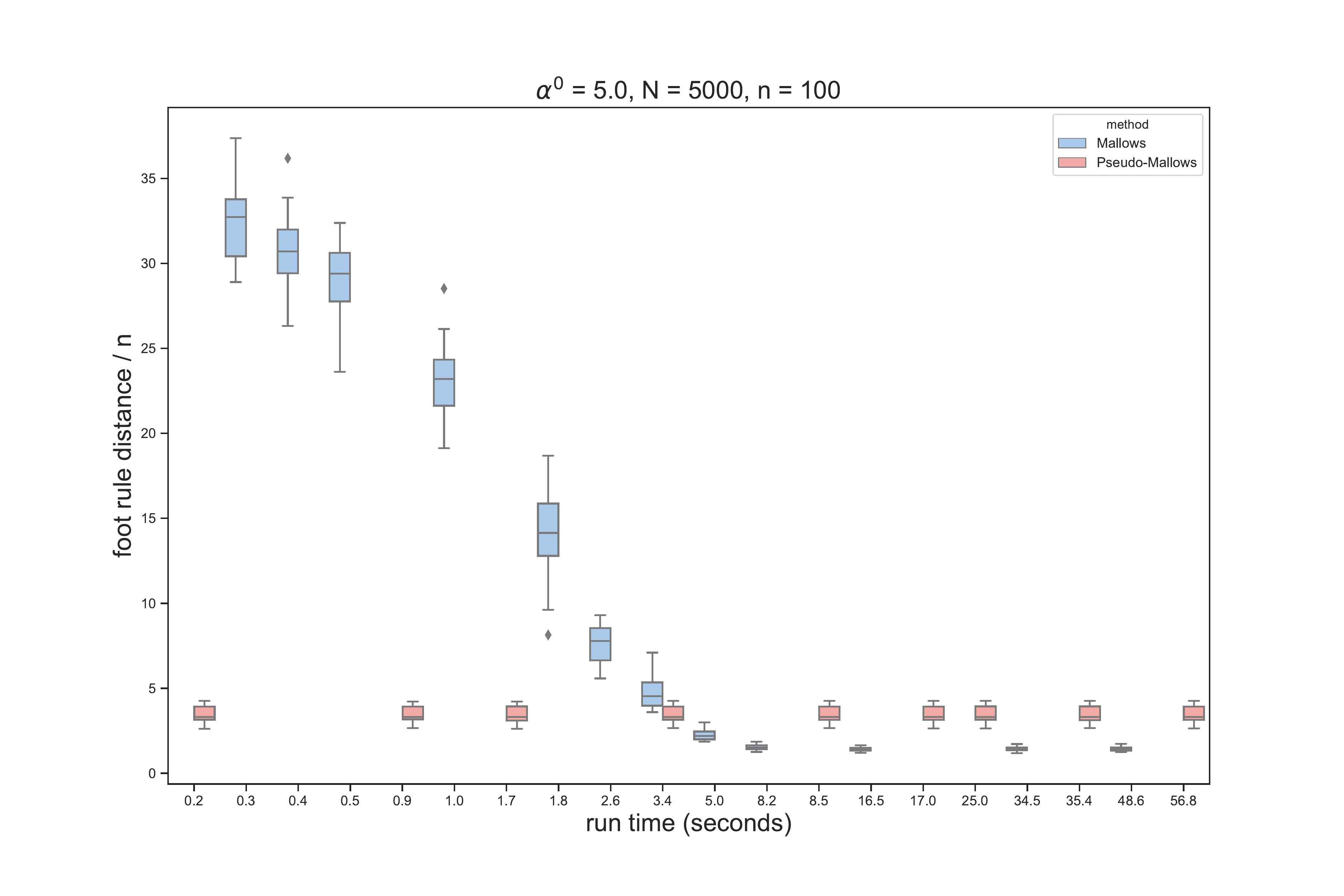}
	\end{minipage} 
	\hfill
	\begin{minipage}[t]{.45\textwidth}
		\centering
		\includegraphics[width=\textwidth]{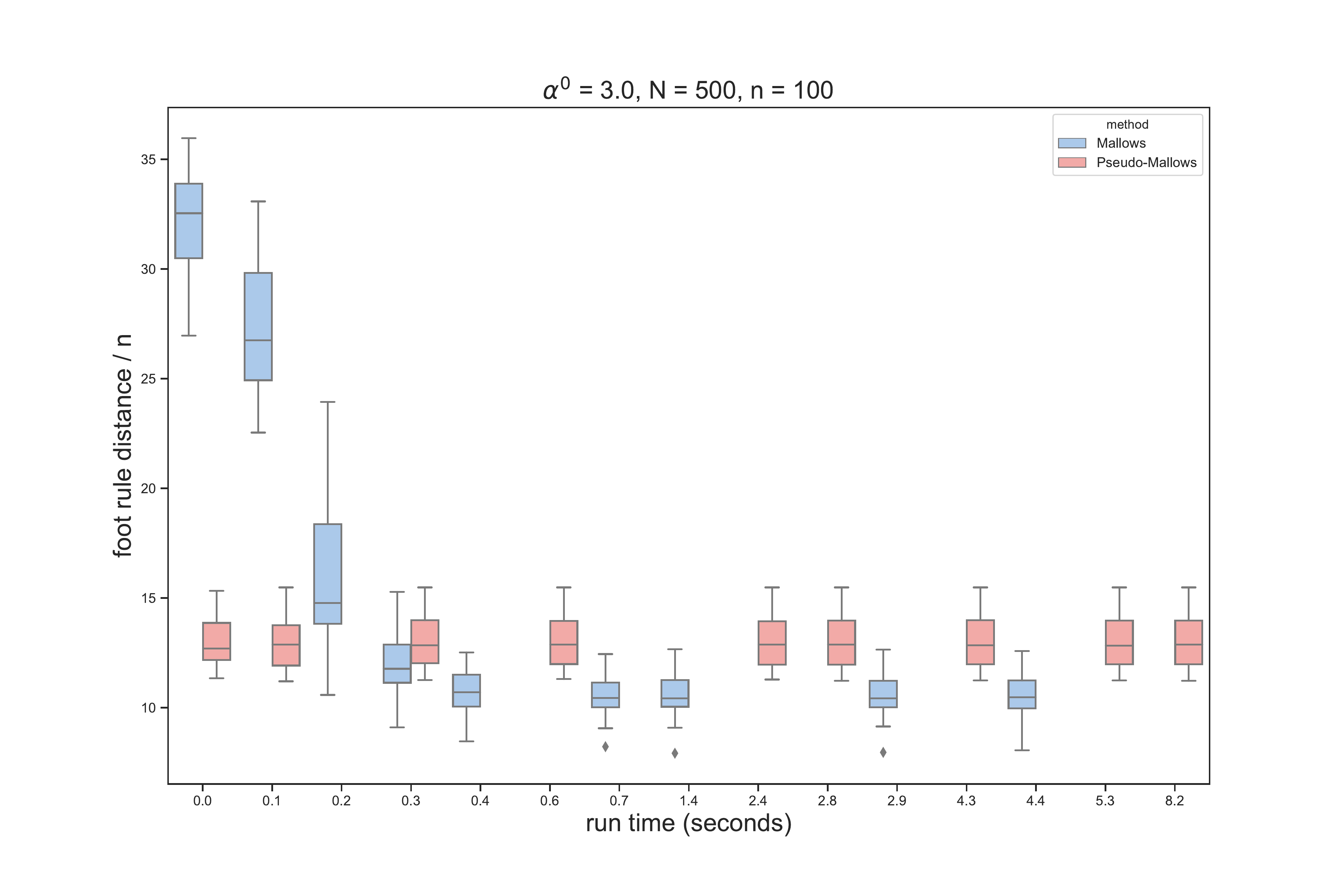}
	\end{minipage} 
	\hfill
	\begin{minipage}[t]{.45\textwidth}
		\centering
		\includegraphics[width=\textwidth]{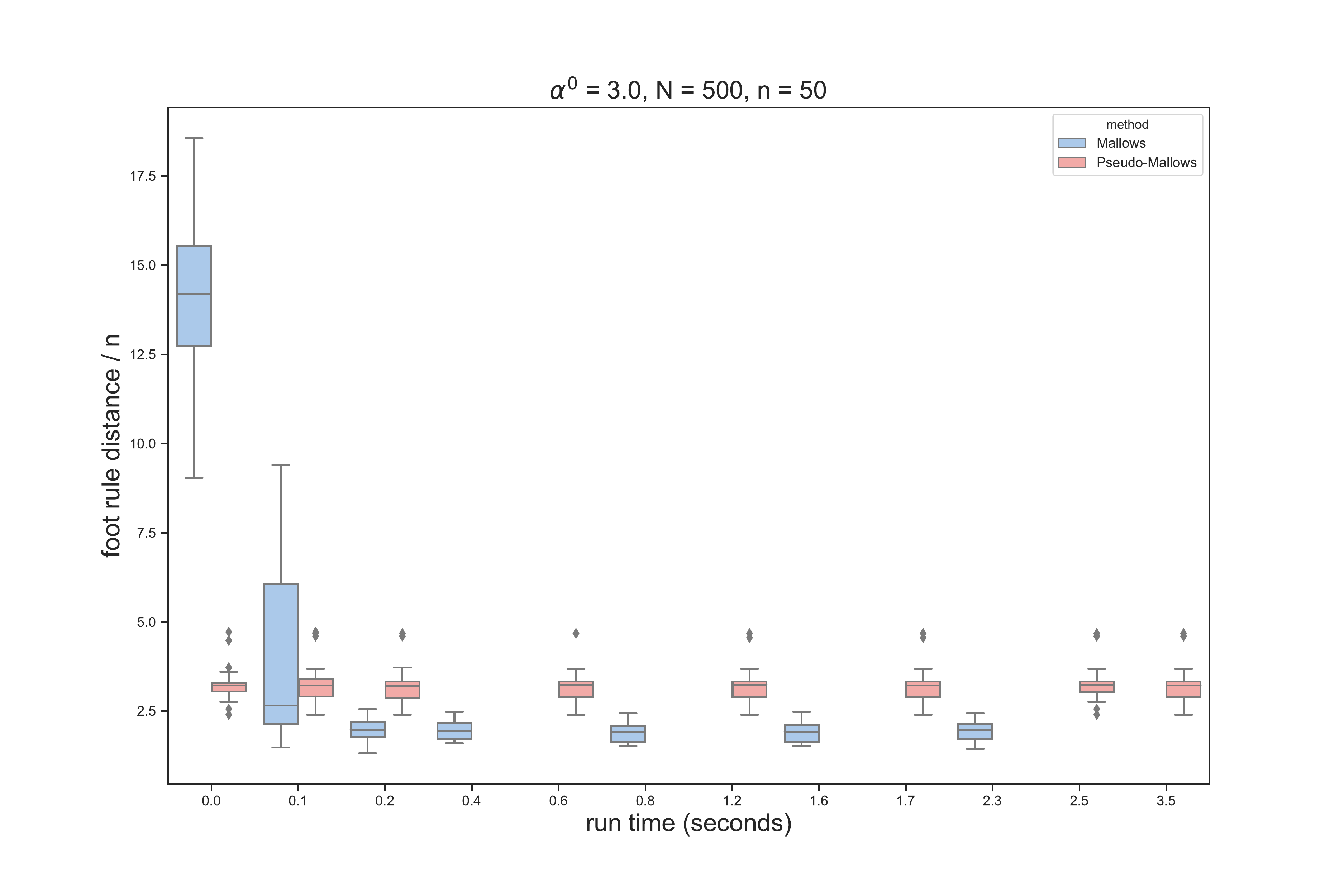}
		
	\end{minipage} 
	\caption{Results of the simulation study of Section \ref{sec: full_sim}: consensus estimation and computation time. $y$-axis: footrule distance between the CP consensus and true consensus $\bm{\rho}^0$, $x$-axis: computation time in seconds.}
	\label{fig:box_full_time}
\end{figure}

Figure \ref{fig:box_full_time} illustrates a comparison between the Mallows MCMC's and the Pseudo-Mallows' estimation accuracy of the consensus parameter $\bm{\rho}$, for varying computation time. On the $y$-axis, we record the footrule distance between the CP-consensus estimate and the true consensus $\bm{\rho}^0$ as the measure of accuracy. Computation time varies along the $x$-axis. 

Consistently through all four scenarios shown in Figure \ref{fig:box_full_time}, the Mallows MCMC's estimation accuracy initially improves as the algorithm runs for a longer period of time, to stabilise after a certain time point, corresponding to the burn-in period, i.e., the time it takes for the algorithm to converge to the target distribution. The Pseudo-Mallows on the other hand, does not have a burn-in period, and accurate estimation can be achieved in a much shorter amount of time. However, once convergence and good mixing is reached by the Mallows MCMC, its estimation of $\bm{\rho}$ is more accurate compared to using the Pseudo-Mallows. This is not surprising as the data comes from the Mallows distribution, and the Pseudo-Mallows is an approximation of the Mallows MCMC.

The efficiency of the Pseudo-Mallows is not due to the computation time per iteration. On the contrary, it takes shorter time to run 1 iteration of the Mallows MCMC. However, it requires many more samples for the Mallows MCMC to reach convergence due to the burn-in period, and even after convergence is reached, many more MCMC samples than the desired sample dimension are needed due to the high autocorrelations between consecutive samples, and to the low acceptance rate. The Pseudo-Mallows' samples, on the other hand, are drawn independently, so that the desired sample dimension corresponds to what one actually needs to sample. As a result, much fewer samples are needed. In summary, due to these reasons, the Pseudo-Mallows' algorithm is much faster compared to the Mallows MCMC. Compared to the Mallows MCMC, the Pseudo-Mallows can provide comparable estimation accuracy of the group consensus parameter in a much shorter period of time. This is especially beneficial when computation time is a priority. 

\subsection{Individual ranking estimation and personalized recommendations based on clicking data} \label{sec:sim_clicking}
In this section, we conduct a study to compare the Mallows MCMC and the Pseudo-Mallows' abilities in learning individual rankings and making personalized recommendations. 

We first create Mallows full ranking datasets of $N = 500$ users and $n = 50$ items, centered on $\bm{\rho}^0 = \{1, 2, ..., n\}$, with known $\alpha^0 = 1, 3, 5, 10$. 20 datasets are generated for each $\alpha^0$ value. The full ranking datasets are then converted to clicking data by using a truncated Poisson distribution, with a mean of $\lambda  = 5$, minimum of 1 and maximum of $n-3$, to decide on the number of clicked items $c^j$ for each user. The top-$c^j$ ranked items in $\mathbf{R}^j$ are regarded as clicked for user $j$. 

To compare the recommendation accuracy for varying computation time, for each dataset we fix the value of $\alpha^0$, and run both the Mallows MCMC and the Pseudo-Mallows  for a set of numbers of iterations, thus leading to a set of different computation times. For the Pseudo-Mallows, the $\mathcal{V}$-set is used to determine the sequence following which each item in the consensus parameter is sampled; while for each user, the uniform distribution is used to decide the sequence following which items are sampled at each iteration, as described in Section \ref{sec:aug_ind_pseudo}. Based on the samples of the full individual rankings, we calculate the posterior probabilities (\ref{eq:pseudolikelihood_ind_Mallows}) and make $k=3$ recommendations for each user.  For each user, we notate the set of $k$ recommendation as ${Rec}^j = \{a_1, ..., a_k\}^j$, and we consider a recommendation $a_i$ is correct if $R^j_{a_i} \in [c_j+1, c_j+k]$.

\begin{figure}[h!]
	
	\begin{minipage}[t]{.47\textwidth}
		\centering
		\includegraphics[width=\textwidth]{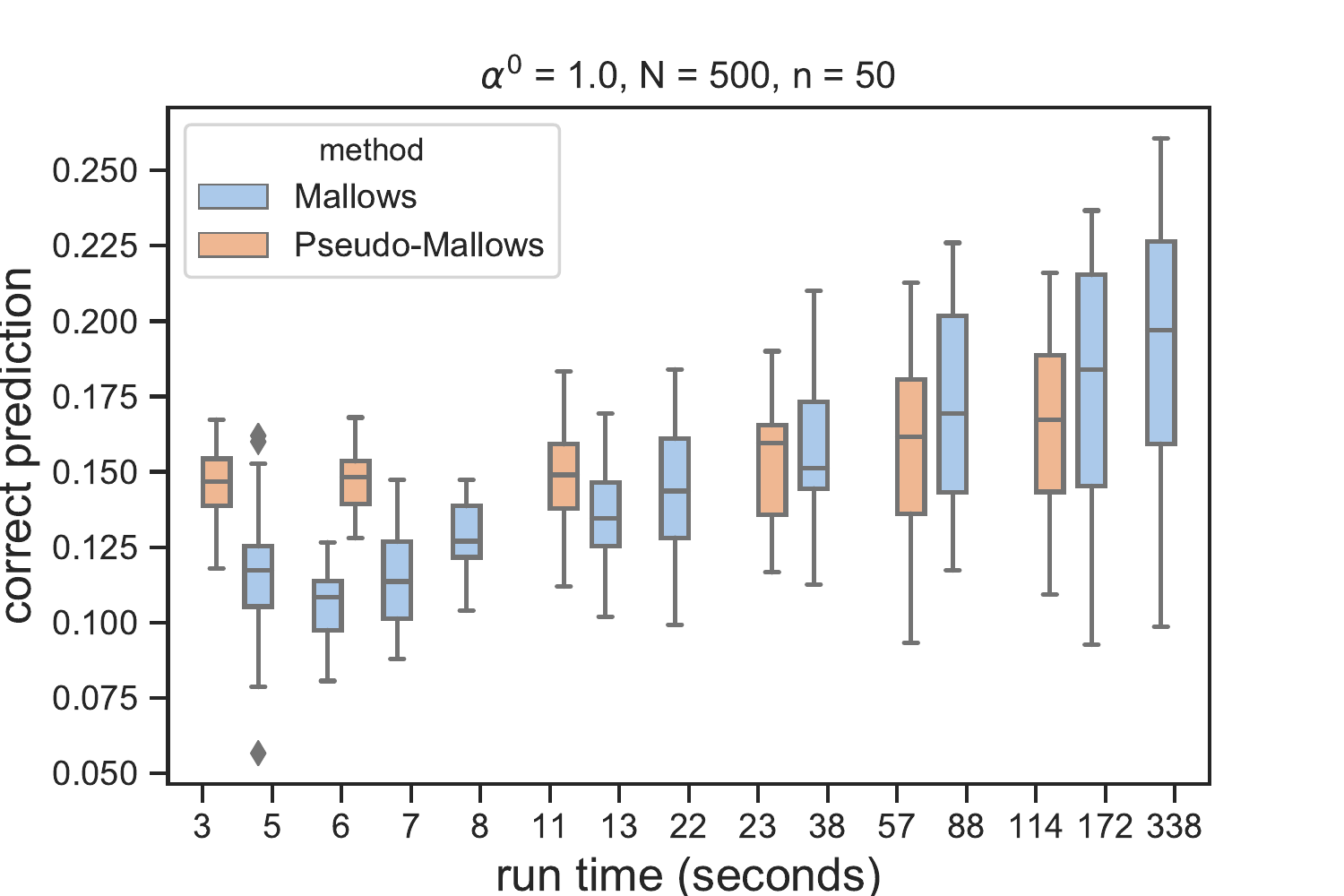}
		
	\end{minipage} 
	\hfill
	\begin{minipage}[t]{.47\textwidth}
		\centering
		\includegraphics[width=\textwidth]{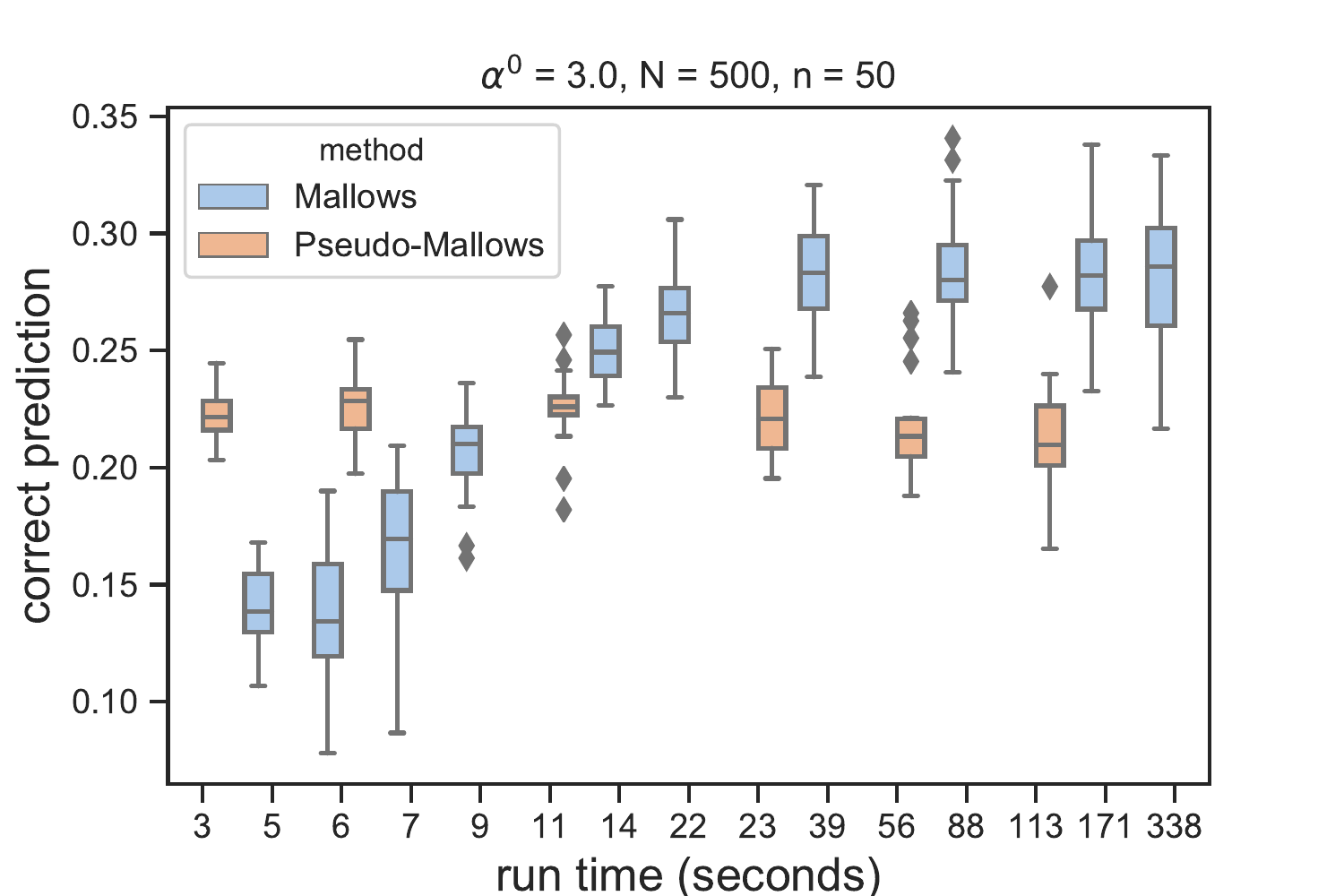}
		
	\end{minipage} 
	\begin{minipage}[t]{.47\textwidth}
		\centering
		\includegraphics[width=\textwidth]{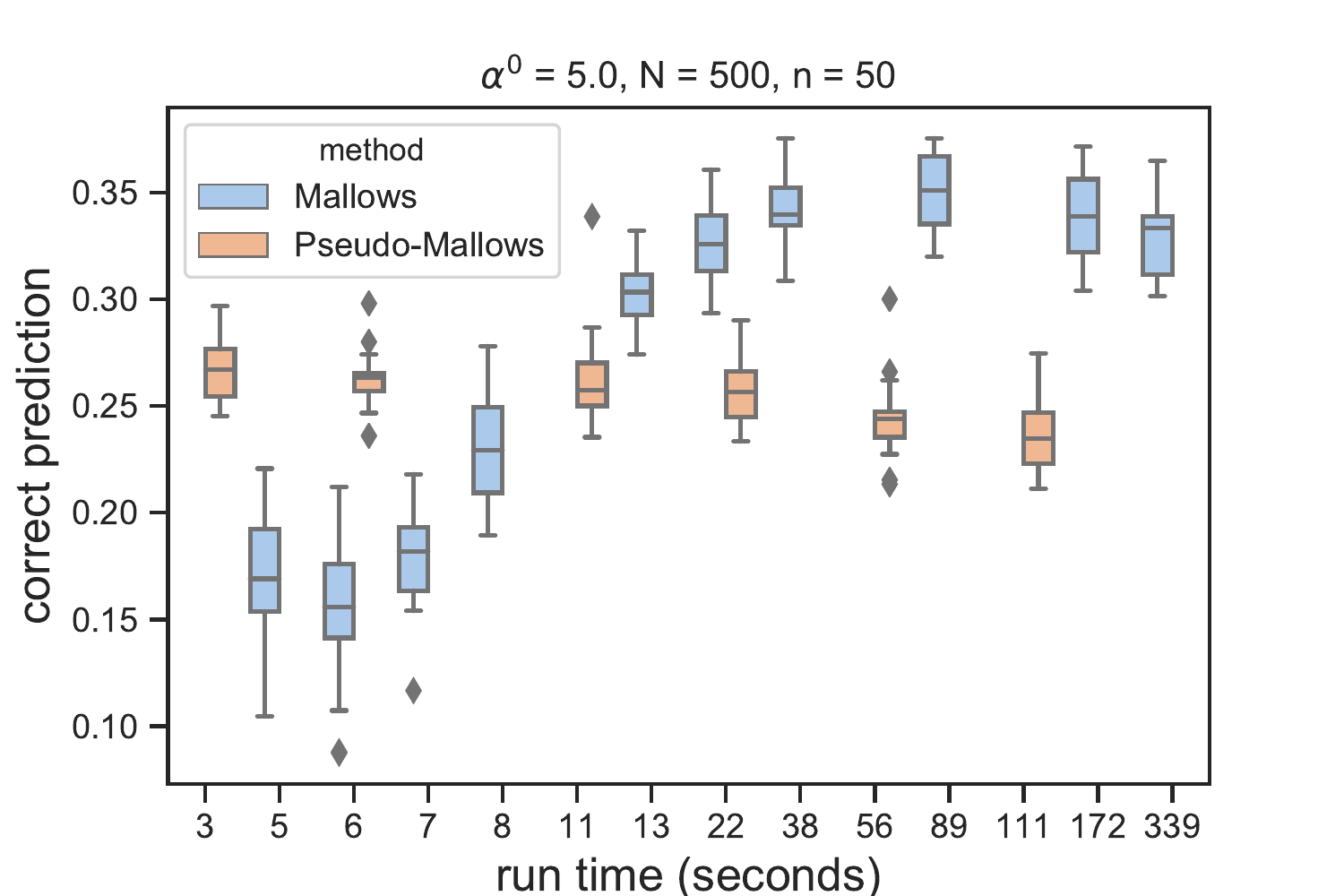}
		
	\end{minipage} 
	\hfill
	\begin{minipage}[t]{.47\textwidth}
		\centering
		\includegraphics[width=\textwidth]{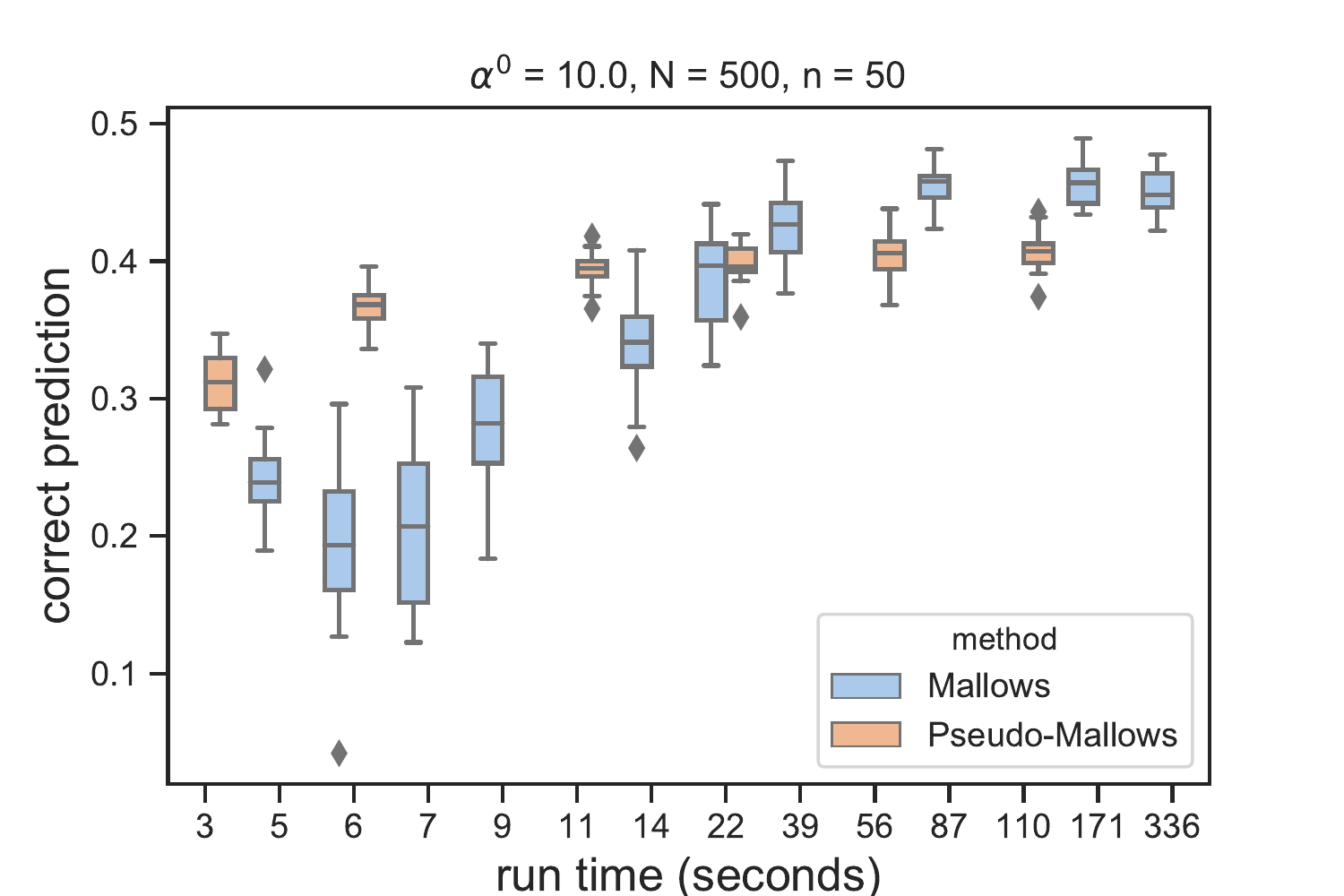}
		
	\end{minipage} 
	\caption{Results of the simulation study of Section \ref{sec:sim_clicking}: recommendation accuracy vs. running time. $y$-axis: percentage of correct recommendations, $x$-axis: computation time in seconds.}
	\label{fig:ind_simulation}
\end{figure}

In Figure \ref{fig:ind_simulation}, the y-axis in each panel shows the percentage of correct recommendations made by either the Mallows MCMC or the Pseudo-Mallows algorithm, against the time it takes to run the algorithms. As with the full ranking experiment described in Section \ref{sec: full_sim}, the burn-in period of the Mallows MCMC can be clearly observed. When not run with enough iterations, the MCMC does not produce estimations that are truly representative of the target posterior distributions. We do not observe a significant initialization effect when using the Pseudo-Mallows, which can provide effective estimation much faster compared to the Mallows MCMC. If we run the Mallows MCMC long enough, eventually, it is able to make individual recommendations that are more accurate than the Pseudo-Mallows, since the data comes from the Mallows distribution. However, when computing time is a constraint, the Pseudo-Mallows' estimates of each user's individual full ranking and recommendations are more reliable and precise than the Mallows MCMC.

\subsection{Personalized recommendations from clicking data - a case study} \label{sec:NRK}
In this section, we make personalized recommendations with the Pseudo-Mallows and the Mallows MCMC in a real life clicking dataset from the Norwegian Broadcasting Corporation (NRK)\citep{liu2019diverse}, and compare recommendation performances. 

The NRK dataset contains $N$ = 7872 anonymous users and $n = $200 different movies, TV programs and news programs, available on the NRK TV website. Different episodes of the same TV series are considered as one item, and multiple clicks on an item by a user are only counted once. Each user has a minimum of 13 clicks. To construct a training dataset, we randomly remove 10 clicks per user, and these 10 clicks per user are later used as the ground truth for recommendation accuracy checking. The training dataset is summarized as follows:

\begin{table}[h!]
	\centering
	\begin{tabular}{||c c c c c||} 
		\hline\hline
		N & n & total clicks & avg clicks per user & density \\ [0.5ex] 
		\hline\hline
		7872 & 200 & 82817 & 10.5 & 5\%\\ 
		\hline
	\end{tabular}
\caption{Summary characteristics of the NRK clicking dataset.}
\label{tab:summary_NRK}
\end{table}

\subsubsection{Pre-processing} \label{sec:NRK_preprocessing}
In this paper we assume that the $N$ users share the same consensus. 
For a large dataset, it can be expected that there exist many user groups with different consensus rankings. Therefore, before applying the methods, we partition the dataset into $K$ clusters using the cluster assignment \citep{liu2019diverse}. The NRK dataset is thus partitioned into 17 clusters, with cluster sizes ranging from 35 users to 2801 users.

For each cluster, we then obtain an estimate of the scale parameter $\alpha$ following the procedures described in Section \ref{sec:alpha_tuning}. We take cluster 1 for example. Each user has clicked on an average of $\lambda = 8$ items. We generate many such datasets of $N = 500$ users and $n = 200$ items, with a grid of known $\alpha^0$ values. These full ranking datasets are then binarized with a Truncated Exponential distribution with a minimum of 3 clicks, and a mean of $\lambda$ clicks per user. For each simulated dataset, or each $\alpha^0$ value, we then calculate the average pairwise individual similarity, as defined in (\ref{eq:mean_pairwise_sim}).

\begin{figure}[h!]
	\centering
	\includegraphics[width=0.45\textwidth]{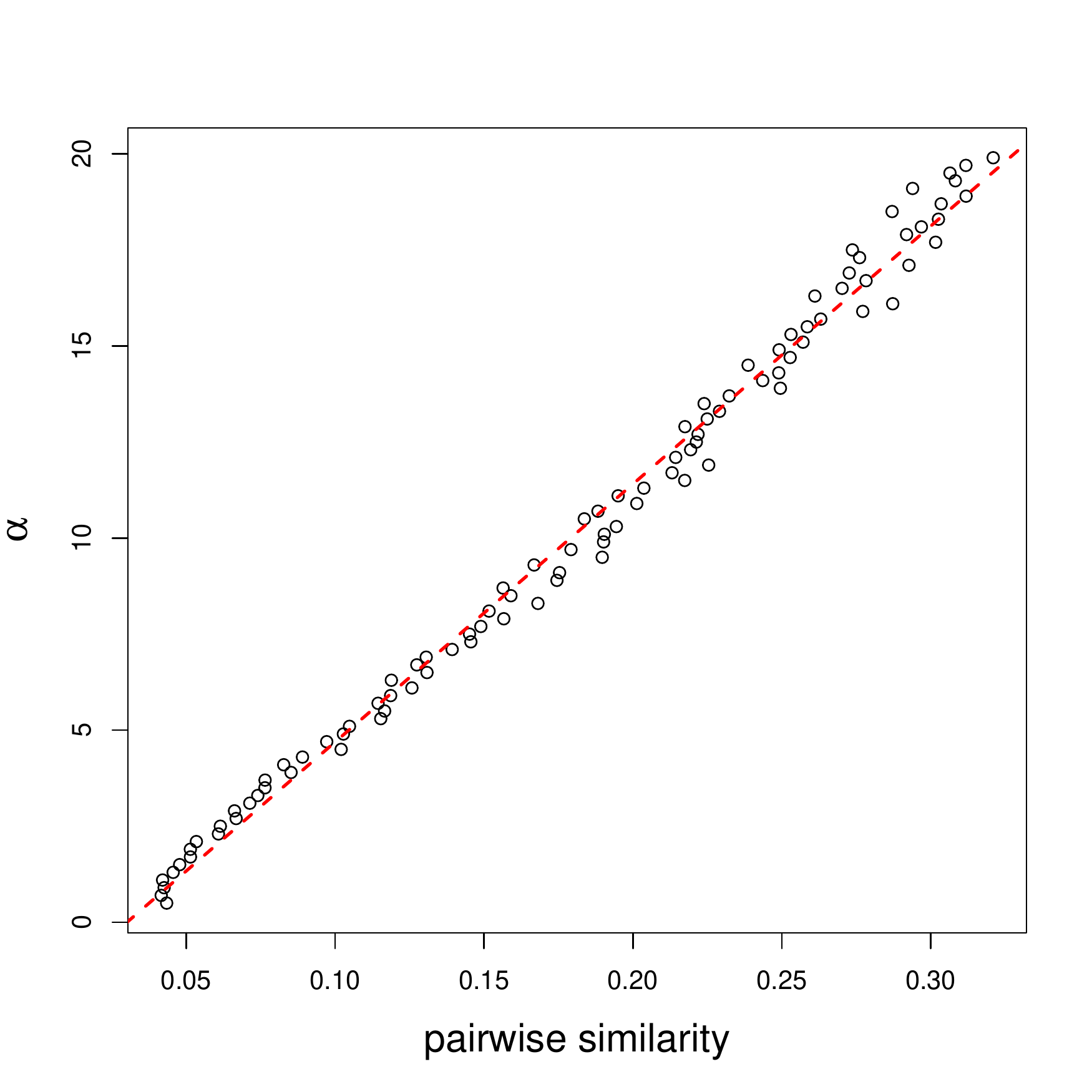}
	\caption{$\alpha$ tuning for cluster 1 of the NRK dataset. $x$-axis: average pairwise user cosine similarities, $y$-axis: corresponding $\alpha^0$ values from the full ranking dataset.}
	\label{fig:NRK_alpha_tuning}
\end{figure}

If we plot the values of $\alpha^0$ and its corresponding pairwise user similarity, as shown in Figure \ref{fig:NRK_alpha_tuning}, it can be observed that it appears to exist a one-to-one, in this case, linear relationship between $\alpha^0$ and the average pairwise user cosine similarity. Pairwise user similarity is thus a good statistics to estimate $\alpha$. Lastly, we compute the pairwise user similarity in NRK cluster 1, and the $\alpha^0$ value of the simulated dataset with the closest pairwise user cosine similarity is chosen. We repeat this procedure for every cluster to obtain an estimate of $\alpha$ for all clusters. 
 
\subsubsection{Comparison of computation speed and recommendation accuracy} \label{sec:NRK_result}
After partitioning the dataset into 17 clusters, we run both the Mallows MCMC and the Pseudo-Mallows algorithm for each cluster independently and in parallel. The overall computing time is determined by the cluster that requires the longest computation time. The Mallows MCMC is run for 13 000 iterations, and the Pseudo-Mallows is run for 200 samples, such that the computing time is similar for both algorithms. To compute for the largest cluster of 2801 users, it takes 649 seconds and 587 seconds for the Mallows MCMC and Pseudo-Mallows respectively. 

We then make $K = 10$ recommendations per user with the samples obtained by both algorithms. We compare the 10 recommendations per user made with the two algorithms with the true 10 clicks previously removed during the pre-processing step, and the intersection of the two sets gives the proportion of correct recommendations.

Recommendation accuracy of the two methods is shown in Table \ref{tab:NRK_accuracy}. It can be observed that, given the computing time limitation, the Pseudo-Mallows's recommendation accuracy is more than 60\% higher compared to that of the Mallows MCMC. This is consistent with the previous simulation studies, showing that the Mallows MCMC suffers from the burn-in period as well as from highly autocorrelated samples, so that huge numbers of iterations are needed in order for the algorithm to reach full convergence. When using the Pseudo-Mallows, on the other hand, all samples are drawn independently, and accurate recommendations can be obtained with very few samples in a much shorter time.

\begin{table}[h!]
	
	\caption{Comparison of recommendation accuracies on the NRK dataset.}
	\label{tab:NRK_accuracy}
	\centering
	
	\begin{tabular}{||c c  c ||} 
		\hline\hline
		Method& Mallows MCMC & Pseudo-Mallows \\ [0.5ex] 
		\hline
		 Accuracy& 13.9\%& 22.6\%  \\
		\hline
	\end{tabular}
\end{table}

\begin{figure}[h!]
	
	\begin{minipage}[t]{.47\textwidth}
		\centering
		\includegraphics[width=\textwidth]{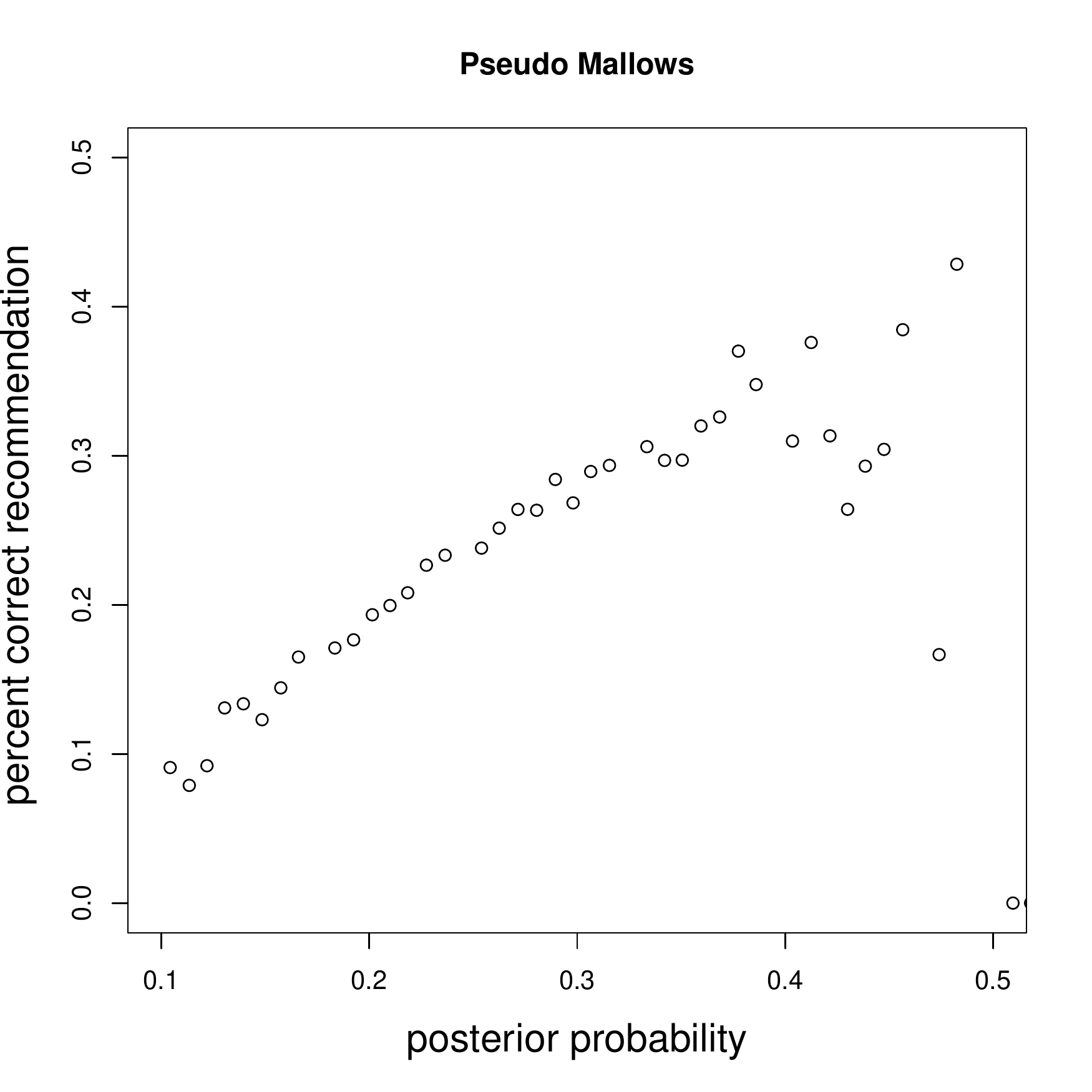}
		
	\end{minipage} 
	\hfill
	\begin{minipage}[t]{.47\textwidth}
		\centering
		\includegraphics[width=\textwidth]{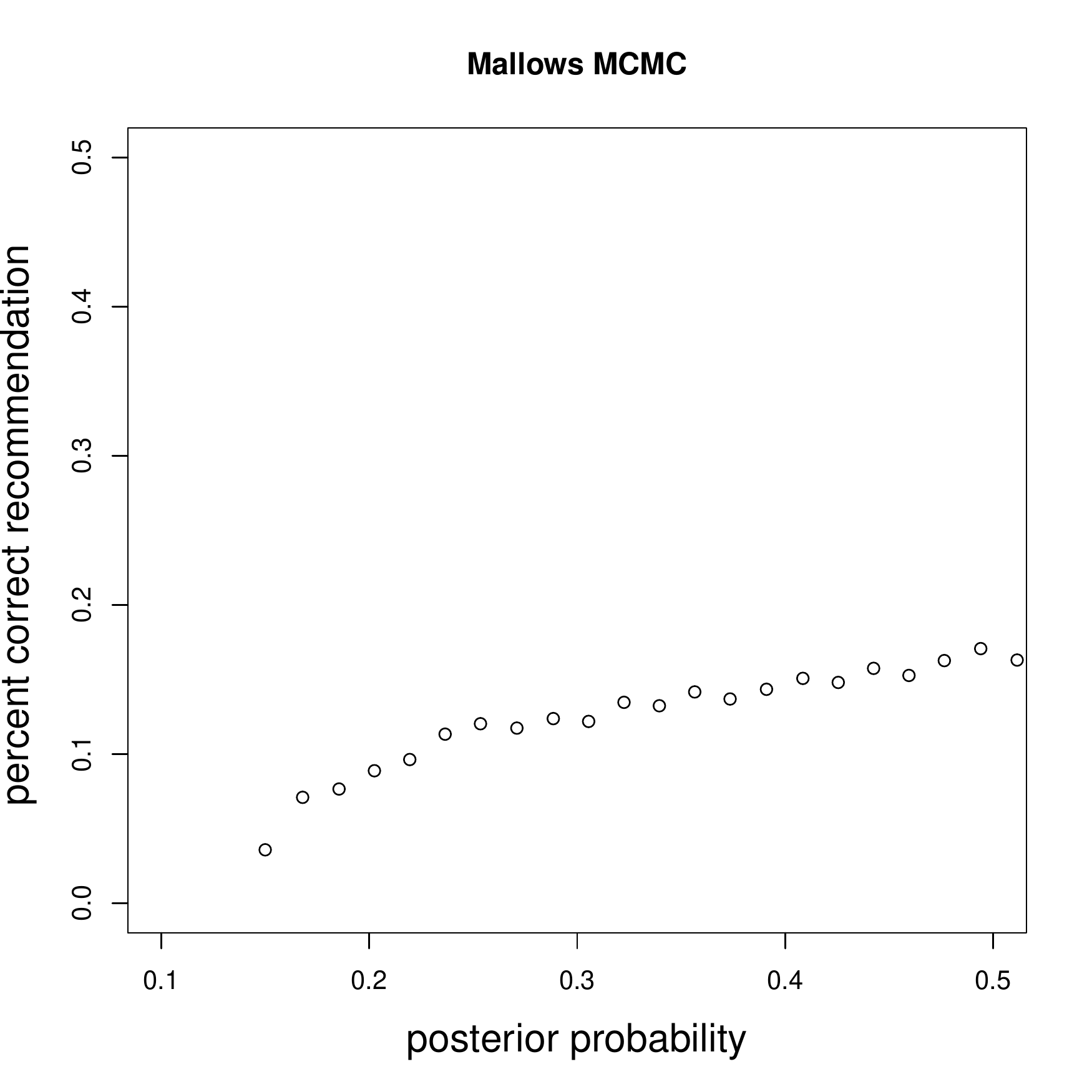}
		
	\end{minipage}

	\caption{Comparison of top-k recommendation's uncertainty estimation for the NRK dataset. $x$-axis: posterior probability of top-k recommendations, $y$-axis: recommendation accuracy }
	\label{fig:NRK_calibrations}
\end{figure}

Not only can the Pseudo-Mallows make accurate recommendations rapidly, but its uncertainty estimation for each recommendation is also well calibrated. Each top-k recommendation has a posterior probability associated with it, as described in (\ref{eq:tpp}). We can bin all the top-$k$ recommendations according to these posterior probabilities, and calculate the recommendation accuracy (estimated with the proportion of correct recommendations) in each bin for both the Pseudo-Mallows and the Mallows MCMC. The right panel in Figure \ref{fig:NRK_calibrations} clearly shows that the Mallows MCMC systematically overestimates the  posterior probabilities of the recommendations. This is another sign that the Mallows MCMC algorithm has yet to reach convergence. As a contrast, the posterior probabilities estimated for each top-k recommendation ($x$-axis) by the Pseudo-Mallows largely reflect the actual recommendation accuracy, as a clear diagonal trend can be observed from the left panel. The accurate uncertainty calibration also shows the effectiveness of the  $\alpha$ tuning method proposed in Section \ref{sec:alpha_tuning}. 

In this sample, the largest cluster contains just under 3000 users and 200 items, and the Pseudo-Mallows algorithm's computing time was under 10 minutes. The computational complexity of the Pseudo-Mallows algorithm is linear in the number of items $n$ and in the number of users $N$, i.e., $\mathcal{O}(N\cdot n)$. We can therefore expect it to be able to work with datasets that contains thousands of items and tens of thousands of users within hours. 

\section{Summary and discussion } \label{sec:summary}
In this work, we have introduced the Pseudo-Mallows distribution to achieve a fast approximation of the Bayesian Mallows posterior distribution. The Pseudo-Mallows distribution is an ordered product of $n$ univariate distributions, where each variable is conditioned on all preceding ones. This ordering is the variational parameter which must be chosen optimally to obtain the best approximation. This is found by minimizing the marginalized version of the KL-divergence between the pseudo-Mallows distribution of all unknown parameters and the posterior Mallows distribution of such parameters given data. When the data are complete rankings of the $n$ items by each user, and the objective is inference on the consensus $\bm{\rho}$,  we have argued that the optimal variational order is the $\mathcal{V}$-set based on the group consensus. We also showed how the $\mathcal{V}$-set can be inferred from the data.

We then extended the Pseudo-Mallows method to learn the group consensus and all the individual rankings from clicking data, which is the most common situation for recommender systems. We introduced a two-step procedure to obtain samples for both the group consensus $\bm{\rho}$ and the individual full rankings $\mathbf{R}^j$. We find that, for each user $j$, the variational order to follow in the step when the $\mathbf{R}^j$ are sampled follows a uniform distribution defined on $\mathcal{P}_n$, and not the $\mathcal{V}$-set. 

Through a simulation study, we demonstrated that the algorithm based on the Pseudo-Mallows distribution can efficiently learn the group consensus parameter with few samples and short computation time. We also showed that the Pseudo-Mallows extension to clicking data is indeed a faster alternative to Bayesian Mallows MCMC, to learn individual rankings. Through a case study using real-life clicking data, we successfully showcased the Pseudo-Mallows' ability to make accurate personalized recommendations given limited computation time. 

There are several areas for further improvement. First, in the presence of heterogeneous groups of preferences, users need to be clustered. One can achieve this by using a mixture of Mallows, i.e., assign users to clusters such that, within each cluster, the users' rankings follow the same Mallows distribution. In the case study described in Section \ref{sec:NRK_preprocessing}, we used the Bayesian Mallows mixture  \citep{vitelli2017probabilistic,liu2019diverse}, available through the BayesMallows package \citep{sorensen2019bayesmallows}, but computationally slow. Faster clustering algorithms, such as K-means clustering, do not always produce enough within-cluster homogeneity. When the clustering is inappropriate, the members within each cluster do not fit well to the Mallows distribution, nor to the Pseudo-Mallows. If we apply the Pseudo-Mallows to such clusters, the uncertainties often will not be well calibrated. Therefore, it would be beneficial to adapt the Pseudo-Mallows approximation to the Bayesian Mallows mixture, so to improve its computational efficiency. The presence of clusters also pose a challenge for $\alpha$ tuning, as the pairwise user cosine similarity reflected in the given dataset may no longer be similar to the user cosine similarity demonstrated by a dataset generated by the Mallows distribution. There is a need to develop faster versions of model based clustering for ranking or clicking data.

In applications such as personalized recommendations and rank aggregation, it is often a subset of items that is of interest, for example, the top-$k$ ranked items.  In our current approach, we estimate the rankings of all items in each iteration, instead of focusing only on a subset of relevant items. In our Pseudo-Mallows algorithm, the ranking for each item is sampled sequentially, and it is possible to sample only a top part of the items, similarly to what has been proposed for the Bayesian Mallows MCMC in \cite{eliseussen2021rank}. Clearly, the overall computation time can be greatly reduced.

Finally, the scale parameter $\alpha$ is tuned by cosine similarity in our work, but instead, we could optimise it as a hyperparamter, for example by Bayesian optimization \citep{alvi2019asynchronous} or by commonly used variational inference optimization schemes such as CAVI or SVI. 

The Pseudo-Mallows method with our proposal of the optimal variational parameters leads to a successful approximation of the Bayesian Mallows posterior model, is computationally convenient and leads to faster inference. This opens for a broader use of the Bayesian Mallows approach to preference learning and recommender systems.

\section*{Implementation}
A GitHub repository (\texttt{https://github.com/sylvia-liu-qinghua/Pseudo-Mallows}) including codes for testing the Pseudo-Mallows algorithm for clicking data, as well as the scripts for running the empirical study, is publicly available.

Pseudo-code schemes of the five algorithms described in the paper are reported below.

%\newpage
\begin{algorithm}[h!]\label{algo:full_data_given_ordering}
	\SetAlgoLined
	\DontPrintSemicolon
	\KwIn{$n\_samples$, $\alpha^0$, \{$\mathbf{R}^1, ..., \mathbf{R}^N$\}, $\{i_1, ..., i_n\}$ or $\{o_1, ..., o_n\}$}
	\KwOut{\{$\bm{\rho}^1, ..., \bm{\rho}^{n\_ samples}$\}}
	\For{$t \gets 1$ to $n\_samples$}{
		set $\bm{s}$ = \{1, ..., $n$\} \;
		\For{$k \gets 1$ to $n$}{
			\For{$r \gets 1$ to $n$}{
				compute $	q_r
				= \frac{\text{exp}\{- \frac{\alpha^0}{n}\sum\limits_{j=1}^{N}d(R^j_{o_k}, r)\}}
				{\sum\limits_{z\in \bm{s}}\text{exp}\{- \frac{\alpha^0}{n}\sum\limits_{j=1}^{N}d(R^j_{o_k}, z)\}} \mathbbm{1}_{\{r \in \bm{s} \}}$
			}
			draw $\rho_{o_k}^t \sim Multinomial (q_1, ..., q_n)$ \;
			$\bm{s} \gets \bm{s} \setminus \rho_{o_k}^t$
		}
	}
	\caption{Pseudo-Mallows Distribution given a fixed ordering}
\end{algorithm}

\newpage
\begin{algorithm}[h!]\label{algo:localsearch}
	\SetAlgoLined
	\DontPrintSemicolon
	\KwIn{$\{i_1, ..., i_n\}^0$, $n\_samples$, $\alpha^0$, \{$\mathbf{R}^1, ..., \mathbf{R}^N$\}}
	\KwOut{$\{i_1, ..., i_n\}$}
 	$\{\bm{\rho}^{MCMC}\} \gets$ \texttt{BayesMallows::compute$\_$mallows}($n\_samples$,  $\alpha^0$, \{$\mathbf{R}^1, ..., \mathbf{R}^N$\})\;
	%$\mathbf{d} \gets \mathbf{0}(max\_iters)$\;
	\For {$l \gets 1$ to $max\_iters$}{
	    %\If{$t > 3$ AND $\sum_{i=t-2}^t d(i-1) > 0$}{
	   	\For {$j \gets \{1,...,n\}$}{
		    \For{$r \gets \{1,...,n\}$}{
		        generate $\{i_1, ..., i_n \}' \gets $ L\&S($\{i_1, ..., i_n\}^{l-1}, j, r$) \;
		        draw $\{\bm{\rho}\}^{l,j,r}$ $\gets$ Algorithm \ref{algo:full_data_given_ordering}($\{i_1, ..., i_n\}', \cdot$)\;
		        Calculate edge $w_{j,r} = \sum\limits_{i=1}^{n}KL(q(\rho_i^{l,j,r})||p(\rho^{MCMC}_i))$
		    }
		
		}
		 Solve the minimal weight perfect matching $M$ via the Hungarian Method \;
		 Compute the new permutation $\{i_1, ..., i_n\}^{*}$ based on $M$\;
% 		 \If{t > 1}{Compute $d(t) = dist(\{i_1, ..., i_n\}^{t-1}, \{i_1, ..., i_n\}^{*})$}
		 Set $\{i_1, ..., i_n\}^{l} \gets \{i_1, ..., i_n\}^{*} $
% 	}
% 	\Else{
% 	    break\;
% 	 }
	}
	\caption{Iterative search algorithm}
	
\end{algorithm}

\newpage
\begin{algorithm}[h!]\label{algo:full_data}
	\SetAlgoLined
	\DontPrintSemicolon
	\KwIn{$n\_samples$, $\alpha^0$, \{$\mathbf{R}^1, ..., \mathbf{R}^N$\}, $\sigma$}
	\KwOut{\{$\bm{\rho}^1, ..., \bm{\rho}^{n\_ samples}$\}}
	initialize $\hat{\bm{\rho}}^0$  = rank($\frac{1}{N}\sum\limits_{j=1}^{N}R^j_1,...,\frac{1}{N}\sum\limits_{j=1}^{N}R^j_n$)\;
	generate $\mathcal{V}_{\hat{\bm{\rho}}^0}$\;
	
	\For{$t \gets 1$ to $n\_samples$}{
		draw $v^t = \{i_1, ..., i_n\}^t $ from $\mathcal{V}_{\hat{\bm{\rho}}^0}$\;
		\For{$m \gets 1$ to $n$}{
			draw $i_m' \sim \mathcal{N}(i_m, \sigma)$ \;
		}
		compute ${v^t}'$ = rank($i_1', ..., i_n'$)$\leftrightarrow \{o_1, ..., o_n\}^t$\;
		set $\bm{s}$ = \{1, ..., $n$\} \;
		\For{$k \gets 1$ to $n$}{
			\For{$r \gets 1$ to $n$}{
				compute $	q_r
				= \frac{\text{exp}\{- \frac{\alpha^0}{n}\sum\limits_{j=1}^{N}d(R^j_{o_k}, r)\}}
				{\sum\limits_{z\in \bm{s}}\text{exp}\{- \frac{\alpha^0}{n}\sum\limits_{j=1}^{N}d(R^j_{o_k}, z)\}} \mathbbm{1}_{\{r \in \bm{s} \}}$
			}
			draw $\rho_{o_k}^t \sim Multinomial (q_1, ..., q_n)$ \;
			$\bm{s} \gets \bm{s} \setminus \rho_{o_k}^t$
		}
	}
	\caption{Pseudo-Mallows for sampling $\bm{\rho}$}
	
\end{algorithm}

\newpage
\begin{algorithm}[h!]\label{algo:sampleOneUser}
	\SetAlgoLined
	\DontPrintSemicolon
	\KwIn{$\bm{B}^j$, $\alpha$, $\bm{\rho}$}
	\KwOut{$\mathbf{R}^j = \{R^j_1, ..., R^j_n\}$}
	generate $\mathcal{A}^j = \{A_i: B^j_i = 1\}$, ${\mathcal{A}^j}^c = \{A_k: B^j_k = 0\}$\;
	compute $\bm{\rho}^t$ = rank($\rho_i: A_i \in \mathcal{A}^j$), $\bm{\rho}^b$ = rank($\rho_k: A_k \in {\mathcal{A}^j}^c$) \;
	draw ordering $\bm{o}^t = \{o_1, ..., o_{c_j}\}$ uniformly from $\mathcal{P}_{c_j}$ \;
	draw ordering $\bm{o}^b = \{o_1, ..., o_{c_j^c}\}$ uniformly from $\mathcal{P}_{c_j^c}$ \;
	set $\bm{s}^t$ = \{1, ..., $c_j$\} \;
	\For {each item $i$ in $\bm{o}^t$}{		
			\For{$r \gets 1$ to $c_j$}{
				compute $	q_r
				= \frac{\text{exp}\{- \frac{\alpha}{n}d(\rho_i^t, r)\}}
				{\sum\limits_{z\in \bm{s}^t}\text{exp}\{- \frac{\alpha}{n}d(\rho_i^t, z)\}} \mathbbm{1}_{\{r \in \bm{s}^t \}}$
			}
			draw $R_i^j \sim Multinomial (q_1, ..., q_{c_j})$ \;
			$\bm{s}^t \gets \bm{s}^t \setminus R_i^j$
		}
		set $\bm{s}^b$ = \{1, ..., $c_j^c$\} \;
	\For {each item $k$ in $\bm{o}^b$}{		
		\For{$r \gets 1$ to ${c_j}^c$}{
			compute $q_r
			= \frac{\text{exp}\{- \frac{\alpha}{n}d(\rho_k^b, r)\}}
			{\sum\limits_{z\in \bm{s}^b}\text{exp}\{- \frac{\alpha}{n}d(\rho_k^b, z)\}} \mathbbm{1}_{\{r \in \bm{s}^b \}}$
		}
		draw $R_k^j \sim Multinomial (q_1, ..., q_{c_j^c})$ \;
		$\bm{s}^b \gets \bm{s}^b \setminus R_k^j$
		
		\For {each item $k$ in $\bm{o}^b$}{
			$R^j_k \gets R^j_k + c_j$			
		}		
	}
	
	\caption{Sample for user $j$ based on clicking data}
	
\end{algorithm}

\newpage
\begin{algorithm}[h!]\label{algo:sampleClicking}
	\SetAlgoLined
	\DontPrintSemicolon
	\KwIn{$\{\bm{B}^1, ..., \bm{B}^N$\}, $\alpha^0$, $\sigma_{\bm{\rho}}$, $n\_samples$}
	\KwOut{$\{\mathbf{R}^1, ...., \mathbf{R}^N\}^{1, ..., n\_samples}$, \{$\bm{\rho}^0, ...,\bm{\rho}^{n\_samples} $\}}
	initialize $\hat{\bm{\rho}}^0$  = rank($\frac{1}{N}\sum\limits_{j=1}^{N}B^j_1,...,\frac{1}{N}\sum\limits_{j=1}^{N}B^j_n$)\;
	\For{$t \gets 1$ to $n\_samples$}{
	 \For{user $j \gets 1$ to $N$ in parallel}{
	 		sample for $\mathbf{R}^{j,t}$ based on $\bm{\rho}^{t-1}$ using Algorithm \ref{algo:sampleOneUser}
 		}
 	draw 1 sample for $\bm{\rho}^t$ based on \{$\mathbf{R}^1, ..., \mathbf{R}^N\}^t$ using Algorithm \ref{algo:full_data}
	}

	\caption{Pseudo-Mallows for clicking data}

\end{algorithm}
% Acknowledgements should go at the end, before appendices and references

% \acks{We would like to acknowledge support for this project
% from the National Science Foundation (NSF grant IIS-9988642)
% and the Multidisciplinary Research Program of the Department
% of Defense (MURI N00014-00-1-0637). }


\newpage
\begin{appendices}
\section{Minimization of the expected L-1 distance}\label{sec:proof_l_1}
\textit{For an item $o_k$, the value of $l^*$ that minimizes the expected value of $\sum\limits_{j=1}^N |R^j_{o_k}-l| $ w.r.t. Mallows($\alpha, \bm{\rho}^0 | R_{o_k} \neq \rho_{o_1}, ..., \rho_{o_{k-1}}))$, i.e., 
$$l^* = \operatorname*{argmin}\limits_{l \in \{1, ..., n\} \setminus \{\rho_{o_1}, ..., \rho_{o_{k-1}}\}} \mathbb{E}[\sum\limits_{j=1}^N |R^j_{o^k}-l|],$$}
is the median of the marginal distribution $R_{o_k}$, given that $R_{o_k} \neq \rho_{o_1}, ..., \rho_{o_{k-1}}$.
\begin{proof}
\begin{align*}
     &\mathbb{E}[\sum\limits_{j=1}^N |R^j_{o^k}-l|] \\
    =&\sum\limits_{j=1}^N\mathbb{E}[|R^j_{o^k}-l|] \\
    =&\sum\limits_{j=1}^N \sum\limits_{a=1}^n |a-l| P_k(a),\\
\end{align*}
where $P_k(a) = \sum\limits_{\bm{r}\in \mathcal{P}_n, r_{o_k} = a,r_{o_k} \neq \rho_{o_1}, ..., \rho_{o_{k-1}}}\frac{1}{Z_n(\alpha)}\text{exp}\{d(\bm{r},\bm{\rho}^0)\}$. The minimization of the expected distance above is equivalent to minimizing $\sum\limits_{a=1, a \neq \rho_{o_1}, ..., \rho_{o_{k-1}} }^n |a-l|P_k(a).$
\begin{align*}
    \sum\limits_{a=1, a \neq \rho_{o_1}, ..., \rho_{o_{k-1}} }^n |a-l|P_k(a) &=\sum\limits_{a=1, a \neq \rho_{o_1}, ..., \rho_{o_{k-1}} }^l (l-a)P_k(a)+\sum\limits_{a=l+1, a \neq \rho_{o_1}, ..., \rho_{o_{k-1}} }^n (a-l)P_k(a) \\
    \frac{d}{dl}\sum\limits_{a=1, a \neq \rho_{o_1}, ..., \rho_{o_{k-1}} }^n |a-l|P_k(a) &=\sum\limits_{a=1, a \neq \rho_{o_1}, ..., \rho_{o_{k-1}} }^l P_k(a) - \sum\limits_{a=l+1, a \neq \rho_{o_1}, ..., \rho_{o_{k-1}} }^n P_k(a) \\
    & = 2 \sum\limits_{a=1, a \neq \rho_{o_1}, ..., \rho_{o_{k-1}} }^l P_k(a) - 1.
\end{align*}
Setting the derivative to be 0, we have $\sum\limits_{a=1, a \neq \rho_{o_1}, ..., \rho_{o_{k-1}} }^l P_k(a) = \frac{1}{2}$, i.e., $l^*$ is the median of the marginal distribution of $R_{o_k}$, given $R_{o_k} \neq \rho_{o_1}, ...,\rho_{o_{k-1}}$.    
\end{proof}

\section{Symmetry and median of the rank of the middle item \label{sec:one}}
\subsection{Symmetry of the rank of the middle item}\label{sec:symmetry}
\textit{Given that a ranking $\mathbf{R} \in \mathcal{P}_n$ and $\mathbf{R} \sim \text{Mallows}(\bm{\rho}^0, \alpha)$. Let $o^0_i$ be such that $\rho^0_{o^0_i} = i$, for $i = 1, ..., n$. For a fixed $\bm{\rho}^0$ and $\alpha$, $\forall \alpha \in (0, \infty)$, the middle-ranked item in $\bm{\rho}^0$, i.e., item $o^0_m$ is symmetrically distributed about $m$.}
\begin{proof}
Recall that the $Mallows(\bm{\rho}^0, \alpha)$ likelihood function is
\begin{center}
$P(R|\bm{\rho}^0, \alpha) = \frac{1}{Z_n(\alpha)}\text{exp} \{-\frac{\alpha}{n} d(\mathbf{R}, \bm{\rho}^0)\}$.    
\end{center}
Given two permutations $\bm{r}$ and $\bm{r}' \in \mathcal{P}_n \sim Mallows (\bm{\rho}^0, \alpha)$, their likelihoods are equal if $d(\bm{r}, \bm{\rho}^0) = d(\bm{r}', \bm{\rho}^0)$. 

For a given $d$, let $\bm{r}$ be any permutation s.t. $d(\bm{r}, \bm{\rho}^0) = d$ and $r_{o^0_m} \neq m$. We define one other permutation $\bm{r}^{'}$ s.t. $r^{'}_{o^0_s}= 2m - r_{o^0_{2m-s}} \text{ for } s = 1, ...., 2m-1$. $\bm{r}$ and $\bm{r}^{'}$ have the same distance, i.e. $d(\bm{r}^{'}, \bm{\rho}^0) = d$, because
 \begin{equation}
     \begin{split}
         d &= \sum\limits_{l=1}^n | r_{o^0_l}- \rho^0_{o^0_l}| \\
             & = \sum\limits_{l=1}^n | r_{o^0_l}- l| \\
             & = \sum\limits_{l=1}^n |\Delta_l|, \\
     \end{split}
 \end{equation} and
 \begin{equation}
     \begin{split}
         d(\bm{r}^{'}, \bm{\rho}^0) & = \sum\limits_{l=1}^n | r^{'}_{o^0_l}- \rho^0_{o^0_l}| \\
                                    & = \sum\limits_{l=1}^n |2m - r_{o^0_{2m-l}}- l| \\
                                    & = \sum\limits_{l=1}^n |-\Delta_{2m-l}| \\
                                    & = d \\
     \end{split}
 \end{equation}
 
 This also implies that for the middle item $o^0_m$, and a given $d$, for some $\bm{r}$ and $k \in \{1, ..., m-1\}$, if $r_{o^0_m} = m+k$ exists, there is an $r^{'}$ with the same distance and $r^{'}_{o^0_m} = m-k$. Thus it can be concluded that for the middle item $o^0_m$, for all $k = 1, ..., m-1$, it holds true that
 $$P(R_{o^0_m} = m - k|\bm{\rho}^0, \alpha) = P(R_{o^0_m} = m + k|\bm{\rho}^0, \alpha),$$ 
 i.e., $R_{o^0_m}$ is symmetrically distributed about $m$.
 \end{proof}
 
 \subsection{Median of the rank of the middle item}
 \textit{Given that a ranking $\mathbf{R} \in \mathcal{P}_n$ and $\mathbf{R} \sim \text{Mallows}(\bm{\rho}^0, \alpha)$. Let $o^0_i$ be such that $\rho^0_{o^0_i} = i$, for $i = 1, ..., n$. For a fixed $\bm{\rho}^0$ and $\alpha$, $\forall \alpha \in (0, \infty)$, for the middle-ranked item in $\bm{\rho}^0$, i.e., item $o^0_m$, we have}

		\begin{equation}\label{eq:expected_middle}
		\mathbb{E}[R_{o^0_m}|\bm{\rho}^{0}, \alpha] = \rho^0_{o^0_m} = m
		\end{equation}

\begin{proof}

For the middle item in $\bm{\rho}^0$, i.e., item $o^0_m$, as $R_{o^0_m}$ is symmetrically distributed about the value $m$, the median of the distribution is therefore equal to the expected value. 
 
Based on the definition of expected value, we can rewrite (\ref{eq:expected_middle}) as 
\begin{equation}\label{eq:expected_2}
\begin{split}
    \mathbb{E}[R_{o^0_m}|\bm{\rho}^{0}, \alpha] &= \sum\limits_{\bm{r} \in \mathcal{P}_{n}} r_{o^0_m}P(\mathbf{R} = \bm{r}|\bm{\rho}^0, \alpha)\\ 
 & =\frac{1}{Z_{n}(\alpha)}\sum\limits_{\bm{r} \in \mathcal{P}_n}r_{o^0_m}\text{exp}\Big\{- \frac{\alpha}{n} \bm{d}(\bm{r}, \bm{\rho}^0)\Big\} \\
 & = \frac{1}{Z_{n}(\alpha)} \sum\limits_{d \in D} \sum\limits_{\bm{r}: \bm{d}(\bm{r}, \bm{\rho}^0) = d}r_{o^0_m}\text{exp} \{- \frac{\alpha}{n} d\};
\end{split}
\end{equation}

The normalizing function can be expressed as 
$$Z_{n}(\alpha) = \sum\limits_{\bm{r} \in \mathcal{P}_n}\text{exp}\Big\{- \frac{\alpha}{n} \bm{d}(\bm{r}, \bm{\rho}^0)\Big\}  = \sum\limits_{d \in D} |L_d| \text{exp}\Big\{- \frac{\alpha}{n} \cdot d \Big\} \\, $$ where 
$$L_d = \{\bm{r} \in \mathcal{P}_n: \bm{d}(\bm{r},\bm{\rho}^0 ) = d\}, $$ and $D$ is the set of all possible distances between $\bm{r}$ and $\bm{\rho}^0$, $D = \{d_1, ..., d_{|D|}\}$.
 
We can further express equation (\ref{eq:expected_2}) as

\begin{equation}\label{eq:expected_3}
\begin{split}
    \mathbb{E}[R_{o^0_m}|\bm{\rho}^{0}, \alpha] = & \frac{1}{Z_{n}(\alpha)} \sum\limits_{d \in D} \sum\limits_{\bm{r}: \bm{d}(\bm{r}, \bm{\rho}^0) = d}r_{o^0_m}\text{exp} \{- \frac{\alpha}{n} d\} \\
        =& \frac{1}{Z_{n}(\alpha)} \sum\limits_{d \in D} \sum\limits_{k=1}^{m-1} \Bigg\{ \Big[(m+k) |l_{d,k}^+| \cdot\text{exp} \{- \frac{\alpha}{n} d\} + (m-k) |l_{d,k}^-| \cdot\text{exp} \{- \frac{\alpha}{n} d\}\big]  \\ 
                                                 &+ (m+0) |l_{d,0}| \cdot\text{exp} \{- \frac{\alpha}{n} d\} \Bigg\} ; \\
\end{split}
\end{equation}
where 

$$l_{d,k}^+ = \{\bm{r} \in \mathcal{P}_n: \bm{d}(\bm{r},\bm{\rho}^0 ) = d \text{ and } r_{o^0_m} = m+k\} $$
 $$l_{d,k}^- = \{\bm{r} \in \mathcal{P}_n: \bm{d}(\bm{r},\bm{\rho}^0 ) = d \text{ and } r_{o^0_m} = m-k\}, \text{ and} $$
 
 $$l_{d,0} = \{\bm{r} \in \mathcal{P}_n: \bm{d}(\bm{r},\bm{\rho}^0 ) = d \text{ and } r_{o^0_m} = m\}$$
 
 It holds true that $\sum\limits_{k=1}^{m-1}\big(|l_{d,k}^+|+|l_{d,k}^-|\big) + |l_{d,0}| = |L_{d}|$
 
 Using the symmetry that we have proven in \ref{sec:symmetry}, we can obtain that for the middle item $o^0_m$, and a given $d$, for some $\bm{r}$ and $k \in \{1, ..., m-1\}$, if $r_{o^0_m} = m+k$ exists, there is an $r^{'}$ with the same distance and $r^{'}_{o^0_m} = m-k$. Therefore, it holds that $|l_{d,k}^+| = |l_{d,k}^-|$, for all $k = 1, ..., m-1$ and for all $d \in D$.
 
 Thus, equation (\ref{eq:expected_3}) can be further expressed as:
 \begin{equation}\label{eq:expected_4}
     \begin{split}
           \mathbb{E}[R_{o^0_m}|\bm{\rho}^{0}, \alpha] = & \frac{1}{Z_{n}(\alpha)} \sum\limits_{d \in D} \sum\limits_{k=1}^{m-1} \Big[m |l_{d,k}^+| \cdot\text{exp} \{- \frac{\alpha}{n} d\} + m |l_{d,k}^-| \cdot\text{exp} \{- \frac{\alpha}{n} d\}\big]  \\ 
                                                 &+ m |l_{d,0}| \cdot\text{exp} \{- \frac{\alpha}{n} d\} \Big) \\
         = &\frac{1}{Z_{n}(\alpha)} \sum\limits_{d \in D} \sum\limits_{k=1}^{m-1}\Big[\big(|l_{d,k}^+|+|l_{d,k}^-|\big) + |l_{d,0}|\Big] \cdot m \cdot \text{exp}\{- \frac{\alpha}{n} d\} \\
         = &\frac{1}{Z_{n}(\alpha)}     \sum\limits_{d\in D} |L_d| \cdot m \cdot \text{exp}\{- \frac{\alpha}{n} d\} \\
         = & m. \\
     \end{split}
 \end{equation}
 As the median and the expected value are equal, we have that the median of $R_{o^0_m} = m$
\end{proof}

\section{Expectations of two ``neighboring'' items}\label{sec:appen_two}

Given that a ranking $\mathbf{R} \in \mathcal{P}_n$ and $\mathbf{R} \sim \text{Mallows}(\bm{\rho}^0, \alpha)$. Let $o^0_i$ be such that $\rho^0_{o^0_i} = i$, for $i = 1, ..., n$. For a fixed $\bm{\rho}^0$ and $\alpha$, $\forall \alpha \in (0, \infty)$. For a fixed $j$, for any given $n$ and $\alpha>0$, we have
$$\mathbb{E}[{R}_{o^0_j}|\bm{\rho}^0, \alpha]<\mathbb{E}[{R}_{o^0_{j+1}}|\bm{\rho}^0, \alpha]$$.

\begin{proof}

For a fixed $j$, and for $a < b$, $a \in \{1, 2, ...,n -1\}$, $b \in \{ 2, ...,n\}$, we first define the set of permutation $\mathcal{F}_{a,b} = \{\bm{r: r_{o^0_j} = a, r_{o^0_{j+1} = b}}\}$. Based on this definition, we can partition the space of permutation into six sets:
\begin{enumerate}
    \item {$\mathcal{P}_A = \bigcup\limits_{b=j+2}^n \bigcup\limits_{a=j+1}^b \mathcal{F}_{a,b}$
    }
    \item {$\mathcal{P}_{A'} = \bigcup\limits_{b=j+2}^n \bigcup\limits_{a=j+1}^b \mathcal{F}_{b,a}$
    }
    \item {$\mathcal{P}_B = \bigcup\limits_{b=1}^j \bigcup\limits_{a=1}^{min\{b, j-1\}} \mathcal{F}_{a,b}$
    }
    \item {$\mathcal{P}_{B'} = \bigcup\limits_{b=1}^j \bigcup\limits_{a=1}^{min\{b, j-1\}} \mathcal{F}_{b,a}$
    }
    \item {$\mathcal{P}_C = \bigcup\limits_{b=j+1}^n \bigcup\limits_{a=1}^j \mathcal{F}_{a,b}$
    }
    \item {$\mathcal{P}_{C'} = \bigcup\limits_{b=j+1}^n \bigcup\limits_{a=1}^j \mathcal{F}_{b,a}$
    },
\end{enumerate}
and it holds true that $\mathcal{P}_n = \mathcal{P}_A \bigcup \mathcal{P}_{A'}\bigcup \mathcal{P}_{B}\bigcup \mathcal{P}_{B'}\bigcup \mathcal{P}_{C}\bigcup \mathcal{P}_{C'} $.

We can rewrite the expectations as \\
\begin{equation}\label{eq:E_j}
    \begin{split}
        \mathbb{E}[R_{o^0_{j}}|\bm{\rho}^0, \alpha] &= \frac{1}{Z_n(\alpha)}\sum\limits_{\bm{r}\in \mathcal{P}_n} r_{o^0_j}\text{exp}\{-\frac{\alpha}{n}d(\bm{r}, \bm{\rho^0})\} \\
        &=\frac{1}{Z_n(\alpha)}\Bigg[ \sum\limits_{\bm{r}\in \mathcal{P}_A} r_{o^0_j}\text{exp}\{-\frac{\alpha}{n}d(\bm{r}, \bm{\rho^0})\} +\sum\limits_{\bm{r}\in \mathcal{P}_{A'}} r_{o^0_j}\text{exp}\{-\frac{\alpha}{n}d(\bm{r}, \bm{\rho^0})\} \Bigg]\\
        & + \frac{1}{Z_n(\alpha)}\Bigg[\sum\limits_{\bm{r}\in \mathcal{P}_B} r_{o^0_j}\text{exp}\{-\frac{\alpha}{n}d(\bm{r}, \bm{\rho^0})\} +\sum\limits_{\bm{r}\in \mathcal{P}_{B'}} r_{o^0_j}\text{exp}\{-\frac{\alpha}{n}d(\bm{r}, \bm{\rho^0})\} \Bigg]\\
        & + \frac{1}{Z_n(\alpha)}\Bigg[\sum\limits_{\bm{r}\in \mathcal{P}_C} r_{o^0_j}\text{exp}\{-\frac{\alpha}{n}d(\bm{r}, \bm{\rho^0})\}+\sum\limits_{\bm{r}\in \mathcal{P}_{C'}} r_{o^0_j}\text{exp}\{-\frac{\alpha}{n}d(\bm{r}, \bm{\rho^0})\}\Bigg],
    \end{split}
\end{equation}
and

\begin{equation}\label{eq:E_j+1}
    \begin{split}
        \mathbb{E}[R_{o^0_{j+1}}|\bm{\rho}^0, \alpha] &= \frac{1}{Z_n(\alpha)}\sum\limits_{\bm{r}\in \mathcal{P}_n} r_{o^0_{j+1}}\text{exp}\{-\frac{\alpha}{n}d(\bm{r}, \bm{\rho^0})\}\\
        & = \frac{1}{Z_n(\alpha)}\Bigg[\sum\limits_{\bm{r}\in \mathcal{P}_A} r_{o^0_{j+1}}\text{exp}\{-\frac{\alpha}{n}d(\bm{r}, \bm{\rho^0})\} +\sum\limits_{\bm{r}\in \mathcal{P}_{A'}} r_{o^0_{j+1}}\text{exp}\{-\frac{\alpha}{n}d(\bm{r}, \bm{\rho^0})\} \Bigg]\\
        &+\frac{1}{Z_n(\alpha)} \Bigg[\sum\limits_{\bm{r}\in \mathcal{P}_B} r_{o^0_{j+1}}\text{exp}\{-\frac{\alpha}{n}d(\bm{r}, \bm{\rho^0})\} +\sum\limits_{\bm{r}\in \mathcal{P}_{B'}} r_{o^0_{j+1}}\text{exp}\{-\frac{\alpha}{n}d(\bm{r}, \bm{\rho^0})\} \Bigg]\\
        &+ \frac{1}{Z_n(\alpha)}\Bigg[\sum\limits_{\bm{r}\in \mathcal{P}_C} r_{o^0_{j+1}}\text{exp}\{-\frac{\alpha}{n}d(\bm{r}, \bm{\rho^0})\}+\sum\limits_{\bm{r}\in \mathcal{P}_{C'}} r_{o^0_{j+1}}\text{exp}\{-\frac{\alpha}{n}d(\bm{r}, \bm{\rho^0})\}\Bigg].
    \end{split}
\end{equation}

Let us first consider $\mathcal{P}_A$ and $\mathcal{P}_{A'}$'s contributions to $\mathbb{E}[R_{o^0_{j}}|\bm{\rho}^0, \alpha]$ and $\mathbb{E}[R_{o^0_{j+1}}|\bm{\rho}^0, \alpha]$.
For a a permutation $\bm{r} \in \mathcal{P}_A$ and its corresponding $\bm{r}'$ s.t. $r_{o^0_j} = r'_{o^0_{j+1}} = a$, $r_{o^0_{j+1}} = r'_{o^0_{j}} = b$, and $r_l = r'_l $  $ \forall i\neq o^0_j, o^0_{j+1}$, it holds that $d(\bm{r}, \bm{\rho}^0)$ =  $d(\bm{r}^{'}, \bm{\rho}^0)$ because \\
\begin{align*}
    d(\bm{r}, \bm{\rho^0}) 
    & = \sum\limits_{i=1}^{j-1}|r_{o^0_i} - \rho^0_{o^0_i}| + |r_{o^0_j} - \rho^0_{o^0_j}| + |r_{o^0_{j+1}} - \rho^0_{o^0_{j+1}}| + \sum\limits_{i = j+2}^{n}|r_{o^0_i} - \rho^0_{o^0_{i}}|\\
    & = \sum\limits_{i \neq j, j+1} |r_{o^0_i} - i| + |a-j| + |b - (j+1)| \\
    & = \sum\limits_{i \neq j, j+1} |r_{o^0_i} - i| + a-j + b - (j+1), %\text{, for } a\geq j+1 \text{ and } b\geq j+2.
\end{align*}
and \\
\begin{align*}
    d(\bm{r'}, \bm{\rho^0}) 
    &= \sum\limits_{i=1}^{j-1}|r'_{o^0_i} - \rho^0_{o^0_i}| + |r'_{o^0_j} - \rho^0_{o^0_j}| + |r'_{o^0_{j+1}} - \rho^0_{o^0_{j+1}}| + \sum\limits_{i = j+2}^{n}|r'_{o^0_i} - \rho^0_{o^0_{i}}|\\
    & = \sum\limits_{i \neq j, j+1} |r'_{o^0_i} - i| + |b-j| + |a - (j+1)| \\
    & = \sum\limits_{i \neq j, j+1} |r_{o^0_i} - i| + b-j + a - (j+1) \\
    & = d(\bm{r}, \bm{\rho^0}) %\text{, for } a\geq j+1 \text{ and } b\geq j+2
\end{align*}

That is to say, for each permutation $\bm{r} \in \mathcal{P}_A$, we can find another permutation $\bm{r}' \in \mathcal{P}_{A'}$ with equal distance to $\bm{\rho}^0$. Therefore, the second row of equation (\ref{eq:E_j}) can be derived as: 

\begin{align*}
    & \textcolor{white}{=}\frac{1}{Z_n(\alpha)}\Bigg[ \sum\limits_{\bm{r}\in \mathcal{P}_A} r_{o^0_j}\text{exp}\{-\frac{\alpha}{n}d(\bm{r}, \bm{\rho^0})\} +\sum\limits_{\bm{r}\in \mathcal{P}_{A'}} r_{o^0_j}\text{exp}\{-\frac{\alpha}{n}d(\bm{r}, \bm{\rho^0})\} \Bigg] \\
    &= \frac{1}{Z_n(\alpha)} 
    \Bigg[\sum\limits_{b = j+1}^n\sum\limits_{a = j+1}^b \sum\limits_{\bm{r} \in \mathcal{P}: r_{o^0_j} = a, r_{o^0_{j+1}} = b} a \exp\{-\frac{\alpha}{n}d(\bm{r}, \bm{\rho}^0)\} \\
    &+\sum\limits_{b = j+1}^n\sum\limits_{a = j+1}^b \sum\limits_{\bm{r} \in \mathcal{P}: r_{o^0_j} = b, r_{o^0_{j+1}} = a} b \exp\{-\frac{\alpha}{n}d(\bm{r}, \bm{\rho}^0)\} \bigg] \\
    & = \frac{1}{Z_n(\alpha)} \sum\limits_{b=j+2}^n\sum\limits_{a=j+1}^b (a+b)\sum\limits_{\bm{r} \in \mathcal{P}: r_{o^0_j} = a, r_{o^0_{j+1}} = b} \exp\{-\frac{\alpha}{n}d(\bm{r}, \bm{\rho}^0)\} \\
    & = \sum\limits_{b=j+2}^n\sum\limits_{a=j+1}^b (a+b) P(\mathcal{F}_{a,b}|\bm{\rho}^0, \alpha)
\end{align*}

Similarly for $E[R_{o^o_{j+1}}]$, the second row of equation (\ref{eq:E_j+1}) can be simplified as
\begin{align*}
    & \textcolor{white}{=}\frac{1}{Z_n(\alpha)}\Bigg[ \sum\limits_{\bm{r}\in \mathcal{P}_A} r_{o^0_{j+1}}\text{exp}\{-\frac{\alpha}{n}d(\bm{r}, \bm{\rho^0})\} +\sum\limits_{\bm{r}\in \mathcal{P}_{A'}} r_{o^0_{j+1}}\text{exp}\{-\frac{\alpha}{n}d(\bm{r}, \bm{\rho^0})\} \Bigg] \\
    &= \frac{1}{Z_n(\alpha)} 
    \Bigg[\sum\limits_{b = j+1}^n\sum\limits_{a = j+1}^b \sum\limits_{\bm{r} \in \mathcal{P}: r_{o^0_j} = a, r_{o^0_{j+1}} = b} b \exp\{-\frac{\alpha}{n}d(\bm{r}, \bm{\rho}^0)\} \\
    &+\sum\limits_{b = j+1}^n\sum\limits_{a = j+1}^b \sum\limits_{\bm{r} \in \mathcal{P}: r_{o^0_j} = b, r_{o^0_{j+1}} = a} a \exp\{-\frac{\alpha}{n}d(\bm{r}, \bm{\rho}^0)\} \bigg] \\
    & = \frac{1}{Z_n(\alpha)} \sum\limits_{b=j+2}^n\sum\limits_{a=j+1}^b (a+b)\sum\limits_{\bm{r} \in \mathcal{P}: r_{o^0_j} = a, r_{o^0_j} = b} \exp\{-\frac{\alpha}{n}d(\bm{r}, \bm{\rho}^0)\} \\
    & = \sum\limits_{b=j+2}^n\sum\limits_{a=j+1}^b (a+b) P(\mathcal{F}_{a,b}|
    \bm{\rho}^0, \alpha)
\end{align*}

The second row in (\ref{eq:E_j}) and (\ref{eq:E_j+1}) are therefore the same.

Let us now consider $\mathcal{P}_B$ and $\mathcal{P}_{B'}$'s contributions to $\mathbb{E}[R_{o^0_{j}}|\bm{\rho}^{0}, \alpha]$ and $\mathbb{E}[R_{o^0_{j+1}}|\bm{\rho}^{0}, \alpha]$.

For any $r \in \mathcal{P}_B$ and its corresponding $r' \in \mathcal{P}_{B'}$ s.t. $r_{o^0_j} = r'_{o^0_{j+1}} = a$, $r_{o^0_{j+1}} = r'_{o^0_{j}} = b$, and $r_i = r'_i \forall i\neq o^0_j, o^0_{j+1}$, we also have $d(\bm{r}, \bm{\rho}^0)$ =  $d(\bm{r}^{'}, \bm{\rho}^0)$, since \\
\begin{align*}
      d(\bm{r}, \bm{\rho^0}) 
    & = \sum\limits_{i=1}^{j-1}|r_{o^0_i} - \rho^0_{o^0_i}| + |r_{o^0_j} - \rho^0_{o^0_j}| + |r_{o^0_{j+1}} - \rho^0_{o^0_{j+1}}| + \sum\limits_{i = j+2}^{n}|r_{o^0_i} - \rho^0_{o^0_{i}}|\\
    & = \sum\limits_{i \neq j, j+1} |r_{o^0_i} - i| + |a-j| + |b - (j+1)| \\
    & = \sum\limits_{i \neq j, j+1} |r_{o^0_i} - i| + j-a + (j+1)-b, \text{ and } \\ 
\end{align*}
\begin{align*}
    d(\bm{r'}, \bm{\rho^0}) 
    & = \sum\limits_{i=1}^{j-1}|r'_{o^0_i} - \rho^0_{o^0_i}| + |r'_{o^0_j} - \rho^0_{o^0_j}| + |r'_{o^0_{j+1}} - \rho^0_{o^0_{j+1}}| + \sum\limits_{i = j+2}^{n}|r'_{o^0_i} - \rho^0_{o^0_{i}}| \\
    & = \sum\limits_{i \neq j, j+1} |r'_{o^0_i} - i| + |b-j| + |a - (j+1)| \\
    & = \sum\limits_{i \neq j, j+1} |r_{o^0_i} - i| + j-b + (j+1)- a \\
    & = d(\bm{r}, \bm{\rho^0})
\end{align*}

Following similar logic, we can obtain that the third row in (\ref{eq:E_j}) and (\ref{eq:E_j+1}) are also the same.

Now let us consider $\mathcal{P}_C$ and $\mathcal{P}_{C'}$'s contributions to $\mathbb{E}[R_{o^0_{j}}]$ and $\mathbb{E}[R_{o^0_{j+1}}]$.

For any given permutation $\bm{r} \in \mathcal{P}_C$ and its corresponding $r' \in \mathcal{P}_{C'}$ s.t. $r_{o^0_j} = r'_{o^0_{j+1}} = a$, $r_{o^0_{j+1}} = r'_{o^0_{j}} = b$, and $r_l = r'_l $  $ \forall i\neq o^0_j, o^0_{j+1}$ , we have
\begin{align*}
    d(\bm{r}, \bm{\rho^0}) 
    &= \sum\limits_{i=1}^{j-1}|r_{o^0_i} - \rho^0_{o^0_i}| + |r_{o^0_j} - \rho^0_{o^0_j}| + |r_{o^0_{j+1}} - \rho^0_{o^0_{j+1}}| + \sum\limits_{i = j+2}^{n}|r_{o^0_i} - \rho^0_{o^0_{i}}| \\
    & = \sum\limits_{i \neq j, j+1} |r_{o^0_i} - i| + |a-j| + |b - (j+1)| \\
    & = \sum\limits_{i \neq j, j+1} |r_{o^0_i} - i| + j-a + b-(j+1) \\
    & = \sum\limits_{i \neq j, j+1} |r_{o^0_i} - i| + b-a -1, \text{ and }
\end{align*}

\begin{align*}
    d(\bm{r'}, \bm{\rho^0}) 
    & = \sum\limits_{i=1}^{j-1}|r'_{o^0_i} - \rho^0_{o^0_i}| + |r'_{o^0_j} - \rho^0_{o^0_j}| + |r'_{o^0_{j+1}} - \rho^0_{o^0_{j+1}}| + \sum\limits_{i = j+2}^{n}|r'_{o^0_i} - \rho^0_{o^0_{i}}|\\
    & = \sum\limits_{i \neq j, j+1} |r'_{o^0_i} - i| + |b-j| + |a - (j+1)| \\
    & = \sum\limits_{i \neq j, j+1} |r_{o^0_i} - i| + b-j + (j+1)- a \\
    & = \sum\limits_{i \neq j, j+1} |r_{o^0_i} - i| + b-a+1 \\
    & = d(\bm{r}, \bm{\rho^0}) +2. \\
\end{align*}

The last row of (\ref{eq:E_j}) can be derived as:
\begin{align*}
    & \textcolor{white}{=}\frac{1}{Z_n(\alpha)}\Bigg[ \sum\limits_{\bm{r}\in \mathcal{P}_C} r_{o^0_j}\text{exp}\{-\frac{\alpha}{n}d(\bm{r}, \bm{\rho^0})\} +\sum\limits_{\bm{r}\in \mathcal{P}_{C'}} r_{o^0_j}\text{exp}\{-\frac{\alpha}{n}d(\bm{r}, \bm{\rho^0})\} \Bigg] \\
    &= \frac{1}{Z_n(\alpha)} 
    \Bigg[\sum\limits_{b = j+1}^n\sum\limits_{a = 1}^j \sum\limits_{\bm{r} \in \mathcal{P}: r_{o^0_j} = a, r_{o^0_{j+1}} = b} a \exp\{-\frac{\alpha}{n}d(\bm{r}, \bm{\rho}^0)\} \\
    &+\sum\limits_{b = j+1}^n\sum\limits_{a = 1}^j \sum\limits_{\bm{r} \in \mathcal{P}: r_{o^0_j} = b, r_{o^0_{j+1}} = a} b \exp\{-\frac{\alpha}{n}d(\bm{r}, \bm{\rho}^0)\} \bigg] \\
    & = \frac{1}{Z_n(\alpha)} \sum\limits_{b=j+1}^n\sum\limits_{a=1}^j \sum\limits_{\bm{r} \in \mathcal{P}: r_{o^0_j} = b, r_{o^0_{j+1}} = a} a\exp\{-\frac{\alpha}{n}d(\bm{r}, \bm{\rho}^0)\}+  b \exp\{-\frac{2\alpha}{n}\}\exp\{-\frac{\alpha}{n}d(\bm{r}, \bm{\rho}^0)\}\\
    & = \sum\limits_{b=j+1}^n\sum\limits_{a=1}^j (a+b\exp\{-\frac{2\alpha}{n}\}) P(\mathcal{F}_{a,b}|\bm{\rho}^0, \alpha).
\end{align*}
Similarly, we can obtain that the last row of (\ref{eq:E_j+1}) is
\begin{align*}
    & \textcolor{white}{=}\frac{1}{Z_n(\alpha)}\Bigg[ \sum\limits_{\bm{r}\in \mathcal{P}_C} r_{o^0_{j+1}}\text{exp}\{-\frac{\alpha}{n}d(\bm{r}, \bm{\rho^0})\} +\sum\limits_{\bm{r}\in \mathcal{P}_{C'}} r_{o^0_{j+1}}\text{exp}\{-\frac{\alpha}{n}d(\bm{r}, \bm{\rho^0})\} \Bigg] \\
    & = \sum\limits_{b=j+1}^n\sum\limits_{a=1}^j (b+a\exp\{-\frac{2\alpha}{n}\}) P(\mathcal{F}_{a,b}|\bm{\rho}^0, \alpha).
\end{align*}

Hence, the difference between $\mathbb{E}[R_{o^0_{j+1}}|\bm{\rho}^0, \alpha]$ and $\mathbb{E}[R_{o^0_{j}}|\bm{\rho}^0, \alpha]$ reduces to:

\begin{align*}
    \mathbb{E}[R_{o^0_{j+1}}|\bm{\rho}^0, \alpha] - \mathbb{E}[R_{o^0_{j}}|\bm{\rho}^0, \alpha] & = \sum\limits_{b=j+1}^{n}\sum\limits_{a=1}^{j} (b-a)(1-\exp\{-\frac{2\alpha}{n}\})P(\mathcal{F}_{a,b}|\bm{\rho}^0, \alpha).
\end{align*}

It can be easily observed that every term in the above equation are positive for all $\alpha > 0$, and we can conclude that $\mathbb{E}[R_{o^0_{j+1}}|\bm{\rho}^0, \alpha] > \mathbb{E}[R_{o^0_{j}}|\bm{\rho}^0, \alpha]$ for all $\alpha >0$.

\end{proof}

\end{appendices}

% Note: in this sample, the section number is hard-coded in. Following
% proper LaTeX conventions, it should properly be coded as a reference:

%In this appendix we prove the following theorem from
%Section~\ref{sec:textree-generalization}:

% In this appendix we prove the following theorem from
% Section~6.2:

% \noindent
% {\bf Theorem} {\it Let $u,v,w$ be discrete variables such that $v, w$ do
% not co-occur with $u$ (i.e., $u\neq0\;\Rightarrow \;v=w=0$ in a given
% dataset $\dataset$). Let $N_{v0},N_{w0}$ be the number of data points for
% which $v=0, w=0$ respectively, and let $I_{uv},I_{uw}$ be the
% respective empirical mutual information values based on the sample
% $\dataset$. Then
% \[
% 	N_{v0} \;>\; N_{w0}\;\;\Rightarrow\;\;I_{uv} \;\leq\;I_{uw}
% \]
% with equality only if $u$ is identically 0.} \hfill\BlackBox

% \noindent
% {\bf Proof}. We use the notation:
% \[
% P_v(i) \;=\;\frac{N_v^i}{N},\;\;\;i \neq 0;\;\;\;
% P_{v0}\;\equiv\;P_v(0)\; = \;1 - \sum_{i\neq 0}P_v(i).
% \]
% These values represent the (empirical) probabilities of $v$
% taking value $i\neq 0$ and 0 respectively.  Entropies will be denoted
% by $H$. We aim to show that $\fracpartial{I_{uv}}{P_{v0}} < 0$....\\

% {\noindent \em Remainder omitted in this sample. See http://www.jmlr.org/papers/ for full paper.}

\vskip 0.2in
\bibliography{reference}

\end{document}